\documentclass[a4paper,10pt]{article}

\usepackage[top=1in,right=0.7in,left=0.7in,bottom=1in]{geometry}

\usepackage{latexsym}
\usepackage{amssymb}
\usepackage{float}
\usepackage{graphics,graphicx}
\usepackage{tabularx}
\usepackage{amsfonts}
\usepackage{amsmath}
\usepackage{empheq}
\usepackage{enumerate}
\usepackage{booktabs}
\usepackage{multirow}
\usepackage{sectsty}
\usepackage[affil-it,auth-sc]{authblk}
\usepackage{subfig}
\usepackage{color}
\usepackage{mathtools}
\usepackage{mathrsfs}
\usepackage{bmpsize}
\usepackage{xcolor}
\usepackage{siunitx}
\usepackage{bm}
\usepackage{breqn}

\sectionfont{\normalsize}
\subsectionfont{\normalsize}

\begin{document}

\newcommand{\singlespace}{\baselineskip=12pt\lineskiplimit=0pt\lineskip=0pt}
\def\ds{\displaystyle}

\newcommand{\beq}{\begin{equation}}
\newcommand{\eeq}{\end{equation}}
\newcommand{\lb}{\label}
\newcommand{\ph}{\phantom}
\newcommand{\beqar}{\begin{eqnarray}}
\newcommand{\eeqar}{\end{eqnarray}}
\newcommand{\barr}{\begin{array}}
\newcommand{\earr}{\end{array}}
\newcommand{\jump}{\parallel}
\newcommand{\Ehat}{\hat{E}}
\newcommand{\That}{\hat{\bf T}}
\newcommand{\Ahat}{\hat{A}}
\newcommand{\chat}{\hat{c}}
\newcommand{\shat}{\hat{s}}
\newcommand{\khat}{\hat{k}}
\newcommand{\muhat}{\hat{\mu}}
\newcommand{\mc}{M^{\scriptscriptstyle C}}
\newcommand{\mei}{M^{\scriptscriptstyle M,EI}}
\newcommand{\mec}{M^{\scriptscriptstyle M,EC}}
\newcommand{\hbeta}{{\hat{\beta}}}
\newcommand{\rec}[2]{\left( #1 #2 \ds{\frac{1}{#1}}\right)}
\newcommand{\rep}[2]{\left( {#1}^2 #2 \ds{\frac{1}{{#1}^2}}\right)}
\newcommand{\derp}[2]{\ds{\frac {\partial #1}{\partial #2}}}
\newcommand{\derpn}[3]{\ds{\frac {\partial^{#3}#1}{\partial #2^{#3}}}}
\newcommand{\dert}[2]{\ds{\frac {d #1}{d #2}}}
\newcommand{\dertn}[3]{\ds{\frac {d^{#3} #1}{d #2^{#3}}}}
\newcommand{\ct}{\captionof{table}}
\newcommand{\cf}{\captionof{figure}}

\def\c{{\circ}}
\def\bob{{\, \underline{\overline{\otimes}} \,}}
\def\ob{{\, \underline{\otimes} \,}}
\def\scalp{\mbox{\boldmath$\, \cdot \, $}}
\def\gdp{\makebox{\raisebox{-.215ex}{$\Box$}\hspace{-.778em}$\times$}}
\def\daa{\makebox{\raisebox{-.050ex}{$-$}\hspace{-.550em}$: ~$}}
\def\mK{\mbox{${\mathcal{K}}$}}
\def\cK{\mbox{${\mathbb {K}}$}}

\def\Xint#1{\mathchoice
   {\XXint\displaystyle\textstyle{#1}}%
   {\XXint\textstyle\scriptstyle{#1}}%
   {\XXint\scriptstyle\scriptscriptstyle{#1}}%
   {\XXint\scriptscriptstyle\scriptscriptstyle{#1}}%
   \!\int}
\def\XXint#1#2#3{{\setbox0=\hbox{$#1{#2#3}{\int}$}
     \vcenter{\hbox{$#2#3$}}\kern-.5\wd0}}
\def\ddashint{\Xint=}
\def\fpint{\Xint=}
\def\dashint{\Xint-}
\def\cpvint{\Xint-}
\def\intl{\int\limits}
\def\cpvintl{\cpvint\limits}
\def\fpintl{\fpint\limits}
\def\ointl{\oint\limits}
\def\bA{{\bf A}}
\def\ba{{\bf a}}
\def\bB{{\bf B}}
\def\bb{{\bf b}}
\def\bc{{\bf c}}
\def\bC{{\bf C}}
\def\bD{{\bf D}}
\def\bE{{\bf E}}
\def\be{{\bf e}}
\def\bbf{{\bf f}}
\def\bF{{\bf F}}
\def\bG{{\bf G}}
\def\bg{{\bf g}}
\def\bi{{\bf i}}
\def\bH{{\bf H}}
\def\bK{{\bf K}}
\def\bL{{\bf L}}
\def\bM{{\bf M}}
\def\bN{{\bf N}}
\def\bn{{\bf n}}
\def\b0{{\bf 0}}
\def\bo{{\bf o}}
\def\bX{{\bf X}}
\def\bx{{\bf x}}
\def\bP{{\bf P}}
\def\bp{{\bf p}}
\def\bQ{{\bf Q}}
\def\bq{{\bf q}}
\def\bbr{{\bf r}}
\def\bR{{\bf R}}
\def\bS{{\bf S}}
\def\bs{{\bf s}}
\def\bT{{\bf T}}
\def\bt{{\bf t}}
\def\bU{{\bf U}}
\def\bu{{\bf u}}
\def\bv{{\bf v}}
\def\bw{{\bf w}}
\def\bW{{\bf W}}
\def\by{{\bf y}}
\def\bz{{\bf z}}
\def\T{{\bf T}}
\def\Te{\textrm{T}}
\def\Id{{\bf I}}
\def\bxi{\mbox{\boldmath${\xi}$}}
\def\balpha{\mbox{\boldmath${\alpha}$}}
\def\bbeta{\mbox{\boldmath${\beta}$}}
\def\bepsilon{\mbox{\boldmath${\epsilon}$}}
\def\bvarepsilon{\mbox{\boldmath${\varepsilon}$}}
\def\bomega{\mbox{\boldmath${\omega}$}}
\def\bphi{\mbox{\boldmath${\phi}$}}
\def\bsigma{\mbox{\boldmath${\sigma}$}}
\def\bfeta{\mbox{\boldmath${\eta}$}}
\def\bDelta{\mbox{\boldmath${\Delta}$}}
\def\btau{\mbox{\boldmath $\tau$}}
\def\tr{{\rm tr}}
\def\dev{{\rm dev}}
\def\div{{\rm div}}
\def\Div{{\rm Div}}
\def\Grad{{\rm Grad}}
\def\grad{{\rm grad}}
\def\Lin{{\rm Lin}}
\def\Sym{{\rm Sym}}
\def\Skw{{\rm Skew}}
\def\abs{{\rm abs}}
\def\Re{{\rm Re}}
\def\Im{{\rm Im}}
\def\capB{\mbox{\boldmath${\mathsf B}$}}
\def\capC{\mbox{\boldmath${\mathsf C}$}}
\def\capD{\mbox{\boldmath${\mathsf D}$}}
\def\capE{\mbox{\boldmath${\mathsf E}$}}
\def\capG{\mbox{\boldmath${\mathsf G}$}}
\def\tcapG{\tilde{\capG}}
\def\capH{\mbox{\boldmath${\mathsf H}$}}
\def\capK{\mbox{\boldmath${\mathsf K}$}}
\def\capL{\mbox{\boldmath${\mathsf L}$}}
\def\capM{\mbox{\boldmath${\mathsf M}$}}
\def\capR{\mbox{\boldmath${\mathsf R}$}}
\def\capW{\mbox{\boldmath${\mathsf W}$}}


\def\salto#1#2{%
	[\mbox{\hspace{-#1em}}[#2]\mbox{\hspace{-#1em}}]}

\def\i{\mbox{${\mathrm i}$}}
\def\mC{\mbox{\boldmath${\mathcal C}$}}
\def\mB{\mbox{${\mathcal B}$}}
\def\mE{\mbox{${\mathcal{E}}$}}
\def\mL{\mbox{${\mathcal{L}}$}}
\def\mK{\mbox{${\mathcal{K}}$}}
\def\mV{\mbox{${\mathcal{V}}$}}
\def\C{\mbox{\boldmath${\mathcal C}$}}
\def\E{\mbox{\boldmath${\mathcal E}$}}

\def\AAM{{\it Advances in Applied Mechanics }}
\def\ACME{{\it Arch. Comput. Meth. Engng.}}
\def\ARMA{{\it Arch. Rat. Mech. Analysis}}
\def\AMR{{\it Appl. Mech. Rev.}}
\def\ASCEEM{{\it ASCE J. Eng. Mech.}}
\def\ACTA{{\it Acta Mater.}}
\def\CMAME {{\it Comput. Meth. Appl. Mech. Engrg.}}
\def\CRAS{{\it C. R. Acad. Sci. Paris}}
\def\CRM{{\it Comptes Rendus M\'ecanique}}
\def\EFM{{\it Eng. Fracture Mechanics}}
\def\EJMA{{\it Eur.~J.~Mechanics-A/Solids}}
\def\IJES{{\it Int. J. Eng. Sci.}}
\def\IJF{{\it Int. J. Fracture}}
\def\IJMS{{\it Int. J. Mech. Sci.}}
\def\IJNAMG{{\it Int. J. Numer. Anal. Meth. Geomech.}}
\def\IJP{{\it Int. J. Plasticity}}
\def\IJSS{{\it Int. J. Solids Structures}}
\def\IngA{{\it Ing. Archiv}}
\def\JAM{{\it J. Appl. Mech.}}
\def\JAP{{\it J. Appl. Phys.}}
\def\JAE{{\it J. Aerospace Eng.}}
\def\JE{{\it J. Elasticity}}
\def\JM{{\it J. de M\'ecanique}}
\def\JMPS{{\it J. Mech. Phys. Solids}}
\def\JSV{{\it J. Sound and Vibration}}
\def\MACRO{{\it Macromolecules}}
\def\MMT{{\it Mech. Mach. Th.}}
\def\MOM{{\it Mech. Materials}}
\def\MMS{{\it Math. Mech. Solids}}
\def\MMT{{\it Metall. Mater. Trans. A}}
\def\MPCPS{{\it Math. Proc. Camb. Phil. Soc.}}
\def\MSE{{\it Mater. Sci. Eng.}}
\def\NATURE{{\it Nature}}
\def\NATUREM{{\it Nature Mater.}}
\def\PHIL{{\it Phil. Trans. R. Soc.}}
\def\PMPS{{\it Proc. Math. Phys. Soc.}}
\def\PNAS{{\it Proc. Nat. Acad. Sci.}}
\def\PRE{{\it Phys. Rev. E}}
\def\PRL{{\it Phys. Rev. Letters}}
\def\PRSL{{\it Proc. R. Soc.}}
\def\RIIT{{\it Rozprawy Inzynierskie - Engineering Transactions}}
\def\ROCK{{\it Rock Mech. and Rock Eng.}}
\def\QAM{{\it Quart. Appl. Math.}}
\def\QJMAM{{\it Quart. J. Mech. Appl. Math.}}
\def\SCIENCE{{\it Science}}
\def\SCRMAT{{\it Scripta Mater.}}
\def\SM{{\it Scripta Metall.}}
\def\ZAMM{{\it Z. Angew. Math. Mech.}}
\def\ZAMP{{\it Z. Angew. Math. Phys.}}
\def\ZVDI{{\it Z. Verein. Deut. Ing.}}

\def\salto#1#2{
[\mbox{\hspace{-#1em}}[#2]\mbox{\hspace{-#1em}}]}

\renewcommand\Affilfont{\itshape}
\setlength{\affilsep}{1em}
\renewcommand\Authsep{, }
\renewcommand\Authand{ and }
\renewcommand\Authands{ and }
\setcounter{Maxaffil}{2}



\title{ Non-holonomic constraints inducing flutter instability in structures under conservative loadings \\
}

\author{Alessandro Cazzolli, Francesco Dal Corso, Davide Bigoni\footnote{Corresponding author: Davide Bigoni fax:
+39 0461 282599; tel.: +39 0461 282507; web-site:
http://www.ing.unitn.it/$\sim$bigoni/; e-mail: bigoni@ing.unitn.it.}}
 \affil[]{DICAM, University of Trento, via~Mesiano~77, I-38123
Trento, Italy.}

\date{}
\maketitle 

\date{}
\maketitle

\begin{abstract}
	
Non-conservative loads of the follower type are usually believed to be the source  of dynamic instabilities such as flutter and divergence. 
It is shown that these instabilities (including Hopf bifurcation, flutter, divergence, and destabilizing effects connected to dissipation phenomena) can be 
obtained in structural systems loaded by conservative forces, as a consequence of the application of non-holonomic constraints. 
These constraints may be realized through a \lq perfect skate' (or a non-sliding wheel), or, more in general, through the slipless contact between two circular rigid cylinders, one of which is free of rotating about its axis. 
The motion of the structure produced by these dynamic instabilities may reach a limit cycle, a feature that can be exploited for soft robotics applications, especially for the realization of limbless locomotion. 
\end{abstract}


\section{Introduction}

Flutter and divergence (in a word, \lq dynamic') instabilities of elastic structures are connected to a number of counterintuitive and surprising features: (i.) they may occur in the absence of 
quasi-static bifurcations, and may or may not degenerate into a limit cycle, determining a so-called \lq Hopf bifurcation' in the former case; (ii.) they are facilitated by a sufficiently small viscosity. Moreover,  (iii.) a vanishing  viscosity leads to a discontinuity in the critical load value, an oddity called  \lq Ziegler paradox', so that (iv.) this value may depend on the direction of the vanishing viscosity limit, when more than one dissipative source is present. 
All these features and the presence itself of flutter and divergence instabilities in structures are believed to be strictly related to the action of non-conservative loads,\footnote{
Bolotin \cite{bolotin} writes \lq The Euler method is [only] applicable if the external forces have a potential (i.e. if they are conservative forces), and in general is not applicable if they do not.'
} which are considered of difficult realization.\footnote{
The attempt by Willems \cite{willems} of producing a tangential follower load was indicated as misleading by Huang et al. \cite{huang}. Elishakoff \cite{elishakoff} reports 
that \lq Bolotin felt --if my
memory serves me well!-- that it should be impossible to produce
Beck’s column experiment via a conservative system of forces.' 
Anderson and Done \cite{anderson} write \lq 
Sometimes, the creation of a force like [a follower force] in
the laboratory presents awkward practical problems, and the
simulation of this force wherever possible by a conservative
force would be very convenient. However, because of the differing nature of the fundamental properties of the conservative and
non-conservative systems, the simulation could only work in a
situation where the two systems behave in similar ways; that is
when the conservative system is not operating in a regime of
oscillatory instability. (The conservative system can not become
dynamically instable since, by definition, it has no energy source
from which to supply the extra kinetic energy involved in the
instability).'
Koiter \cite{koiter3} states that \lq [...] it appears impossible to achieve any non-conservative loading conditions in the laboratory by purely mechanical means', because  
\lq non-conservative external loads always require an external energy source, much as a fluid flow or an interaction with electro-dynamic phenomenon'. 
Koiter was  
strongly convinced that follower forces were a sort of \lq physical non-sense' (Koiter, \cite{koiter2}, \cite{koiter}), so that 
Singer et al. \cite{singer} write \lq An example in the field of elastic stability of what Drucker
referred to as {\it playing useless games} was presented by Koiter,
in his 1985 Prandtl lecture, where he discussed the physical significance of instability due to non-conservative, purely
configuration-dependent, external loads.' Several years after these negative views, Bigoni and Noselli \cite{bigoninoselli} and Bigoni and Misseroni \cite{bigonimisseroni} respectively showed how to realize a tangentially follower force 
(Ziegler, 1952 \cite{ziegpara}) and a fixed-line force (Reut, 1939 \cite{reut}) with devices involving Coulomb friction. Friction is a dissipative and follower force, in the present article we show how this force can be eliminated thus realizing a purely conservative system of load. 
}
Indeed, quoting Anderson and Done \cite{anderson}, a \lq conservative system can not become
dynamically unstable since, by definition, it has no energy source
from which to supply the extra kinetic energy involved in the
instability'. 

We show in the present article that the last sentence is wrong, as a supply of \lq extra kinetic energy' can be provided by means of a mass falling within a gravitational field, or through a release of the elastic energy initially stored in a spring and therefore:  

\begin{quote}
{\it We introduce two ways to induce dynamical instablities (such as flutter and divergence), Hopf bifurcation, destabilizing effect of dissipation, Ziegler paradox and its directionality dependence in visco-elastic mechanical structures under purely conservative loading conditions. The key to this behaviour is the use of non-holonomic constraints.} 
\end{quote}

\begin{figure}[!htb]
	\centering
		\includegraphics[width=0.4\textwidth]{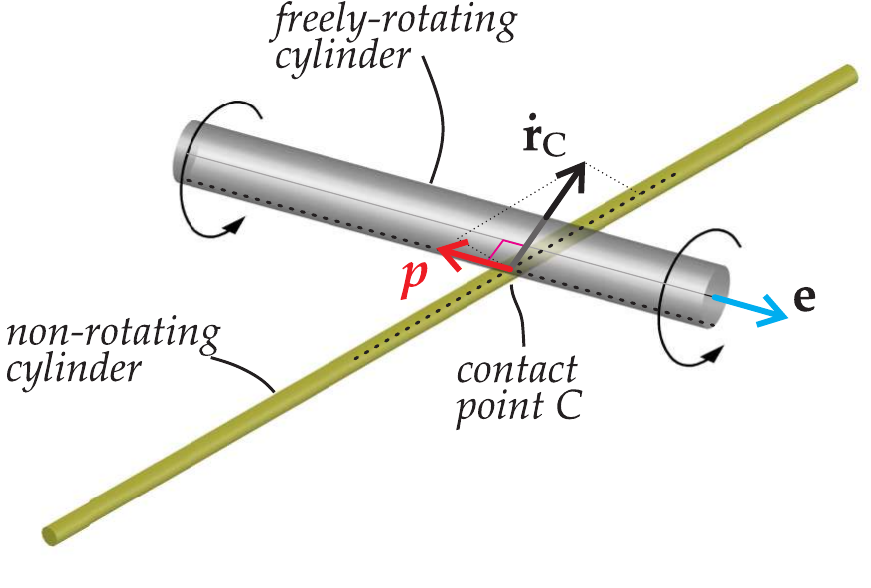}
	\caption{A non-holonomic constraint forbidding relative velocity $\dot{\bbr}_C$ of the instantaneous contact point $C$  along direction $\be(t)$, Eq.(\ref{numerouno}). This constraint is obtained through slipless contact between two rigid (massless and circular) cylinders, one of which is free of rotating about its axis while the other is not. 
		Note that the reaction $p(t)$ transmitted between the two cylinders is aligned parallel to $\be$, so that it provides null work during every motion and the conservativeness of a system is not altered.
	}
	\label{figotta}
\end{figure}

Recently, flutter and divergence instabilities have been shown to occur in structures loaded through non-conservative forces produced with frictional devices 
by Bigoni and Noselli \cite{bigoninoselli},
 realizing the tangentially follower load postulated by Ziegler \cite{ziegpara}, and by Bigoni and Misseroni \cite{bigonimisseroni}, obtaining the fixed-line load introduced by Reut \cite{reut}. The same type of instabilities is disclosed here in structural systems subject to conservative loadings when non-holonomic constraints are applied to the structure. 
More specifically, similarly to \cite{jar,kuleshov,neimark}, the non-holonomic constraint is realized through two rigid (massless) cylinders in {\it slipless contact}, where one of these cylinders can freely rotate about its axis while the other can not.
The non-existence of a potential \cite{lanczos,neimark} characterizes both a frictional device and a non-holonomic constraint, but only the latter is conservative. Indeed, the non-holonomic constraint introduces a kinematic condition prescribed in terms of velocity, thus realizing  a (reaction) force acting on the structure and having a modulus varying in time during motion \cite{ruina}. In particular, the velocity $\dot{\bbr}_C$ of the instantaneous contact point $C$ between the two cylinders is constrained to have a null component along the axis of the freely rotating cylinder (Fig. \ref{figotta}) 
\begin{equation}
\label{numerouno}
\dot \bbr_C\scalp \be=0, 
\end{equation}
where a superimposed dot represents the derivative in the time variable $t$ and $\be(t)$ is the unit vector aligned parallel to the rotating cylinder's axis.
The reaction $p(t)$ transmitted by the constraint is parallel to $\be$, so that equation (\ref{numerouno}) makes clear that this reaction does not work for every possible displacement. This means that {\it every mechanical system composed of elastic elements, dead loads, and (time-independent) non-holonomic constraints  is conservative}.

\begin{figure}[!htb]
	\begin{center}
		\includegraphics[width=0.95\textwidth]{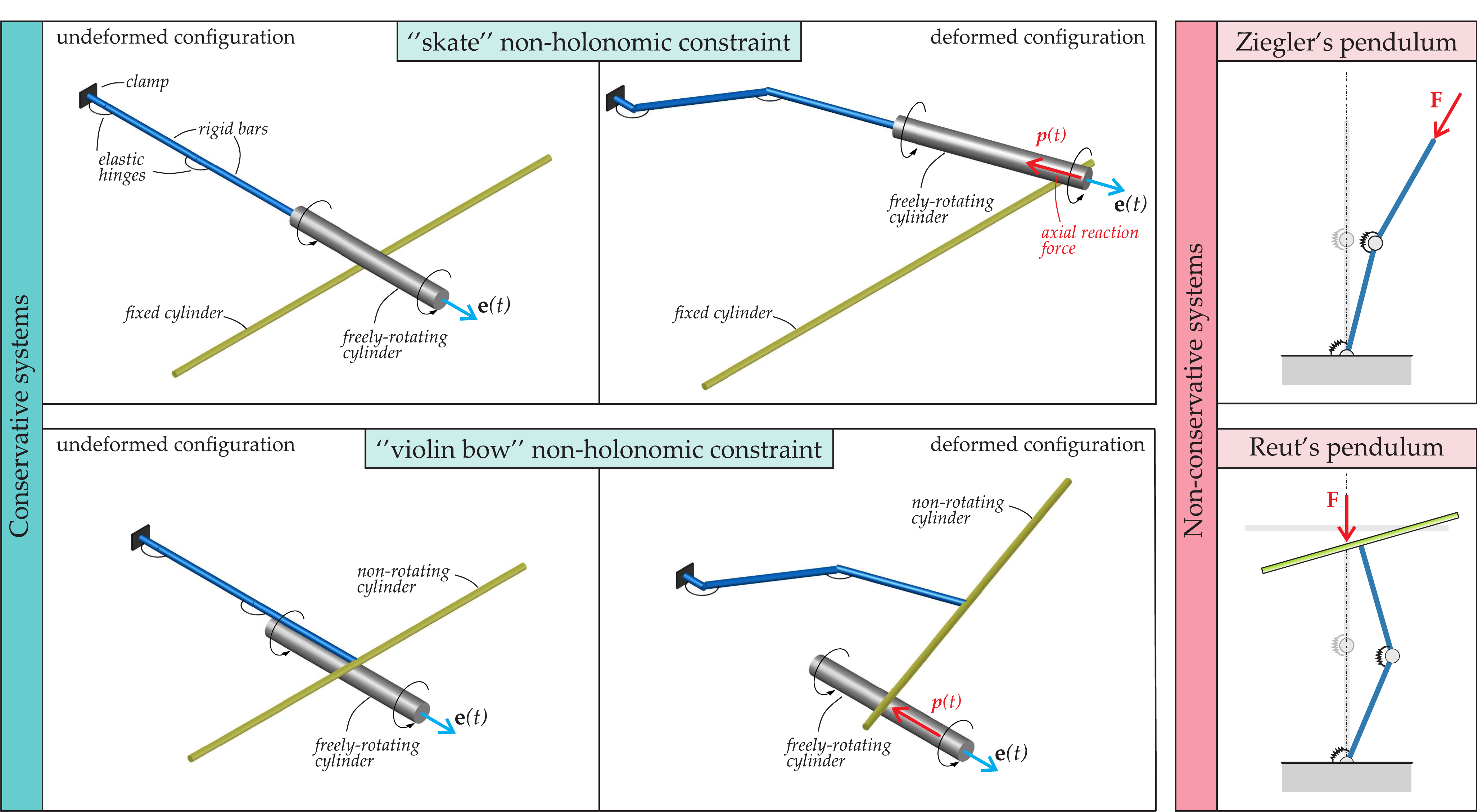}
	\end{center}
	\caption{Left: the two rigid circular cylinders (of which only one is free of rotating about its axis) in slipless contact, Fig.\ref{figotta}, and connected to a double pendulum, realize two types of non-holonomic constraints, called in the following  
		\lq skate' and \lq violin bow' constraints, the former (the latter) transmitting a reaction tangential to the structure (a reaction on a fixed line) similar to the force postulated by Ziegler (Reut). Right: the original version of the Ziegler's and Reut's structures, loaded through non-conservative forces.}
	\label{figuna}
\end{figure}

When properly connected to the end of a structure, for instance a visco-elastic double pendulum, the two cylinders imposing condition (\ref{numerouno}) may be exploited to realize either a \lq skate', or a \lq violin bow' constraint, the former transmitting to the structure a tangential follower reaction similar to the Ziegler's load and 
the latter a reaction acting on a fixed line similar to the Reut's load, but now varying in their modulus during motion. In Fig.\ref{figuna}, the two proposed structures with non-holonomic constraints (which will be subject to conservative loads) are shown on the left, while the non-conservative counterparts are shown on the right.
In particular, the \lq skate' constraint can be realized (i.) with a freely-rotating, but non-sliding wheel, or (ii.) with a perfect skate, or (iii.) by connecting the freely-rotating cylinder to the end of the structure, with the other cylinder initially orthogonal to it and fixed in space. On the other hand, the \lq violin bow' constraint can be obtained by connecting the fixed cylinder to the end of the structure, with the freely-rotating cylinder initially orthogonal to it and fixed in space, Fig.\ref{figuna}. Note that the structural systems subject to non-holonomic constraints are conservative, when loaded with dead forces or springs and if the viscosity is set to be zero.

In the present article, discrete structural systems made up of $N$ rigid bars and subject to one of the two proposed types of non-holonomic constraints (Fig.\ref{figuna}, left) are investigated. Such structures are loaded in a conservative way, namely, either with a dead force $F$, or with a deformed linear spring (see Figs.\ref{fig_discrete0} and \ref{fig_discrete} and details presented in the next Sections).

The non-holonomic constraint acting on these structures permits the existence of infinite equilibrium configurations \cite{neimark} and of Hopf bifurcations connected to stable limit cycles in the presence of dissipation, as partially anticipated by the motion of the non-holonomic Chaplygin's sleigh on an inclined plane \cite{neimark} and by the problem of shimmy instability \cite{beregi1,beregi2,facchini,ziegler}.

With reference to a column made up of $N$ rigid segments and subject to a perfectly aligned non-holonomic constraint (realized for $\beta_0=0$ in Figs.\ref{fig_discrete0} and \ref{fig_discrete}), the trivial configuration becomes the unique quasi-static solution. It is shown that: 

\begin{quote}
	{\it 
	A visco-elastic column subject to the \lq skate' or the \lq violin bow' constraint 
	and the same structure subject to non-conservative load, respectively of the Ziegler or Reut type,  
		evidences exactly the same critical loads for flutter (Hopf bifurcation) and divergence, and the same Ziegler paradox, with the same directionality dependence occurring when multiple sources of dissipation are considered.
	} 
\end{quote}
More specifically, the directionality dependence in the limit of null viscosity has been analyzed by Bolotin \cite{bolotin} and Kirillov \cite{kirillov_1} for non-conservative systems subject to follower forces, showing that the ideal critical load for the undamped system may be recovered only for special ratios between different viscosities.

In addition, a \emph{\lq viscosity-independent Ziegler paradox'} is found, in which the flutter load  becomes independent of the viscosity, but cannot become higher than that evaluated for the corresponding system assumed without viscosity \lq from the beginning'. In particular, such a surprising behaviour is shown to be related to the presence of two specific damping parameters acting on the proposed structures when subject to non-holonomic constraints.

The stability of Hopf bifurcation, and therefore the achievement of limit cycles in the neighbourhood of the critical point \cite{agostinelli, bigonikiri, bigonimiss22, jenkins}, is influenced by the considered mechanical system, so that a perfect match between the mechanical behaviour of the proposed non-holonomic systems and of their non-conservative counterparts is lost, particularly when non-linearities dominate.

\begin{figure}[!htb]
	\begin{center}
		\includegraphics[width=1\textwidth]{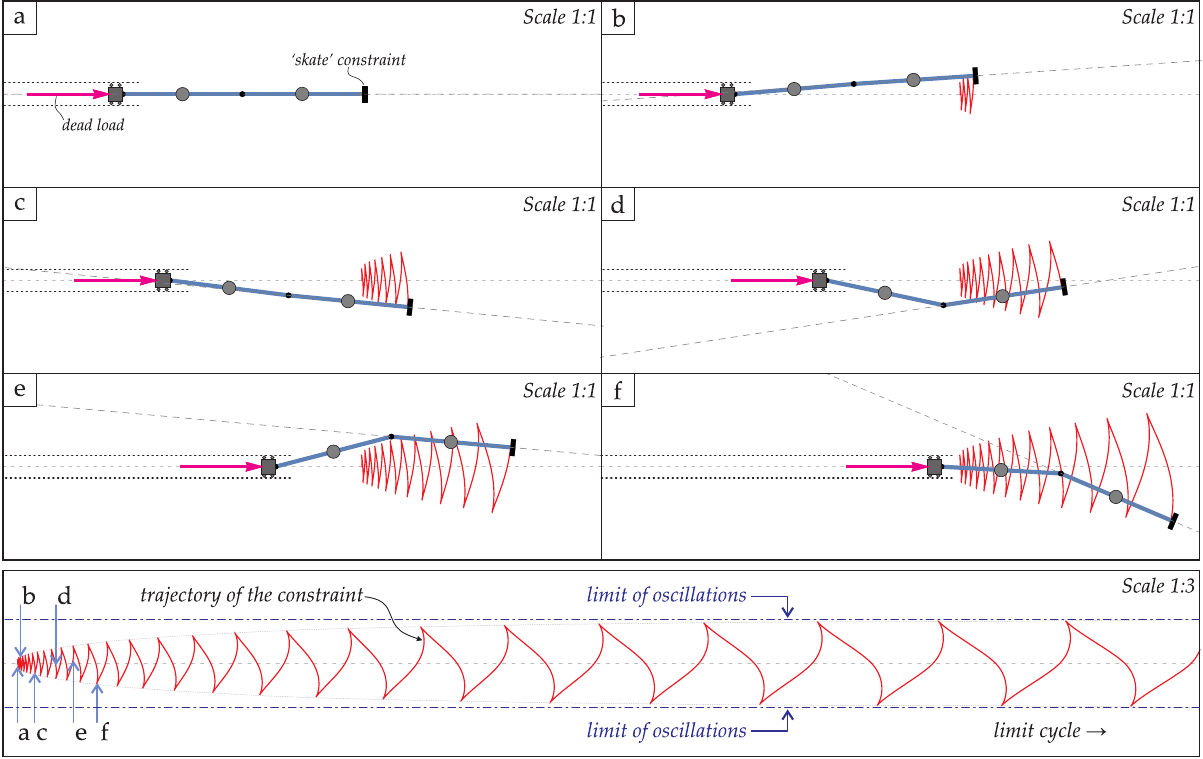}
	\end{center}
	\caption{Dynamics of a visco-elastic double pendulum subject to a \lq skate' non-holonomic constraint (realized with a non-sliding wheel) at one end and to a dead load at the other. 
		When the applied dead force lies within  the flutter region, a complex motion is generated (parts a-f), leading the structure to reach a limit cycle. Such behaviour is visible in the trajectory of the structure's end shown in the lower part, exhibiting sharp corners induced by the non-holonomic constraint.  
		Note  that instability permits motion to a structure that would be at rest in the trivial configuration as the friction parallel to the wheel axis is assumed to be infinite (non-slip condition).
}
	\label{fig_evolution}
\end{figure}

A typical dynamic evolution of a double visco-elastic column connected to a non-holonomic \lq skate' constraint at its final end (represented as a non-sliding wheel) is shown in Fig.\ref{fig_evolution}. The structure is loaded on the left end through a dead force of constant magnitude (so that the system is conservative when the viscosity is set to zero), selected within the flutter region. 
The occurrence of flutter is shown, through evaluation 
of the first Lyapunov coefficient \cite{kutsnetsov,marsden}, to correspond to a 
supercritical Hopf bifurcation, so that the complex motion taking place after bifurcation [sketched in parts (a)-(f) of Fig.\ref{fig_evolution}] reaches a periodic orbit in the neighbourhood of the bifurcation point (lowest  part of Fig.\ref{fig_evolution}). It may be interesting to note that the structure shown in Fig. \ref{fig_evolution} is subject to a constant force, so that it would suffer a varying acceleration in the absence of viscosity. However, {\it the presence of viscosity (even the rotational viscosity at the hinges is enough) is sufficient to allow the mechanical system to reach, after a transient phase, a steady motion with a constant mean velocity,} as detailed in Section \ref{stokaz}. 
	
The presented results provide a new key to theoretically interpret and experimentally realize dynamic instabilities until now believed to be possible only as connected to 
non-conservative loads. Several applications in energy harvesting and soft robotics can be envisaged, but two in particular merit a special attention, namely, frictional contact and locomotion. 
In particular, our results show how a micromechanism might act at a sliding surface between two solids to reduce friction through instability. 
Moreover, it will be shown that imposing {\it rotations} (instead than applying forces) to the structures that will be analyzed, a motion is induced, which means that our results have implications in the problem of {\it limbless locomotion}.

\section{A first insight on flutter instability of a conservative system from a double pendulum with non-holonomic constraint and dead load}

\begin{figure}[!htb]
	\begin{center}
		\includegraphics[width=0.5\textwidth]{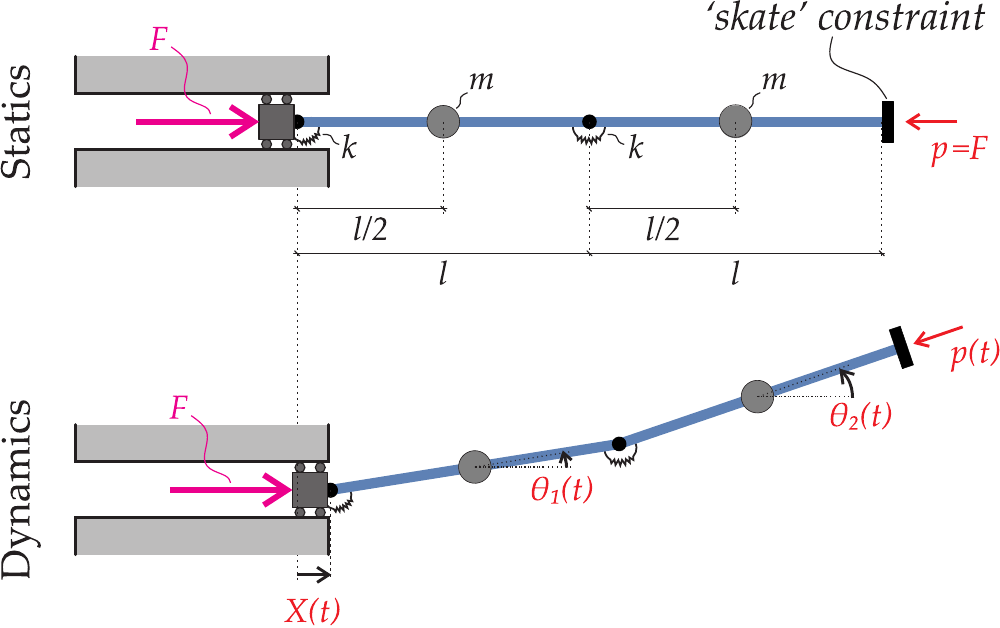}
	\end{center}
	\caption{An elastic double-pendulum (in its quasi-static trivial equilibrium, upper part, and in a deformed configuration during motion, lower part)  subject to a \lq skate' non-holonomic constraint at its right end and a dead load $F$ applied on its left end. The mechanical system is conservative because the sum of elastic and kinetic energies as well as the potential of the load $F$ remains constant in time. The non-holonomic constraint at the final end transmits an unknown time-dependent reaction force $p(t)$, which is maintained tangential to the second rod, as in the Ziegler case.}
	\label{def0}
\end{figure}

With the purpose to provide  a first insight on flutter instabilities in conservative non-holonomic systems, an elastic double pendulum (two rigid bars of equal length $l$, each equipped with a centred mass $m$ and elastic hinges of stiffness $k$) is considered as sketched in Fig.\ref{def0}. The structure is loaded on its left end with a dead force $F$,  while at its right end is subject to a \lq skate' non-holonomic constraint, which can be realized for instance with a freely rotating and massless wheel. 

The planar kinematics of the system is completely described by the two bars' rotations $\theta_1$ and $\theta_2$ and the horizontal displacement $X$ of the left end (Fig.\ref{def0}, lower part). The reaction $p$ of the \lq skate' remains tangential to the second bar. In the quasi-static trivial equilibrium, the structure is straight (Fig.\ref{def0}, upper part), so that the \lq skate' constraint  reacts with an horizontal force equal to the applied load $F$.\\ The  motion is governed by the following non-linear ordinary differential equations\footnote{
\textcolor{red}{Equations (\ref{conserv}) are highlighted in red because they correct the corresponding equations (2) in the published version of the present article.
The different terms, belonging to an early incorrect version, are found to be irrelevant for the results and conclusion drawn in the paper.
}
}
\textcolor{red}{
\begin{subequations}
\label{conserv}
\begin{empheq}[left=\empheqlbrace]{align}
&k\left(2\theta_1-\theta_2\right)+\frac{ m \,l}{4}\left[l \left(5 \ddot{\theta}_1+2 \ddot{\theta}_2 \cos (\theta_1-\theta_2)+2 \dot{\theta}_2^2 \sin (\theta_1-\theta_2)\right)-6 \ddot{X}\sin \theta_1\right] -p\,l \sin (\theta_1-\theta_2)=0,\\[1ex]
&k \left(\theta_2- \theta_1\right)+\frac{m\, l}{4}  \left[l \left(2 \ddot{\theta}_1 \cos (\theta_1-\theta_2)-2 \dot{\theta}_1^2 \sin (\theta_1-\theta_2)+\ddot{\theta}_2\right)-2 \sin \theta_2 \ddot{X}\right]=0,\\[1ex]
&2\, m \ddot{X}-\frac{m\, l}{2} \left(3 \ddot{\theta}_1 \sin \theta_1+3 \dot{\theta}_1^2 \cos\theta_1+\ddot{\theta}_2 \sin \theta_2+\dot{\theta}_2^2 \cos\theta_2\right)+p \cos \theta_2 -F=0, 
\end{empheq}
\end{subequations}
}
which are complemented by the non-holonomic constraint 
\beq
\label{wlf}
\cos \theta_2\, \dot{X}-\dot{\theta}_1 \, l\sin (\theta_1-\theta_2)=0,
\eeq
implying that the {\it velocity} of the right end of the structure remains orthogonal to the second bar, equivalent to Eq.(\ref{numerouno}). 
Imposing $\dot{X}=0$ and replacing $F$ and $p(t)$ respectively with an unknown reaction, say $R$, and by a follower force, say $P$, of constant modulus, Eqs.(\ref{conserv}a) and (\ref{conserv}b) reduce to those governing the dynamics of the Ziegler double pendulum, plus 
Eq.(\ref{conserv}c) providing the horizontal reaction $R$ of the elastic hinge on the left of that structure. 

In the Ziegler double pendulum, the follower force $P$ does work during the dynamics and may provide a non-null work during a closed cycle of deformation. 
Differently from that behaviour, {\it the reaction $p(t)$ of the non-holonomic constraint never works, as the latter always moves in a direction orthogonal to $p(t)$}. Therefore, the structure shown in Fig. \ref{def0} is fully-conservative, as the total mechanical energy (kinetic, elastic and potential of $F$) is always conserved.

It is crucial now to perform a linearization of equations (\ref{conserv}) and (\ref{wlf}) around the trivial, straight $\{\theta_1,\theta_2,X\}=\textbf{0}$, configuration. To this purpose, 
the equation expressing the non-holonomic constraint (\ref{wlf}) provides $\dot{X}=0$, so that a null perturbation in  the reaction force $p$ is obtained from Eq.(\ref{conserv}c). It follows that the remaining two equations (\ref{conserv}a) and (\ref{conserv}b) reduce to the same equations governing the linearized dynamics of the Ziegler double pendulum which, expressed in   matrix form, are
\begin{equation}
\label{ziozig}
\underbrace{\frac{m\,l^2}{4}\left[\begin{array}{cc}
5&2\\
2&1
\end{array}\right]}_{\text{Mass}\,\in\, Sym}
\left[\begin{array}{c}
\ddot{\theta}_1\\
\ddot{\theta}_2
\end{array}\right]
+
\Biggl\{
\underbrace{k\left[\begin{array}{cc}
2&-1\\
-1&1
\end{array}\right]}_{\text{Elastic stiffness}\,\in\, Sym}
+
\underbrace{F\,l\left[\begin{array}{cc}
-1&1\\
0&0
\end{array}\right]}_{\text{Geometric stiffness}\,\notin\, Sym}\Biggr\}\left[\begin{array}{c}
\theta_1\\
\theta_2
\end{array}\right]=\textbf{0},
\end{equation}
showing that the non-holonomic reaction provides a symmetry-breaking in the stiffness matrix. The linearized equations (\ref{ziozig}) should not surprise, as they reflect the fact that the mechanical system shown in Fig. \ref{def0} under small amplitude dynamics involves a negligible change in the inclination of the non-holonomic reaction. In conclusion:
\begin{center}
	{\it The structure shown in Fig. \ref{def0} obeys the same linearized dynamics found for the Ziegler double pendulum subject to a follower force, so that the same flutter and divergence instabilities are detected in the two structures. Similarly, the introduction of viscosity in the two structures leads to the same linearized dynamics,\\ so that the same behaviour will be found, including the so-called Ziegler paradox.}
\end{center}
Moreover:
\begin{center}
{\it Although the two structural systems share the same equations governing the linearized dynamics,\\ the equations governing the non-linear evolution of the motions are strongly different}. 
\end{center}
The mentioned features are thoroughly proven in the next Sections, where governing equations, critical conditions, non-linear dynamics are obtained for an extended version of the double pendulum shown in Fig. \ref{def0} by considering a generic number $N$ of rigid bars,  two  different non-holonomic constraints and loading conditions, and four different sources of viscosity.

\section{Discrete elastic columns with non-holonomic constraints}

\subsection{The column and its loadings}\label{sec21}

\begin{figure}
\begin{center}
\includegraphics[width=0.6\textwidth]{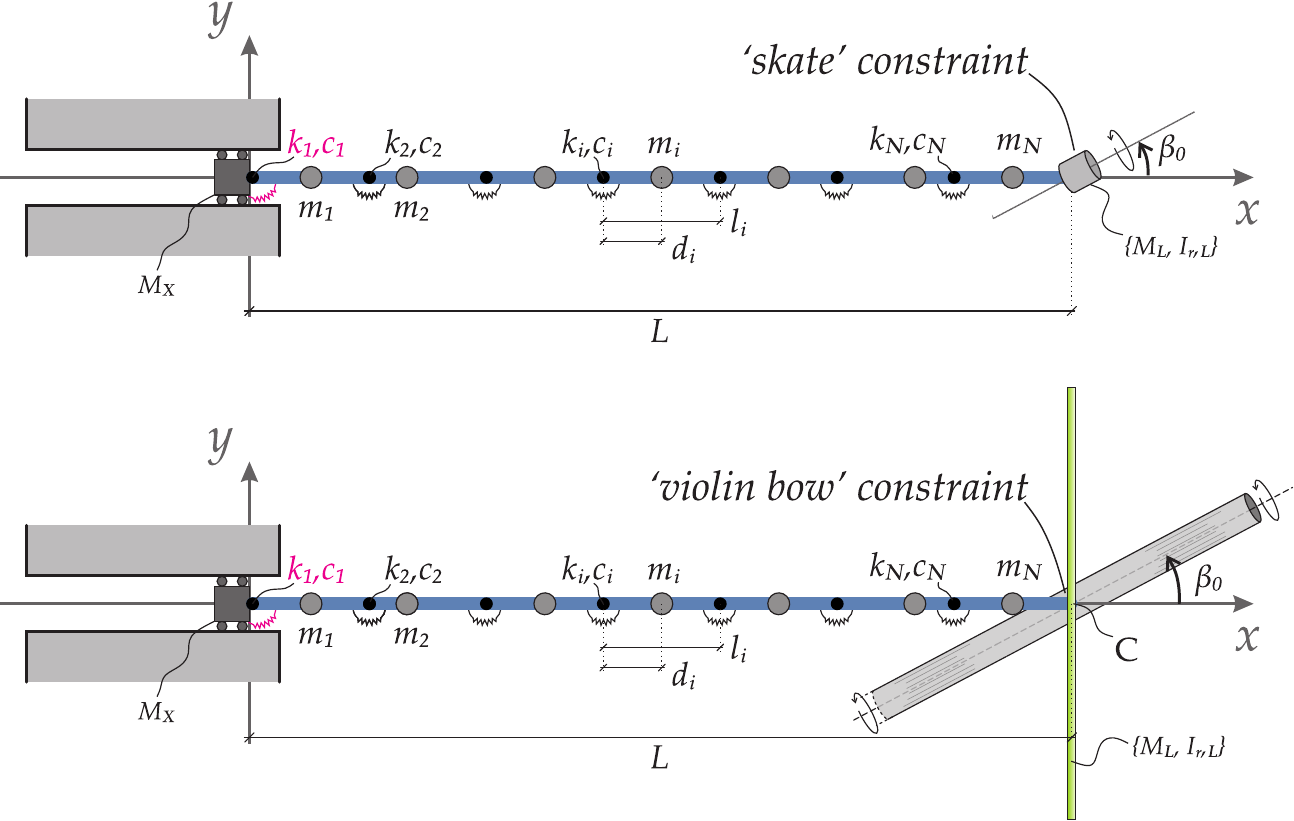} 
\end{center}
\caption{
Two undeformed structures differing only in the non-holonomic constraint applied at the right end. Each structure is discretized as a chain made up of $N$ rigid bars of mass $m_i$ and connected to each other through rotational visco-elastic springs (with viscosity parameter $c_i$ and stiffness $k_i$, $i=1, ..., N$). The first rigid bar on the left ($i=1$) is connected to a rigid block sliding along the $x-$axis. The two non-holonomic constraints are inclined at an  angle $\beta_0$ with respect to the last bar. These constraints are a \lq skate' (upper part) or a \lq violin bow' (lower part) and constrain the velocity to have a null component in the direction orthogonal to the skate (realized for instance with a non-sliding wheel) or parallel to a non-sliding freely rotating cylinder (which is in turn in contact with the \lq violin bow'). In the case $\beta_0=0$, the two structures reduce to the non-holonomic counterpart of the non-conservative Ziegler and Reut columns, respectively. Note that the last rigid element of the column reported in the lower part is \lq T--shaped', so that the freely-rotating cylinder can continue to transmit a force to the structure during motion.
}
\label{fig_discrete0}
\end{figure}

\begin{figure}[!htb]
\begin{center}
\includegraphics[width=0.9\textwidth]{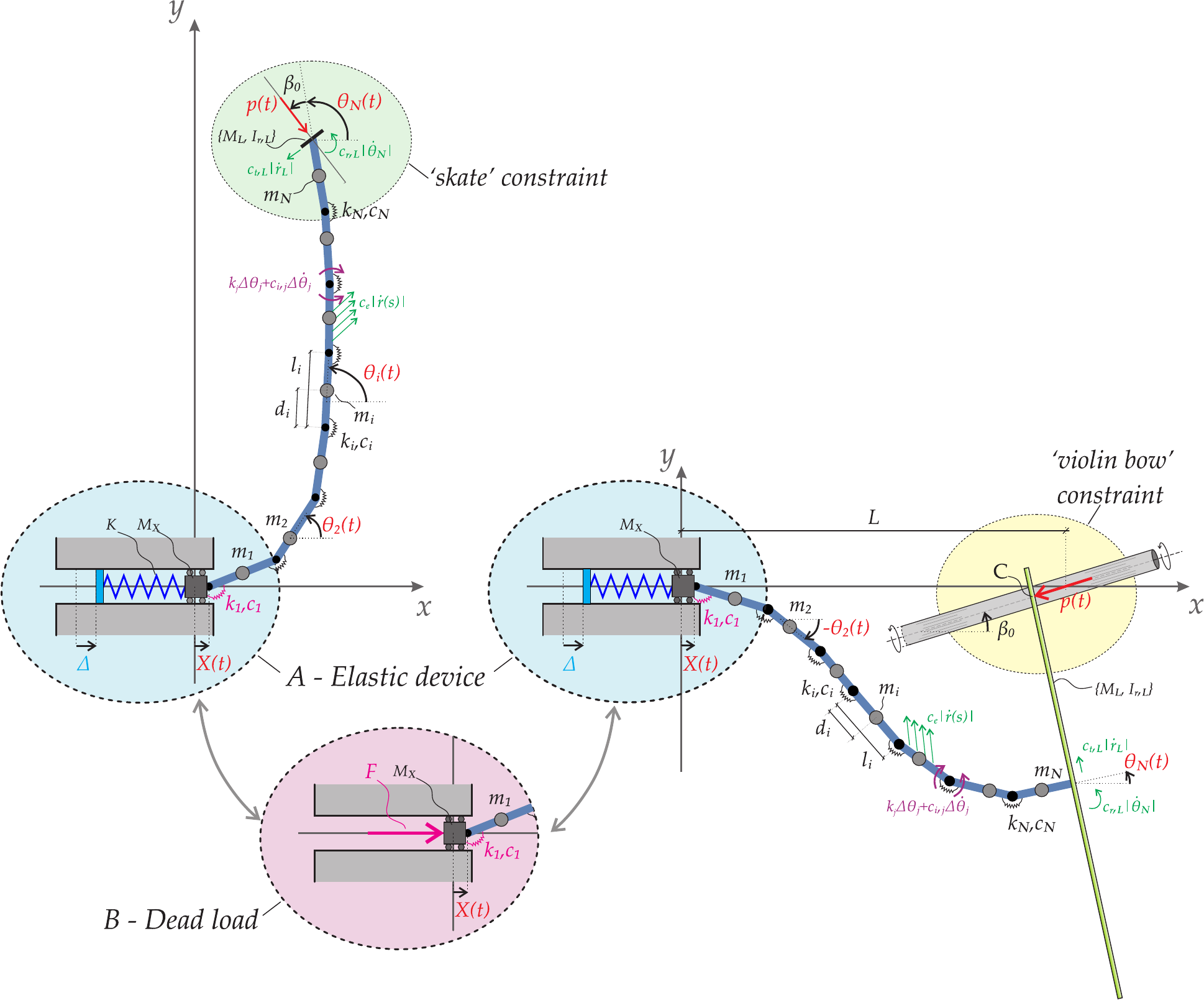} 
\end{center}
\caption{Deformed configuration for the elastic column subject  to the non-holonomic \lq skate' (left) and \lq violin bow' (right)  constraints. The {\it conservative} load is applied via
(A) a prescribed compressive displacement $\Delta$ at the left end of a linear elastic spring of stiffness $K$ attached to the rigid block or through (B) a dead load $F$. The particular condition $\beta_0=0$ leads to uniqueness of the trivial quasi-static solution.  
}
\label{fig_discrete}
\end{figure}

The motion in the $x$--$y$ plane is analyzed for an elastic column of total length $L$, discretized as a chain of $N$ rigid bars of length $l_i$ ($i=1,..., N$, so that $L=\sum_{i=1}^N l_i$), 
connected to each other with visco-elastic hinges of elastic stiffness $k_i$ and viscous parameter $c_i$, Fig.\ref{fig_discrete0}.
The first hinge is connected to a rigid block which may only slide along the $x$-direction and is loaded in one of the following {\it conservative} ways : 
\begin{itemize}
\item \textbf{A - elastic device}: the rigid block is loaded by imposing a compressive  displacement $\Delta$ at the left end of a  linear spring of stiffness $K$, whose right end is attached to the rigid block (Fig.\ref{fig_discrete});
\item \textbf{B - dead load}: the rigid block is subject to a force $F$ (for instance produced by gravity) of fixed direction parallel to the $x$-axis  (Fig.\ref{fig_discrete}).
\end{itemize} 
Denoting with $\theta_i(t)$  the anti-clockwise inclination  of the $i$-th bar with respect to $x$-axis and with $X(t)$ the position of the sliding block along the $x$-axis (Fig.\ref{fig_discrete}), the configuration of the column  at generic instant of time $t$ is fully described through the generalized coordinates vector $\bq(t)$ with the following $N+1$ components
\beq
\bq(t)=[ \theta_1(t), \quad ...\, , \quad \theta_i(t), \quad ... \, , \quad \theta_N(t), \quad X(t)].
\eeq
The undeformed trivial configuration of the system ($\bq(t)=\b0$) is given by all the rigid bars aligned with the $x$-axis ($\theta_i=0$, $i=1,...,N$) and $X=0$ (thus neglecting horizontal rigid motions). 

The mass of the column is discretized in $N$ masses $m_i$, one for each rigid bar and located at a distance $d_i \leq l_i$ from the  $i$-th hinge, so that the position in time for each mass $m_i$ is given by the coordinates $x_i(t)$ and $y_i(t)$, namely, 
\begin{equation}
x_i=X+d_i\cos{\theta_i}+\sum_{j=1}^{i-1} l_j \cos{\theta_j}, \qquad y_i(t)=d_i\sin{\theta_i}+\sum_{j=1}^{i-1} l_j \sin{\theta_j}.
\end{equation}

Moreover, the sliding block is considered to have mass $M_X$. Generic deformed configurations are reported in Fig.\ref{fig_discrete} for the two considered structures subject to conservative loads and to the non-holonomic \lq skate' and \lq violin bow' constraints (detailed in the next section).

\subsection{The non-holonomic constraints}

Two types of non-holonomic constraints (both sketched in Figs.\ref{fig_discrete0} and \ref{fig_discrete}) are assumed to be acting at the end of the elastic column, singled out by the coordinates 
\begin{equation}
\label{xywheel}
x_L=X+\sum_{i=1}^N l_i \cos{\theta_i}, \qquad y_L=\sum_{i=1}^N l_i \sin{\theta_i}. 
\end{equation}
Each one of these two constraints generates an unknown reaction $p(t)$ (positive when compressive), through the imposition of the linear kinematic restriction on the velocity, Eq.(\ref{numerouno}), which can be expressed in terms of generalized velocities $\dot{q}_k$ ($k=1,...,N+1$) as
\begin{equation}
\label{nhol2}
\sum_{k=1}^{N+1} a_{k}(\textbf{q},\,\beta_0)\dot{q}_k=0,
\end{equation}
where $a_{k}(\textbf{q},\beta_0)$ is the vector of coefficients expressing the specific type of non-holonomic constraint (\lq skate' or \lq violin bow'). Taking into account the reaction force $p$, the problem is characterized by $N+2$ unknowns.  
Each of the non-holonomic constraints is characterized by a mass $M_L$ and a rotational moment of  inertia $I_{r,L}$, both applied at the end of the final  bar ($i=N$).

As sketched in Figs.\ref{fig_discrete0} and \ref{fig_discrete}, the two non-holonomic constraints can be obtained from condition (\ref{numerouno}) as follows.
\begin{itemize}
	\item A {\bf \lq skate'  type constraint} applied with inclination $\beta_0$, representing the angle between the \lq skate' and the last bar of the column; the skate can freely rotate about its vertical axis and is constrained to slide parallely to its cutting edge, so that the only non-null component of the velocity is parallel to it: 
	\begin{equation}
	\label{nhol}
	\begin{split}
	\begin{bmatrix}
	\dot{x}_L & \dot{y}_L 
	\end{bmatrix}
	\cdot
	\begin{bmatrix}
	\cos{(\theta_N+\beta_0)} \\
	\sin{(\theta_N+\beta_0)}
	\end{bmatrix}=
	\end{split} 
	\dot{X}\cos{(\theta_N+\beta_0)}+\sum_{i=1}^{N} l_i \dot{\theta}_i \sin{(\theta_N+\beta_0-\theta_i)}=0. 
	\end{equation}
The vector of the non-holonomic coefficients $\ba(\bq,\,\beta_0)$ shown in the Eq.(\ref{nhol2}) for  this type of constraint is therefore given by 
\begin{equation}
\label{nhol2zieg}
\ba(\bq,\,\beta_0)=\left[l_1 \sin{\left(\theta_N+\beta_0-\theta_1\right)},\dots,\,l_{N-1} \sin{\left(\theta_N+\beta_0-\theta_{N-1}\right)},\,l_N \sin{\left(\beta_0\right)},\,\cos{\left(\theta_N+\beta_0\right)}\right].
\end{equation}
Finally, the reaction force $p(t)$ is always orthogonal to the sliding direction of the \lq skate'.

		\item A {\bf \lq violin bow' type constraint} applied with inclination $\beta_0$, representing the angle between the freely-rotating cylinder and the $x-$axis; the \lq T-shaped' rigid element (attached to the $N$-th rigid bar of the column) can freely rotate about the contact point with the cylinder and can translate orthogonally to the cylinder's axis, such that the velocity component parallel to it must vanish. The coordinates $x_C$ and $y_C$ of the contact point $C$ (see Fig.\ref{fig_discrete}) between the \lq T-shaped' rigid element and the freely rotating cylinder is singled out by the coordinates 
	\begin{equation}
	\label{posA}
	x_C=x_L+\zeta\tan \theta_N,\qquad y_C=y_L-\zeta, \qquad \mbox{where}\,\,\,\,\zeta=\frac{y_L+(L-x_L)\tan \beta_0}{1+\tan \beta_0\, \tan \theta_N} ,
	\end{equation}
so that its velocity can be obtained through the Poisson's theorem as 
\begin{equation}
\label{velC}
\begin{bmatrix}
\dot{x}_C \\ \dot{y}_C 
\end{bmatrix}
=
\begin{bmatrix}
\dot{x}_L \\ \dot{y}_L 
\end{bmatrix}
+
\dot\theta_N\be_3\times
\begin{bmatrix}
x_C-x_L \\
y_C-y_L
\end{bmatrix}=
\begin{bmatrix}
\dot{x}_L + \dot\theta_N \zeta \\ \dot{y}_L +\dot\theta_N \zeta \tan \theta_N,
\end{bmatrix},
\end{equation} 
where $\be_3$ is the unit vector orthogonal to the $x-y$ plane. The non-holonomic constraint is therefore expressed by the vanishing of the component of the velocity along the cylinder's axis at the point $C$, namely
\begin{equation}
\label{stocazreut}
\left(\dot{x}_L+\dot\theta_N\, \zeta\right)\cos \beta_0+\left(\dot{y}_L+\dot\theta_N\, \zeta\tan \theta_N\right)\sin \beta_0=0
\end{equation}
which, recalling Eq.(\ref{xywheel}), can be rewritten as
\begin{equation}
\label{nholreut}
\dot{X}\cos \beta_0+\dot\theta_N \sin \beta_0 L+\sum_{i=1}^{N-1} l_i
\sin (\beta_0-\theta_i)(\dot\theta_i-\dot\theta_N)=0.
\end{equation}
Therefore, the vector $\ba(\bq,\,\beta_0)$, representing  the non-holonomic constraint as introduced in the Eq.(\ref{nhol2}), is given by
\begin{equation}
\label{nhol2reut}
\ba(\bq,\,\beta_0)=\left[l_1 \sin{\left(\beta_0-\theta_1\right)},
\,\,\,
\dots,\,\,\,l_{N-1} \sin{\left(\beta_0-\theta_{N-1}\right)},\,\,\,l_N \left(\sin{\beta_0}L-\sum_{i=1}^{N-1} l_i
\sin (\beta_0-\theta_i)\right),\,\,\,\cos{\beta_0}\right] .
\end{equation}
Finally, the reaction force $p(t)$ is always directed along the axis of the freely-rotating cylinder.

\end{itemize}

\subsection{Energies and dissipation}

The total potential energy $\mathcal{V}(\bq)=\mathcal{E}(\bq)-\mathcal{W}(\bq)$, equal to the difference  between the elastic energy $\mathcal{E}$ and the work  $\mathcal{W}$  done by the external loads, is given for the considered structural systems by
\begin{equation}
\label{elasticenergy}
\mathcal{V}(\bq)=\Xi(\bq)+\frac{1}{2} k_1 \theta_1^2+\frac{1}{2}\sum_{i=2}^N k_i \left( \theta_i-\theta_{i-1} \right)^2 ,
\end{equation}
where (note that $\Delta$ is independent of time)
\begin{equation}
\label{plof}
\Xi(\bq)=\left\{
\begin{array}{ll}
\dfrac{1}{2} K \left[X-\Delta\right]^2\qquad\qquad & \text{for elastic device A,}\\[3mm]
-F\, X &\text{for dead loading B.}
\end{array}
\right.
\end{equation}

The expression for the kinetic energy $\mathcal{T}$ is the same for both systems A and B and given by 
\begin{equation}
\label{kineticenergy}
\mathcal{T}(\bq,\dot{\bq})=\frac{1}{2} M_X \dot{X}^2+\frac{1}{2} M_L \left(\dot{x}_L^2+\dot{y}_L^2\right)+\frac{1}{2}\sum_{i=1}^N m_i \left(\dot{x}_i^2+\dot{y}_i^2\right)+\frac{1}{2} I_{r,L} \dot{\theta}_N^2.  
\end{equation}
Dissipative effects  are taken into account from different viscous sources through the following Rayleigh dissipation function
\begin{equation}
\label{dissipfun}
\mathcal{F}_d(\dot{\bq})=\frac{1}{2} c_{e} \sum_{i=1}^{N} \int_0^{l_i} (\dot{X}_i(s_i)^2+\dot{Y}_i(s_i)^2) \text{d}s_i + \frac{1}{2} c_{t,L} (\dot{x}_L^2+\dot{y}_L^2)+ \frac{1}{2} c_{r,L} \dot{\theta}_N^2 
+\frac{1}{2} c_{i}\dot{\theta}_1^2+\frac{1}{2} c_{i} \sum_{i=2}^N(\dot{\theta}_i-\dot{\theta}_{i-1})^2 ,
\end{equation}
where  $X_i(s_i)$ and $Y_i(s_i)$ are coordinates of the generic point located at the curvilinear coordinate $s_i\in\left[0,\,l_i\right]$ along the $i$-th bar, which can be written as 
\begin{equation}
X_i(s_i)=X+s_i\cos{\theta_i}+\sum_{j=1}^{i-1} l_j \cos{\theta_j}, \qquad Y_i(s_i)=s_i\sin{\theta_i}+\sum_{j=1}^{i-1} l_j \sin{\theta_j} .
\end{equation} 

Each (non-negative) linear damping  coefficient $c_e$, $c_i$, $c_{t,L}$, and $c_{r,L}$ introduced in Eq.(\ref{dissipfun}) is related to a specific viscous source:
\begin{itemize}
\item \textit{external translational damping} $c_{e}$, providing a distributed viscous force along each bar, which may model the air drag during motion;
\item \textit{internal rotational damping} $c_i$, modelling a viscous dissipation induced by torsion at the hinges (and identical for all the hinges);
\item \textit{translational damping of the non-holonomic constraint} $c_{t,L}$, modelling the effects of the air drag generated by the device realizing the constraint;
\item \textit{rotational damping of the non-holonomic constraint} $c_{r,L}$, modelling the damping generated by pivoting movements at the contact of the device realizing the constraints, for instance the rotational resistance of a wheel on a plane. 
\end{itemize}

The presence of these four viscous sources allows to analyze their different influences on the stability of the structure, an influence which is often stronger than one might expect \cite{tommy}. 
In the quasi-static case, the velocity is null and so the dissipation function $\mathcal{F}_d$ vanishes.

It is remarked that both the loading systems A and B are conservative, so that the supply of energy is provided for System A by the elastic energy initially stored in the spring through the compressive displacement $\Delta$ and for system B by the potential energy of the dead load $F$. In the stability evaluations concerning the loading system A, perturbations on the distance $\Delta$ will not be  considered.
Finally, functionals (\ref{elasticenergy}), (\ref{kineticenergy}) and (\ref{dissipfun}) are independent of the choice of the non-holonomic constraint [which acts on the system via Eq.(\ref{nhol2})] and of its inclination $\beta_0$.

\subsection{Equations of motion}

The equations of motion for the visco-elastic column subject to non-holonomic constraints can be derived  from the d'Alembert-Lagrange equation \cite{neimark} expressed for virtual displacements $\delta q_k$ as
\begin{equation}
\label{dallag}
\sum_{k=1}^{N+1}\left(\frac{d}{d t}\frac{\partial \mathcal{T}(\bq,\dot{\bq})}{\partial \dot{q}_k}-\frac{\partial \mathcal{T}(\bq,\dot{\bq})}{\partial q_k}-Q_k(\bq,\dot{\bq})+p(t)\,a_{k}(\bq,\beta_0)\right)\delta q_k=0,
\end{equation}
where the generalized forces $Q_k$  are given by
\begin{equation}\label{generalizedforces}
Q_k(\bq,\dot{\bq})=-\frac{\partial \mathcal{V}(\bq)}{\partial q_k}-\frac{\partial \mathcal{F}_d(\dot{\bq})}{\partial \dot{q}_k} ,
\qquad k=1,..., N+1,
\end{equation}
and account for the dissipative forces obtained from the dissipation function $\mathcal{F}_d$, Eq.(\ref{dissipfun}), and for the presence of the external energy supply, 
contained in the total potential energy $\mathcal{V}$ in terms of parameter $\Delta$ for the elastic device A, or force $F$ for the dead load B, Eq.(\ref{plof}). 

In the d'Alembert-Lagrange equation (\ref{dallag}), 
the unknown reaction force  $p(t)$ transmitted to the column by the non-holonomic constraint plays the role of a Lagrangian multiplier enforcing condition (\ref{nhol2}).

Invoking arbitrariness of the virtual displacements $\delta q_k$, the d'Alembert-Lagrange equation (\ref{dallag}) yields
\begin{equation}
\label{ELeqs}
\frac{d}{d t}\frac{\partial \mathcal{T}(\bq,\dot{\bq})}{\partial \dot{q}_k}-\frac{\partial \mathcal{T}(\bq,\dot{\bq})}{\partial q_k}-Q_k(\bq,\dot{\bq})+p(t)\,a_{k}(\bq,\beta_0)=0, \qquad \forall \,\ k=1,2,...,N+1,
\end{equation}
which, together with the non-holonomic constraint, Eq.(\ref{nhol2}), provides a system of $N+2$ equations in $N+2$ unknowns, namely, the $N+1$ generalized coordinates $q_k(t)$ and the non-holonomic constraint reaction $p(t)$.

\subsection{Quasi-static response of the column with non-holonomic constraints}

Under quasi-static conditions ($\dot{\bq}(t)=\b0$) the kinetic energy and the dissipation  are null ($\mathcal{T}=\mathcal{F}_d=0$, so that $Q_k=\partial{\mathcal{V}/\partial q_k}$) and the equations of motion (\ref{ELeqs}) simplify to the $N+1$ equilibrium equations
\begin{equation}
\label{QSeqs}
\frac{\partial \mathcal{V}( \bq^{\mbox{\tiny{QS}}})}{\partial q_{k}^{\mbox{\tiny{QS}}}}+p^{\mbox{\tiny{QS}}}\, a_{k}(\bq^{\mbox{\tiny{QS}}},\beta_0)=0 , \qquad \forall\,\, k=1,2,...,N+1,
\end{equation}
to be solved for the unknown reaction force $p^{\mbox{\tiny{QS}}}$ at prescribed values of the spring displacement $\Delta$ or dead load $F$. 
In particular, the total potential energy (\ref{elasticenergy}) is a polynomial of degree 2 in the generalized coordinates, so that Eq.(\ref{QSeqs}) can be rewritten as
\begin{equation}
\label{QSeqsextended}
\left\{
\begin{array}{cc}
\mathbb{K}^{(N)}_{ik} q^{\mbox{\tiny{QS}}}_k+p^{\mbox{\tiny{QS}}}\, a_{k}( \bq^{\mbox{\tiny{QS}}},\,\beta_0)=0&\qquad \forall\,\, i,k=1,2,...,N\\[3mm]
\dfrac{\partial\, \Xi(\bq^{\mbox{\tiny{QS}}})}{\partial X}+p^{\mbox{\tiny{QS}}}\, a_{N+1}( \bq^{\mbox{\tiny{QS}}},\,\beta_0)=0&
\end{array}
\right.
\end{equation}
where the dependence of the vector $a_k$ on the considered quasi-static solution $q^{\mbox{\tiny{QS}}}_k$ and on the inclination angle $\beta_0$ is highlighted and, recalling Eq.(\ref{plof}), 
\begin{equation} 
\frac{\partial\, \Xi(\bq)}{\partial X}=\left\{
\begin{array}{ll}
K \left[X(t)-\Delta\right]\qquad & \text{elastic device A,}\\[3mm]
-F &\text{dead load B,}
\end{array}
\right.
\end{equation}
while the $N\times N$ elastic stiffness matrix $\mathbb{K}^{(N)}$ 
\begin{equation}
\label{K2stiff}
\mathbb{K}^{(N)}=\left[
\begin{array}{cccccc}
k_1+k_2 & -k_2 &  &  &  &\\
-k_2 & k_2+k_3 & -k_3 &  &  &\\
 & -k_3 & k_3+k_4 & \ddots &  &\\
 &  & \ddots & \ddots & -k_{N-1} & \\
 &  &  &  -k_{N-1} & k_{N-1}+k_N & -k_N\\
 &  &  &  &-k_N & k_N\\
\end{array}
\right], 
\end{equation}
is tridiagonal, symmetric and positive-definite. Moreover, it is also independent of the type of non-holonomic constraint applied at the end of the structure and of the loading condition. It can be proven (this delicate proof is deferred to Appendix \ref{appendixA}) that the determinant of matrix (\ref{K2stiff}) is given by the product of the $N$ rotational stiffness parameters $k_j$, so that
\begin{equation}
\label{detkNprod}
\det\mathbb{K}^{(N)}=\prod\limits_{j=1}^{N}k_j .
\end{equation}

Note that the loading condition in the specific case of a straight column with $\beta_0=0$ is given by the $(N+1)$--th equation of system (\ref{QSeqsextended})
where $a_{N+1}(\textbf{0},\,0)=1$, yielding
\begin{equation}
p^{\mbox{\tiny{QS}}}=\left\{
\begin{array}{ll}
K \left[\Delta-X\right]\qquad & \text{elastic device A,}\\[3mm]
F &\text{dead load B.}
\end{array}
\right.
\end{equation}

Under the quasi-static assumption the velocities are null and the non-holonomic constraint (\ref{numerouno}) is automatically satisfied but  
the equilibrium equations (\ref{QSeqs}), which still depend on the non-holonomic constraint  through the coefficients $a_{k}$, 
contain $N+2$ unknowns. As a consequence, the value of one of the generalized coordinates $q_{k}^{\mbox{\tiny{QS}}}$ or the reaction force $p^{\mbox{\tiny{QS}}}$ can be selected within a one-dimensional \textit{manifold of equilibrium states} \cite{neimark} and the remaining unknowns solved using Eqs.(\ref{QSeqs}). 
In other words, the columns admit an infinite set of quasi-static solutions. 

\begin{figure}[!htb]
\begin{center}
\includegraphics[width=0.7\textwidth]{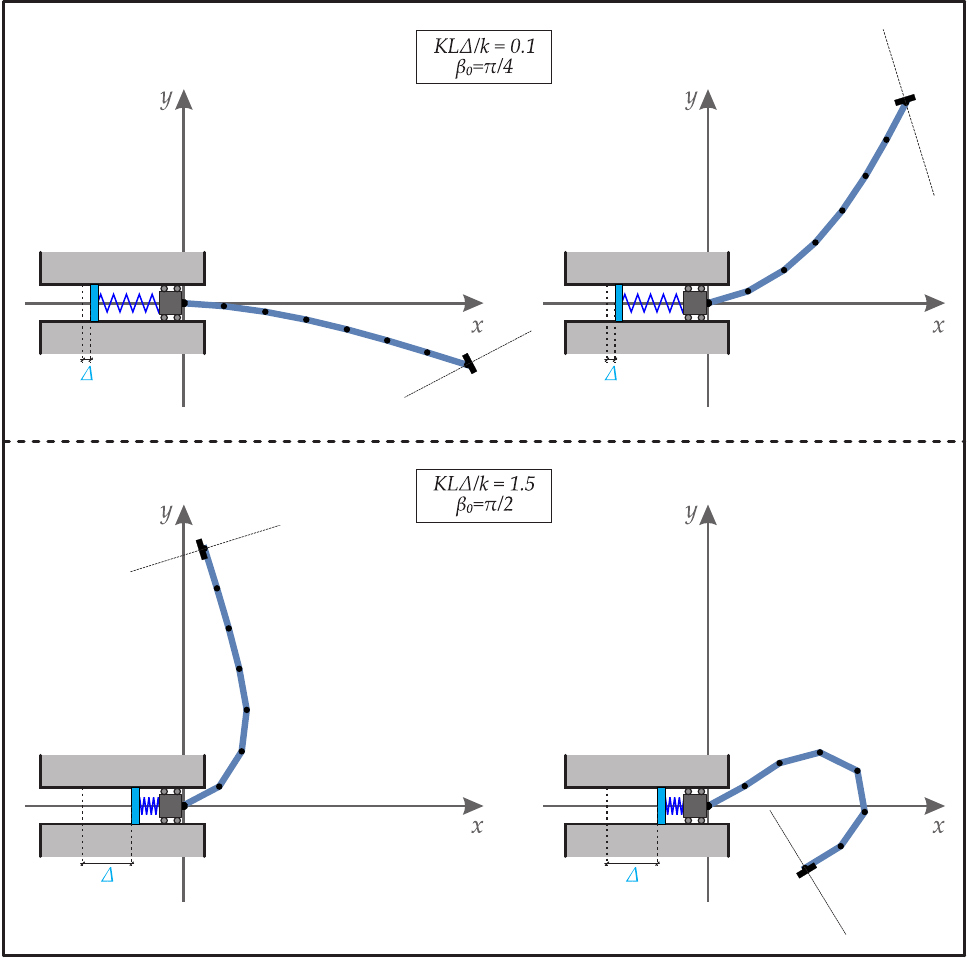}
\end{center}
\caption{Examples of non-unique (and non-trivial) equilibrium configurations for a column made up of $N=7$ rigid bars and subject to the  non-holonomic \lq skate' constraint. Equilibrium has been solved by fixing both $X^{\mbox{\tiny{QS}}}=0$ and $\Delta$, in particular, two pairs of non-unique solutions are reported for $K L\Delta /k  = 0.1$ and $\beta_0=\pi/4$ (upper part), and $K L \Delta/k  = 1.5$ and $\beta_0=\pi/2$ (lower part). }
\label{fig_4deformed}
\end{figure}

As examples of multiple solutions, non-trivial equilibrium configurations are shown in Fig.\ref{fig_4deformed} for two columns made up of $N=7$ rigid bars and loaded through the horizontal spring (loading condition A). 
The two columns differ in the dimensionless parameter $K L \Delta/k$, being equal to $0.1$ (above) or $1.5$ (below), and in the inclination of the \lq skate', $\beta_0=\pi/4$ (above) or $\pi/2$ (below). The two columns are loaded by a force $K(X^{\mbox{\tiny{QS}}}-\Delta)$, which is prescribed through the application of both $\Delta$ and $X^{\mbox{\tiny{QS}}}=0$.

The straight configuration (with null rotation for all the bars, $\theta_{k}^{\mbox{\tiny{QS}}}=0$) can be a solution of the equilibrium equations (\ref{QSeqs}) for both non-holonomic \lq skate' and \lq violin' bow  constraints, at  
every inclination $\beta_0$, in the following cases:
\begin{itemize}
\item When $\beta_0\neq 0$, the straight configuration is a solution whenever the reaction force at the non-holonomic constraint is  null ($p^{\mbox{\tiny{QS}}}=0$), corresponding to an unstressed column and which occurs when 
$X^{\mbox{\tiny{QS}}}=\Delta$ for the elastic device A or when $F=0$ for every value of $X^{\mbox{\tiny{QS}}}$ in the case of the dead loading B. 

This statement can be proven  by considering that the first $N$ equations of the system (\ref{QSeqsextended}) at $\bq^{\mbox{\tiny{QS}}}=0$ are characterized by a non-null vector $a_{k}(\textbf{0},\,\beta_0)\neq 0$, so that the trivial configuration corresponds to a null value of the reaction force $p^{\mbox{\tiny{QS}}}=0$, and consequently to null external load $\partial\, \Xi /\partial X=0$ (namely $X^{\mbox{\tiny{QS}}}=\Delta$ for the device A or $F=0$ for the device B), as shown by the $(N+1)$-th equation of the system (\ref{QSeqsextended});

\item When $\beta_0=0$, the straight configuration is the unique equilibrium solution admitted by the system (\ref{QSeqs}). The proof of this statement can be 
demonstrated with equilibrium consideration; a formal proof is difficult and deferred to Appendix \ref{appendixB}. 

In contrast with the case $\beta_0\neq 0$, now the  reaction force provided by the non-holonomic constraint $p^{\mbox{\tiny{QS}}}$ may be different from zero, so that the elastic chain in its straight configuration may be \emph{axially-stressed} as the result of the imposition of $F$ or $\Delta$ (when $X^{\mbox{\tiny{QS}}}\neq \Delta$).\footnote{
The fact that the straight configuration for $\beta_0=0$ can be arbitrarily stressed is obvious. Formally, since $a_{k}(\textbf{0},\,0)=0$ but $a_{N+1}(\textbf{0},\,0)\neq0$ for both the non-holonomic constraints, the first $N$ equations of system (\ref{QSeqsextended}) particularized for $\bq^{\mbox{\tiny{QS}}}=0$ are always satisfied for any value of the reaction force $p^{\mbox{\tiny{QS}}}$, which is given by $\partial \,\Xi /\partial X$ and $a_{N+1}$ through the $(N+1)$-th equation in (\ref{QSeqsextended}).}

\end{itemize}

In the next Section, the stability is analyzed for a generic configuration of the elastic column subject to both loading conditions A and B and both the non-holonomic constraints. The condition $\beta_0=0$ finally restricts the analysis to the trivial configuration $\bq^{\mbox{\tiny{QS}}}=\b0$ only, where rigid body motions are neglected by setting $X^{\mbox{\tiny{QS}}}=0$.

\section{Linearized dynamics and stability of the column with non-holonomic systems}
\label{sec3}

A first-order expansion of Eqs.(\ref{ELeqs}) about a generic quasi-static solution in both the generalized coordinates $\textbf{q}^{\mbox{\tiny{QS}}}$ and the Lagrangian multiplier $p^{\mbox{\tiny{QS}}}$, therefore satisfying Eqs.(\ref{QSeqs}), can be performed by assuming 
\begin{equation}
\label{pertschain}
\textbf{q}(t)=\textbf{q}^{\mbox{\tiny{QS}}}+\epsilon \hat{\textbf{q}}(t),\quad p(t)= p^{\mbox{\tiny{QS}}}+\epsilon \hat p(t),
\end{equation}
where $\epsilon$ is an arbitrarily small real parameter and $\left\{\hat{\textbf{q}}(t),\,\hat{p}(t)\right\}$ is the set denoting the perturbation fields. From the quasi-static nature of the configuration $\textbf{q}^{\mbox{\tiny{QS}}}$, it follows that
\begin{equation}
\label{pertschain2}
 \dot{\textbf{q}}(t)=\epsilon \dot{\hat{\textbf{q}}}(t),\quad \ddot{\textbf{q}}(t)=\epsilon \ddot{\hat{\textbf{q}}}(t).
\end{equation}

A Taylor series expansion of the Eqs.(\ref{ELeqs}) about $\epsilon=0$ yields the following linearized equations of motion for the perturbed fields, which, removing the superscript \lq  $\,\widehat{~~}\,$', are expressed by
\beq
\label{ELlineqs}
\left\{
\begin{array}{lll}
\textbf{M}\ddot{\textbf{q}}+\textbf{C}\dot{\textbf{q}}+ \left(\textbf{K}+p^{\mbox{\tiny{QS}}} \, \textbf{G}\right) \textbf{q} + p \, \textbf{w} =0,\\[3mm]
\textbf{w} \scalp \dot{\textbf{q}}=0 ,
\end{array}
\right.
\eeq
as functions of the time-dependent \lq load' perturbation $p$.  $\textbf{M}$, $\textbf{C}$, and  $\textbf{K}$ are, respectively, the mass, damping 
and elastic stiffness (symmetric and positive definite) matrices of dimension $N+1$, while the geometric stiffness matrix $\textbf{G}$ is non-symmetric as influenced by the presence of the non-holonomic constraint reaction $p^{\mbox{\tiny{QS}}}$.
The vector $\textbf{w}$ is related to the linearized version of the non-holonomic constraint (\ref{nhol}) or (\ref{nholreut}) for the \lq skate' and \lq violin bow', respectively. A comparison with Eqs.(\ref{QSeqsextended}) and (\ref{K2stiff}) shows that the elastic  stiffness matrices $\textbf{K}$ and $\mathbb{K}^{(N)}$ are related through the second derivative of the function $\Xi$ with respect to the generalized coordinate $X$ [see Eqs.(\ref{elasticenergy}) and (\ref{plof})]
as
\begin{equation}
\label{kblock}
\textbf{K}=\left[
\begin{array}{c|c}
\mathbb{K}^{(N)} & \textbf 0 \\
\hline
\\[-1.5 mm]\textbf 0 & \dfrac{\partial^2 \,\Xi(\bq)}{\partial X^2}
\end{array}
\right]\qquad \text{where} \quad \frac{\partial^2 \,\Xi(\bq)}{\partial X^2}=\left\{
\begin{array}{ll}
K \qquad & \text{for elastic device A,}\\[3mm]
0 &\text{for dead load B}.
\end{array}
\right.
\end{equation}

The geometric stiffness matrix $\textbf{G}$ is the gradient of vector $\ba(\bq,\,\beta_0)$ calculated with respect to the generalized coordinates vector $\bq$ and evaluated at the generic quasi-static solution $\bq^{\mbox{\tiny{QS}}}$
\begin{equation}
\textbf{G}=\left.\frac{\partial \,\ba(\bq,\,\beta_0)}{\partial\, \bq}\right|_{\bq=\bq^{\mbox{\tiny{QS}}}} ,
\end{equation}
and is strictly related to the choice of the non-holonomic constraint through coefficients $a_k$ given by Eqs.(\ref{nhol2zieg}) and (\ref{nhol2reut}) for the \lq skate' or \lq violin bow' constraint, respectively. Following the same notations adopted for $\textbf{K}$, the matrix $\textbf{G}$ can be represented as
\begin{equation}
\label{gblock}
\textbf{G}=\left[
\begin{array}{c|c}
\mathbb{G}^{(N)} & \textbf 0 \\
\hline
\bg & 0
\end{array}
\right]\qquad \text{where} \qquad \bg=\left\{
\begin{array}{ll}
[0,\,0,\,....,\,0,\,-\sin{\left(\theta_N^{\mbox{\tiny{QS}}}+\beta_0\right)}] &\text{for the skate constraint,}\\[3mm]
\textbf{0} \qquad & \text{for the violin bow constraint.}
\end{array}
\right.
\end{equation}
where $\mathbb{G}^{(N)}$ is a $N\times N$ matrix and  $\bg$ is a $N$-component vector. The expressions for the $N\times N$ {\it non-symmetric and singular} geometric stiffness matrices for the \lq skate' and \lq violin bow' constraints, respectively denoted by $\mathbb{G}^{(N)}_S$ and $\mathbb{G}^{(N)}_V$, are
\begin{equation}
\label{gN}
\mathbb{G}^{(N)}_S=\left[
\begin{array}{cccccc}
-\gamma_{S,1} &  &  &  & \gamma_{S,1}\\
 & -\gamma_{S,2} &  &  &\gamma_{S,2}\\
 &  & \ddots &  &\vdots\\
 &  &  & -\gamma_{S,N-1} & \gamma_{S,N-1}\\
 &  &  &  & 0
\end{array}\right],\qquad \mathbb{G}^{(N)}_V=\left[
\begin{array}{cccccc}
-\gamma_{V,1} &  &  &  & \\
 & -\gamma_{V,2} &  &  &\\
 &  & \ddots &  &\\
 &  &  & -\gamma_{V,N-1} & \\
 \gamma_{V,1} & \gamma_{V,2} & \dots & \gamma_{V,N-1} & 0
\end{array}\right]
\end{equation}
where the only non-vanishing components lie on the diagonal and on the $N$-th column ($N$-th row) for the \lq skate' (\lq violin bow') constraint. The coefficients $\gamma_{S,i}$ and $\gamma_{V,i}$ are given by
\begin{equation*}
\gamma_{S,i}=l_i \cos\left(\theta_N^{\mbox{\tiny{QS}}}+\beta_0-\theta_i^{\mbox{\tiny{QS}}}\right),
\qquad \gamma_{V,i}=l_i \cos\left(\beta_0-\theta_i^{\mbox{\tiny{QS}}}\right), \qquad \forall i=1,...,N-1,
\end{equation*}
while $\gamma_{S,N}=\gamma_{V,N} = 0$.

A property that will be useful later, holding for every (null or non-null) value of $\beta_0$, is the following (the proof is deferred to Appendix \ref{appendixDetK})
\begin{equation}
\label{dekSimp}
\det\left(\mathbb{K}^{(N)}+p^{\mbox{\tiny{QS}}}\mathbb{G}^{(N)}\right)=\det\left(\mathbb{K}^{(N)}\right)=\prod\limits_{j=1}^{N}k_j. 
\end{equation}

The symmetric and positive-definite mass and damping matrices defined in Eq.(\ref{ELlineqs}) can be rewritten as
\begin{equation}
\label{mcblock}
\textbf{M}=\left[
\begin{array}{c|c}
\mathbb{M}^{(N)} & \textbf{m} \\
\hline
\textbf{m}^\intercal & \hat{m}
\end{array}
\right],\qquad \qquad 
\textbf{C}=\left[
\begin{array}{c|c}
\mathbb{C}^{(N)} & \bc \\
\hline
\bc^\intercal & \hat{c}
\end{array}
\right]
\end{equation}
where $\mathbb{M}^{(N)}$ and $\mathbb{C}^{(N)}$ are $N\times N$ matrices, $\textbf{m}$ and $\bc$ are $N-$component column vectors, both dependent on vector $\bq^{\mbox{\tiny{QS}}}$, and $\hat m$ and $\hat c$ are positive real numbers depending on the mass and damping coefficients acting along the $X$ generalized coordinate. Moreover, the superscript \lq $~^\intercal \,$' denotes the transpose operator. It is obvious that $\textbf{M}$ and $\textbf{C}$ are independent of the choice of both the non-holonomic constraint (skate or violin bow) and the loading condition (elastic device A or dead load B).

Finally, the $N-$component column vector $\bw$ expressing the non-holonomic constraint, linearized about a generic quasi-static solution $\bq^{\mbox{\tiny{QS}}}$, is given by
\begin{equation}
\bw=\left\{
\begin{array}{ll}
[l_{1}\sin\left(\theta_N^{\mbox{\tiny{QS}}}+\beta_0-\theta_1^{\mbox{\tiny{QS}}}\right),\,l_{2}\sin\left(\theta_N^{\mbox{\tiny{QS}}}+\beta_0-\theta_2^{\mbox{\tiny{QS}}}\right),.....,\,l_{N}\sin\left(\beta_0\right),\,\cos \left(\theta_N^{\mbox{\tiny{QS}}}+\beta_0\right)]^\intercal \qquad & \text{skate,}\\[3mm]
[l_{1}\sin\left(\beta_0-\theta_1^{\mbox{\tiny{QS}}}\right),.....,\,l_{N-1}\sin\left(\beta_0-\theta_{N-1}^{\mbox{\tiny{QS}}}\right),\,L\sin \beta_0-\sum_{i=1}^{N-1}l_i\sin\left(\beta_0-\theta^{\mbox{\tiny{QS}}}_i\right),\,\cos \beta_0]^\intercal &\text{violin bow}.
\end{array}
\right.
\end{equation}
For convenience, the vector $\bw$ is rewritten in the following form
\begin{equation}
\label{wblock}
\bw=[\bw^{(N)},w_{N+1}]^\intercal
\end{equation}
where $\bw^{(N)}$ is the vector collecting the first $N$ components of $\bw$.

Assuming time-harmonic perturbations $q_k=A_k e^{\omega t}$ and  $p=A_p e^{\omega t}$, the differential system (\ref{ELlineqs}) reduces to
\begin{equation}
\label{mat}
\left[
\begin{array}{c|c}
\textbf{M}\omega^2+\textbf{C}\omega+\textbf{K}+p^{\mbox{\tiny{QS}}} \, \textbf{G} & \textbf{w}  \\
\hline
\omega\textbf{w}^\intercal & 0
\end{array}
\right] \textbf{A}=0 ,
\end{equation}
 where vector $\textbf{A}$ collects the unknown  $N+2$ amplitudes
\begin{equation}
\textbf{A}=[A_1,\,A_2,\,....,\,A_N,\,A_{N+1},\,A_p]^\intercal,
\end{equation}
and the singularity of the matrix defined by Eq.(\ref{mat}) provides the value of $\omega$ for which non-null perturbations $\textbf{A}\neq 0$ exist.

Because all the coefficients  in the last row of the matrix in Eq.(\ref{mat}) are linear in $\omega$,  the trivial eigenvalue $\omega=0$ is always a solution of the problem. According to Neimark and Fufaev \cite{neimark}, this peculiarity is usual in non-holonomic systems, so that the trivial eigenvalue is omitted in order to perform the stability analysis of the configurations belonging to the manifold of equilibrium states. The stability is lost when the determinant  of the relevant matrix in Eq.(\ref{mat}) vanishes,
\begin{equation}
\label{det}
\det{\left[
\begin{array}{c|c}
\textbf{M}\omega^2+\textbf{C}\omega+\textbf{K}+p^{\mbox{\tiny{QS}}} \, \textbf{G} & \textbf{w} \\
\hline
\textbf{w}^\intercal & 0
\end{array}
\right]}=0 ,
\end{equation}
a condition depending on the non-holonomic reaction $p^{\mbox{\tiny{QS}}}$,   related through the $(N+1$)--th equation of the system (\ref{QSeqsextended}) to the displacement $\Delta$ (loading condition A) or the dead load $F$ (loading condition B).

Due to the fact that the last row and column of the matrix appearing in Eq.(\ref{det}) are independent of the eigenvalue $\omega$, the vanishing of its determinant provides a polynomial equation in $\omega$ of order $m=2N$, with real coefficients $\rho_j$ ($j=0,...,m$) in the form 
\begin{equation}
\label{pol}
\rho_0\, \omega^{m}+\rho_1\, \omega^{m-1}+\,...\,+\rho_{m-1}\, \omega+\rho_{m}=0,
\end{equation} 
where in the specific case of null damping forces ($\textbf{C}=\textbf 0$) all the terms $\rho_j$ with odd index $j$ are equal to zero.

Two equivalent techniques can be exploited to assess the  stability of configurations belonging to the manifold of equilibrium states. The first technique is based on the direct evaluation of the eigenvalues $\omega$, so that a configuration is stable if the real part of every eigenvalue $\omega$ is non-positive.  The other technique is the Routh-Hurwitz criterion, which, for $\rho_0>0$ in Eq.(\ref{pol}), relates stability to the positiveness of every determinant $D_i$  of the square matrix $\bold{H}_i$ ($i=1,...,m$) defined as the upper-left corner square matrix of rank $i$ of the Hurwitz matrix $\bold{H}_m$, defined as
\begin{equation}
\label{hurwitzmat}
\bold{H}_m=\left[
\begin{array}{ccccccccc}
\rho_1 & \rho_0 & 0 & \dotsb &  &  &  &  & 0 \\
\rho_3 & \rho_2 & \rho_1 & \rho_0 & 0 & \dotsb &  &  & 0 \\
\rho_5 & \rho_4 & \rho_3 & \rho_2 & \rho_1 & \rho_0 & 0 & \dotsb & 0 \\
\dotsb &  &  &  &  &  &  &  & \dotsb\\
\dotsb &  &  &  &  &  &  &  & \dotsb\\
\dotsb &  &  &  &  &  &  &  & \dotsb\\
\rho_{2m-1} & \rho_{2m-2} & \dotsb &  &  &  &  &  & \rho_{m}\\
\end{array}
\right].
\end{equation}

It is worth noting that the direct inspection of the eigenvalues is of difficult use, as analytical expressions for eigenvalues are hardly obtained for systems with many degrees of freedom $N$.

The determinant condition (\ref{det}) is strongly influenced by both the choice of the non-holonomic constraint, through the geometric stiffness matrix $\textbf{G}$ and vector $\bw$, and the choice of the loading conditions, through the term $\Xi$ appearing in the stiffness matrix (\ref{kblock}). Therefore, critical loads for $\beta_0 \neq0$ depend on the quasi-static configuration under consideration, on the type of non-holonomic constraint, and on the loading conditions.

\subsection{Buckling} 

Buckling is a quasi-static bifurcation of the equilibrium path driven by a slow increase in either $\Delta$ or $F$, providing a slow variation in the non-holonomic reaction $p^{\mbox{\tiny{QS}}}$.  
Therefore, buckling can be determined by setting the inertial terms to zero, $\omega=0$, in Eq.(\ref{pol}), thus imposing the condition $\rho_{m}=0$. 
Once $p^{\mbox{\tiny{QS}}}_{cr}$ is evaluated, $\Delta_{cr}$ or $F_{cr}$  follow from the last equation of the system (\ref{QSeqsextended}).

The buckling conditions $\rho_m=0$ for a generic quasi-static configuration is expanded in the Appendix \ref{segona} for both the non-holonomic constraints and both the loading conditions at the initial end. An interesting result is found, namely, that  {\it  buckling loads are excluded for all possible trivial or non-trivial solutions for the \lq violin bow' constraint with prescribed force $F$}, case (B), 
a result which is not obvious for $\beta_0 \neq 0$. 

For $\beta_0=0$ and for all considered non-holonomic constraints and loading conditions, the trivial configuration ($\bq^{\mbox{\tiny{QS}}}=\b0$) is the unique quasi-static solution (see Appendix \ref{appendixB}). Such statement leads to the obvious conclusion that buckling loads are absent, which is confirmed by the fact that the components of the vector $\textbf{w}$ always reduce to
\begin{equation}\label{wb0}
\textbf{w}^{(N)}=\textbf{0},\qquad w_{N+1}=1, 
\end{equation}
so that the buckling condition $\rho_m=0$ becomes
\begin{equation}
\label{detb0}
\det\left(\mathbb{K}^{(N)}+p^{\mbox{\tiny{QS}}}\mathbb{G}^{(N)}\right) =0,
\end{equation}
which is never satisfied as condition (\ref{dekSimp}) holds true.

Buckling will not be further considered in the following and the stability analysis will be restricted to the straight configuration of the column with $\beta_0=0$.

\subsection{Flutter and divergence instabilities for the straight column with $\beta_0=0$}

For every specific chain composed by $N$ rigid bars, it is proven now that all the critical loads (for flutter and divergence instability) are the same for both considered non-holonomic constraints (of the `skate' or `violin bow' type) in the case $\beta_0=0$. This statement can be proven by considering the Eq.(\ref{gN}), where in the case $\beta_0=0$ the coefficients $\gamma_i$ for both \lq skate' and `violin bow' constraints become the same $\gamma_{S,i}=\gamma_{V,i}=l_i$, so that
\beq\label{gtransp}
\textbf{G}_S=\textbf{G}_V^{\intercal}\qquad \text{and}\qquad \mathbb{G}_S^{(N)}=\mathbb{G}_V^{(N)\intercal}, 
\eeq
and therefore the geometric stiffness matrix for the \lq skate' constraint is equal to the transpose of the geometric stiffness matrix for the \lq violin bow' constraint. Due to the Eqs.(\ref{wb0}), the determinant condition (\ref{det}) reduces to 
\begin{equation}
\label{detb0-1}
\det{\left[
\mathbb{M}^{(N)}\omega^2+\mathbb{C}^{(N)}\omega+\mathbb{K}^{(N)}+p^{\mbox{\tiny{QS}}}\mathbb{G}^{(N)}
\right]}=0, 
\end{equation}
where the symmetric matrices $\mathbb{M}^{(N)}$, $\mathbb{C}^{(N)}$ and $\mathbb{K}^{(N)}$ do not depend on the type of non-holonomic constraint. The property of the determinant $\det \textbf{B}=\det \textbf{B}^\intercal$ closes the proof. In fact, the critical loads for both \lq skate' and `violin bow' constraints in the case $\beta_0=0$ are the same because the related characteristic polynomial is provided by the same determinant condition (\ref{detb0-1}). 

In the following, the analysis will be restricted to a double pendulum $N=2$.

\section{The double pendulum subject to the \lq skate' and \lq violin bow' constraints}

The geometric and inertial properties  are considered for simplicity coincident for the two bars, so that $\{l_i,\,m_i,\,d_i\}=\{l,\, m,\,d\}$ ($i=1,2$)
and $L=2l$, with a total mass $M=2m$.  By introducing the characteristic time  $T=L\sqrt{M/k}$ of the structure and the stiffness  $k=k_2$ for  the rotational springs, the following dimensionless quantities can be introduced
\begin{equation}
\begin{split}
&\chi=\frac{X}{L},\quad\tilde \Delta=\frac{\Delta}{L},\quad\tilde d=\frac{d}{l},\quad\tilde l=\frac{l}{L}=\frac{1}{2},\quad \tau=\frac{t}{T},\quad\tilde p=\frac{p L}{k},\quad\tilde F=\frac{F L}{k},\\
&\tilde M_X=\frac{M_X}{M},\quad\tilde M_L=\frac{M_L}{M},\quad \tilde m=\frac{m}{M}=\frac{1}{2},\quad \tilde I_{r,L}=\frac{I_{r,L}}{L^2 M}, \quad\tilde K=\frac{K L^2}{k}, \quad\tilde k_1=\frac{k_1}{k},
\\
&\tilde c_e=\frac{c_e L^2}{\sqrt{k M}},\quad \tilde c_{t,L}=\frac{c_{t,L} L}{\sqrt{k M}},\quad\tilde c_{r,L}=\frac{c_{r,L}}{L\sqrt{k M}},\quad\tilde c_{i}=\frac{c_{i}}{L\sqrt{k M}},
\end{split}
\end{equation}
and
\begin{equation}
\label{plofdimless}
\tilde\Xi=\left\{
\begin{array}{ll}
\dfrac{1}{2} \tilde K \left[\chi(\tau)-\tilde\Delta\right]^2\qquad\qquad & \text{for elastic device A,}\\[3mm]
-\tilde F\, \chi(\tau) &\text{for dead loading B.}
\end{array}
\right.
\end{equation} 
Henceforth, the superimposed dot \lq$\,\,\dot{ }\,\,$' denotes the derivative with respect to the dimensionless time  $\tau$. For both cases of the \lq skate' and \lq violin bow' non-holonomic constraints, the dimensionless expression for the non-linear equations of motion (\ref{ELeqs})  is given by\footnote{
\textcolor{red}{Equations (\ref{ZiegN2}a-c) are highlighted in red because they correct the corresponding equations (52a-c) in the published version of the present article.
The different terms, belonging to an early incorrect version, are found to be irrelevant for the results and conclusion drawn in the paper.
}
}
\textcolor{red}{
\begin{subequations}
\label{ZiegN2}
\begin{empheq}[left=\empheqlbrace]{align}
	0\,=\quad &\frac{\tilde c_e}{8} \left(\frac{4}{3} \,\dot\theta_1+\frac{1}{2} \, \dot\theta_2 \cos(\theta_1-\theta_2)-3 \sin \theta_1 \dot \chi\right)+          \notag\\
+&\tilde c_i \left(2 \dot\theta_1-\dot\theta_2\right)+\frac{\tilde c_{t,L}}{4} \left(\dot\theta_1+\dot\theta_2 \cos (\theta_1-\theta_2)-2\sin\theta_1 \dot \chi\right)+\notag\\
+&\frac{1}{8} \left[(\tilde d^2+1) \ddot\theta_1+\tilde d \,\ddot\theta_2 \cos (\theta_1-\theta_2)+\tilde d \,\dot\theta_2^2 \sin(\theta_1-\theta_2)-2\ddot \chi\sin \theta_1\,\left(\tilde d + 1\right)\right]+\\
+&\frac{\tilde M_L}{4} \left(\ddot\theta_1+\ddot\theta_2 \cos (\theta_1-\theta_2)+\dot\theta_2^2 \sin (\theta_1-\theta_2)-2\sin \theta_1 \ddot \chi\right)+ \tilde p \,\tilde a_1+(1+\tilde k_1) \theta_1-\theta_2\,,\notag \\[2ex]
	0\,=\quad &\frac{\tilde c_e}{8} \left(\frac{1}{2} \,\dot\theta_1 \cos (\theta_1-\theta_2)+\frac{1}{3}\, \dot\theta_2-\sin \theta_2 \dot \chi\right)+\notag\\\
+&\tilde c_{r,L} \,\dot\theta_2+\tilde c_i \left(\dot\theta_2-\dot\theta_1\right)+\frac{\tilde c_{t,L}}{4} \left(\dot\theta_1 \cos (\theta_1-\theta_2)+\dot\theta_2-2 \sin \theta_2 \dot \chi\right)+\notag\\
+&\frac{1}{8}\tilde d \Bigl(\tilde d \ddot\theta_2+\ddot\theta_1 \cos (\theta_1-\theta_2)-\dot\theta_1^2 \sin (\theta_1-\theta_2)-2 \sin \theta_2 \ddot \chi\Bigl)+\tilde I_{r,L} \ddot\theta_2+\\
+&\frac{\tilde M_L}{4} \left(\ddot\theta_1 \cos (\theta_1-\theta_2)-\dot\theta_1^2 \sin (\theta_1-\theta_2)+\ddot\theta_2-2 \sin \theta_2 \ddot \chi\right)+\tilde p \, \tilde a_2+\theta_2-\theta_1\,,\notag\\[2ex]
	0\,=\quad &\frac{\tilde c_e}{8} \left(8\dot \chi-3 \, \dot\theta_1 \sin \theta_1-\dot\theta_2 \sin \theta_2\right)+\frac{\tilde c_{t,L}}{2} \left(2\dot \chi-\dot\theta_1 \sin \theta_1-\dot\theta_2 \sin \theta_2\right)+\tilde M_X \ddot \chi +\notag\\
&+\frac{1}{4}  \left(4 \ddot \chi-(\tilde d+1) \,\ddot\theta_1 \sin \theta_1-(\tilde d+1) \,\dot\theta_1^2 \cos \theta_1-\tilde d\,\ddot\theta_2 \sin \theta_2-\tilde d\,\dot\theta_2^2 \cos \theta_2\right)+\\
&+\frac{\tilde M_L}{2} \left(2\ddot \chi-\ddot\theta_1 \sin \theta_1-\dot\theta_1^2 \cos \theta_1-\ddot\theta_2 \sin \theta_2-\dot\theta_2^2 \cos \theta_2\right)+\tilde p \, \tilde a_3 +\dfrac{\partial \,\tilde \Xi}{\partial\, \chi}\,,\notag\\[2ex]
\textcolor{black}{0\,=}\quad &\textcolor{black}{\tilde a_1\,\dot\theta_1 +\tilde a_2\, \dot\theta_2+ \tilde a_3\dot \chi\,,}
\end{empheq}
\end{subequations}
}
where $\{\chi,\,\theta_1,\,\theta_2\}$ are functions of the dimensionless time $\tau$. Note that Eqs. (\ref{ZiegN2}) represent the (dimensionless) generalization of the Eqs.(\ref{conserv}) when viscosities are taken into account. Moreover, Eqs.(\ref{ZiegN2}) can be directly applied to both the introduced non-holonomic constraints and loading conditions.

The dimensionless non-holonomic constraint (\ref{ZiegN2}d) is expressed through coefficients $\tilde a_i$ representing the dimensionless coefficients of Eqs.(\ref{nhol2zieg}) and (\ref{nhol2reut}). In particular, the coefficients $\tilde a_i$ are expressed
\begin{itemize}
\item for the \lq skate' non-holonomic constraint by
\begin{equation}
\label{zorro1}
\tilde a_1=\frac{1}{2} \,\sin (\beta_0-\theta_1+\theta_2),\quad
\tilde a_2=\frac{1}{2} \,\sin (\beta_0),\quad
\tilde a_3=\cos (\beta_0+\theta_2),
\end{equation}
\item for the \lq violin bow' non-holonomic constraint by
\begin{equation}
\label{zorro2}
\tilde a_1=\frac{1}{2} \,\sin (\beta_0-\theta_1),\quad
\tilde a_2=\sin (\beta_0)-\frac{1}{2} \,\sin (\beta_0-\theta_1),\quad
\tilde a_3=\cos (\beta_0).
\end{equation}
\end{itemize}

Henceforth, the analysis is restricted to the case when $\beta_0=0$, corresponding to the non-holonomic counterpart of the Ziegler and Reut double pendulums.  
The mass $\textbf{M}$, damping $\textbf{C}$ and elastic stiffness $\textbf{K}$ matrices in Eqs.(\ref{ELlineqs}) can easily be evaluated for the trivial equilibrium solution $\textbf{q}^{\mbox{\tiny{QS}}}=\b0$ and $\beta_0=0$.  Their dimensionless counterparts can be written as
\begin{equation}
\label{M2}
\tilde{\textbf{M}}^{(2)}=\left[
\begin{array}{ccc}
\frac{1}{8} \left(\tilde d^2 +1+2\tilde M_L\right) & \frac{1}{8}(\tilde d +2\tilde M_L) & 0 \\
\frac{1}{8} (\tilde d +2\tilde M_L) & \tilde I_{r,L} + \frac{1}{8}(\tilde d^2 + 2\tilde M_L) & 0 \\
0 & 0 & \tilde M_X + 1 + \tilde M_L
\end{array}
\right],
\end{equation}
\begin{equation}
\label{C2}
\tilde{\textbf{C}}^{(2)}=\left[
\begin{array}{ccc}
2\, \tilde c_i + \frac{1}{4} \tilde c_{t,L} + \frac{1}{6} \,\tilde c_e  & \frac{1}{4} \tilde c_{t,L} +\frac{1}{16}\tilde c_e-\tilde c_i & 0 \\[2mm]
\frac{1}{4} \tilde c_{t,L} + \frac{1}{16}\tilde c_e -\tilde c_i & \tilde c_{r,L} + \tilde c_i + \frac{1}{4} \tilde c_{t,L} + \frac{1}{24}\tilde c_e & 0 \\[1mm]
0 & 0 & \tilde c_{t,L} + \tilde c_e
\end{array}
\right],
\end{equation}
\begin{equation}
\label{K2}
\tilde{\textbf{K}}^{(2)}=\left[
\begin{array}{ccc}
1+\tilde k_1 & -1 & 0 \\
-1 & 1 & 0 \\
0 & 0 & \dfrac{\partial^2\,\tilde\Xi}{\partial \, \chi^2}
\end{array}
\right]. 
\end{equation}

The difference between the \lq skate' and the \lq violin bow' non-holonomic constraints lies in their respective geometric stiffness matrices $\bG_S$ and $\bG_V$, which for $\beta_0=0$ are one the transpose of the other, Eq.(\ref{gtransp}). The dimensionless expression for these matrices is
\begin{equation}
\label{G2}
\tilde{\textbf{G}}^{(2)}_S= \tilde{\textbf{G}}^{(2)\,\intercal}_V =\frac{1}{2}\left[
\begin{array}{ccc}
-1 & 1  & 0 \\
0 & 0 & 0 \\
0 & 0 & 0
\end{array}
\right].
\end{equation}

The characteristic polynomial defining the stability properties for both the \lq skate' and \lq violin bow' non-holonomic constraints, Eq.(\ref{detb0-1}), for a system of two rods ($N=2$) with $\beta_0=0$  becomes
\begin{equation}
\label{detbif2}
\det{\left[
\tilde{\mathbb{M}}^{(2)}\Omega^2+\tilde{\mathbb{C}}^{(2)}\Omega+\tilde{\mathbb{K}}^{(2)}+\tilde p^{\mbox{\tiny{QS}}}\tilde{\mathbb{G}}^{(2)}
\right]}=0 ,
\end{equation}
where $\Omega=\omega\,T$ is the dimensionless eigenvalue and $\tilde{\mathbb{M}}^{(2)}$, $\tilde{\mathbb{C}}^{(2)}$, $\tilde{\mathbb{K}}^{(2)}$ and $\tilde{\mathbb{G}}^{(2)}$ are the $2 \times 2$ upper-left corner partitions of matrices (\ref{M2}), (\ref{C2}), (\ref{K2}) and (\ref{G2}), respectively.

As a consequence of the determinant property $\det \bB=\det \bB^\intercal$ (holding for every matrix $\bB$), the same eigenvalues leading to the vanishing of the determinant (\ref{detbif2}), and therefore the same critical loads, are obtained for both non-holonomic constraints of the \lq skate' and \lq violin bow' type.

Note that the dimensionless mass $\tilde M_X$ of the sliding block does not appear in the mass matrix $\tilde{\mathbb{M}}^{(2)}$, Eq.(\ref{M2}), so that it does not influence the value of the critical loads.

Eq.(\ref{detbif2}) leads to a 4$^{th}$-order polynomial equation in the dimensionless eigenvalues $\Omega$,
\begin{equation}
\label{polyN2}
\begin{split}
\rho_0 \,\Omega ^4 
+ \rho_1 \Omega ^3+
\rho_2\Omega ^2 +
\rho_3 \Omega  
+\rho_4=0 ,
\end{split}
\end{equation}
where 
coefficients $\rho_i$ ($i=0,...,4$) can be evaluated through the Cayley-Hamilton theorem as (details are deferred to Appendix \ref{appendixC}, while the explicit expression for $\rho_i$ is included in the Supplementary Material)
\begin{equation}
\label{polyN22}
\begin{array}{cc}
\rho_0=\det\tilde{\mathbb{M}}^{(2)}, \qquad
\rho_1= {\text{tr}}\tilde{\mathbb{M}}^{(2)}\,{\textup{tr}}\tilde{\mathbb{C}}^{(2)}-\textup{tr}\left[\tilde{\mathbb{M}}^{(2)}\tilde{\mathbb{C}}^{(2)}\right], \qquad
\rho_2=
\textup{tr}\tilde{\mathbb{M}}^{(2)}\,\textup{tr}\tilde{\mathbb{K}}^{(2)}_{\textup{Tot}}-\textup{tr}\left[\tilde{\mathbb{M}}^{(2)}\tilde{\mathbb{K}}^{(2)}_{\textup{Tot}}\right]+\det \tilde{\mathbb{C}}^{(2)},
\\[4mm]
\rho_3=
\textup{tr}\tilde{\mathbb{C}}^{(2)}\,\textup{tr}\tilde{\mathbb{K}}^{(2)}_{\textup{Tot}}-\textup{tr}\left[\tilde{\mathbb{C}}^{(2)}\tilde{\mathbb{K}}^{(2)}_{\textup{Tot}}\right],  \qquad
\rho_4=\det \tilde{\mathbb{K}}^{(2)}_{\textup{Tot}}=\det \tilde{\mathbb{K}}^{(2)},
\end{array}
\end{equation}
where \lq $\tr$' defines the trace operator and the following definition has been used
\begin{equation}
\tilde{\mathbb{K}}^{(2)}_{\text{Tot}}=\tilde{\mathbb{K}}^{(2)}+\tilde p^{\mbox{\tiny{QS}}}\tilde{\mathbb{G}}^{(2)}.
\end{equation}

The coefficients $\rho_0$ and $\rho_4$ are assumed to be always greater than zero, a condition expressing a non-vanishing determinant for the matrices $\tilde{\mathbb{M}}^{(2)}$ and $\tilde{\mathbb{K}}^{(2)}$, respectively. Such assumption is automatically satisfied when $\tilde k_1 >0$ and at least one of the positive input parameters $\tilde d$, $\tilde M_L$ and $\tilde I_{r,L}$ is different from zero. Moreover, the coefficient $\rho_1$ is always greater than zero when at least one viscosity parameter takes a non null value.

The  eigenvalues $\Omega$ of the characteristic equation(\ref{polyN2}) can be therefore analyzed as functions of the reaction $\tilde p^{\mbox{\tiny{QS}}}$ of the non-holonomic constraint.

It is important to remark that the polynomial (\ref{polyN2}), obtained for the visco-elastic double pendulum subject to the non-holonomic constraint, coincides with the corresponding equation holding when a non-conservative load (of the Ziegler or Reut type) is applied, so that all the instability thresholds --flutter and divergence-- are shared between the non-conservative systems and their conservative counterparts subject to non-holonomic constraint. 
Therefore, 
\begin{center}
\emph{the presence of non-holonomic constraints may induce flutter and divergence instabilities in mechanical systems subject to conservative loads, 
a circumstance usually considered impossible.}
\end{center}

\paragraph{Flutter instability.} According to Ziegler \cite{ziegler}, the Hurwitz  matrix $\bold{H}_4$ for $N=2$, Eq.(\ref{hurwitzmat}) with $m=4$, is equal to
\begin{equation}
\label{hurwmatm4}
\begin{split}
\bold{H}_4&=\left[
\begin{array}{cccc}
 \rho_1 & \rho_0 & 0 & 0 \\
 \rho_3 & \rho_2 & \rho_1 & \rho_0 \\
 0 & \rho_4 & \rho_3 & \rho_2 \\
 0 & 0 & 0 & \rho_4 \\
\end{array}
\right].\\
\end{split}
\end{equation}

Because $\rho_1=D_1$ and $\rho_4$ are strictly positive, and considering that $D_4=\rho_4 D_3$, that the stability condition is given by the two following inequalities
\begin{equation}
\label{ineqsN2}
\left\{
\begin{array}{ll}
D_2=\rho_1\, \rho_2 - \rho_0\, \rho_3>0,\\
D_3=\left(\rho_1\, \rho_2 - \rho_0\, \rho_3\right)\, \rho_3 - \rho_1^2\, \rho_4>0.
\end{array}\right.
\end{equation}
The imposition of the two inequalities (\ref{ineqsN2}) is equivalent to impose only that related to the determinant $D_3$ subject to the condition of positive values for the coefficients $\rho_2$ and $\rho_3$. Therefore, the critical flutter load $\tilde p^{\mbox{\tiny{QS}}}_{\text{flu}}$ is given by the condition
\begin{equation}
\label{detflutter}
\left(\rho_1\, \rho_2 - \rho_0\, \rho_3\right)\, \rho_3 - \rho_1^2\, \rho_4=0,\qquad  \rho_2>0,\qquad \rho_3>0.
\end{equation}

\paragraph{Divergence instability.} 
Considering the real part of the eigenvalues $\Omega$, the bifurcation for divergence can be obtained as the roots of a fifth-order polynomial in the non-dimensional reaction $\tilde p^{\mbox{\tiny{QS}}}$ (see Appendix \ref{appendixD} for details) given by
\begin{equation}
\label{bifsN2}
\begin{split}
\left[27\left(\rho_1^2\, \rho_4+\rho_0\, \rho_3^2 \right) -9\, \rho_2 (8 \,\rho_0 \,\rho_4+\rho_1 \,\rho_3)+2\, \rho_2^3\right]^2-4 \left(\rho_2^2-3 \,\rho_1\, \rho_3+12\, \rho_0\, \rho_4\right)^3 =0.
\end{split}
\end{equation}
 The real roots $\tilde p^{\mbox{\tiny{QS}}}$ of Eq.(\ref{bifsN2}) can be exploited to investigate the divergence load corresponding to vanishing imaginary parts of the eigenvalues. The divergence load $\tilde p^{\mbox{\tiny{QS}}}_{\text{div}}$ can be therefore analytically computed as a root of Eq.(\ref{bifsN2}) corresponding to at least a couple of positive eigenvalues $\Omega$. Moreover, the Eq.(\ref{bifsN2}) is a quintic in the non-holonomic constraint reaction for the general case and a quartic in the special case $\tilde c_e=\tilde c_{t,L}=\tilde c_{r,L}=0$ and $\tilde c_i\neq 0$.

\subsection{Critical load for flutter and divergence in the ideal case of null dissipation}

In the ideal case of null dissipation, when all the damping forces are absent ($\tilde{\mathbb{C}}^{(2)}=\textbf{0}$) \lq from the beginning', the eigenvalues $\Omega$  can be evaluated as
\begin{equation}
\label{eig2nodamp}
\Omega=\pm\sqrt{\frac{\pm \sqrt{\left(\text{tr}\tilde{\mathbb{M}}^{(2)}\,\text{tr}\tilde{\mathbb{K}}^{(2)}_{\text{Tot}}-\text{tr}\left[\tilde{\mathbb{M}}^{(2)}\tilde{\mathbb{K}}^{(2)}_{\text{Tot}}\right]\right)^2-4 \det\tilde{\mathbb{M}}^{(2)} \det \tilde{\mathbb{K}}^{(2)}}-\left(\text{tr}\tilde{\mathbb{M}}^{(2)}\,\text{tr}\tilde{\mathbb{K}}^{(2)}_{\text{Tot}}-\text{tr}\left[\tilde{\mathbb{M}}^{(2)}\tilde{\mathbb{K}}^{(2)}_{\text{Tot}}\right]\right)}{2 \det\tilde{\mathbb{M}}^{(2)}}},
\end{equation}
so that the critical load $\tilde p^{\mbox{\tiny{QS}}}$ for flutter and divergence instabilities can be computed through the Eq.(\ref{bifsN2}) in the case $\rho_1=\rho_3=0$ (for which the Eq.(\ref{detflutter}) is automatically satisfied)
\begin{equation}
\label{bif2nodamp}
\left(\text{tr}\tilde{\mathbb{M}}^{(2)}\,\text{tr}\tilde{\mathbb{K}}^{(2)}_{\text{Tot}}-\text{tr}\left[\tilde{\mathbb{M}}^{(2)}\tilde{\mathbb{K}}^{(2)}_{\text{Tot}}\right]\right)^2-4 \det\tilde{\mathbb{M}}^{(2)} \det \tilde{\mathbb{K}}^{(2)}=0,
\end{equation}

The instability loads corresponding to flutter $\tilde p^{\mbox{\tiny{QS}}}_{\text{flu}}$ and divergence $\tilde p^{\mbox{\tiny{QS}}}_{\text{div}}$ can be written 
introducing the following notation
\begin{equation}\label{p0d0}
 \tilde p^{\mbox{\tiny{QS}}}_{\text{flu}}=\mathcal{P}_0(\bm{\xi}),\qquad
  \tilde p^{\mbox{\tiny{QS}}}_{\text{div}}=\mathcal{D}_0(\bm{\xi}) ,
\end{equation}  
where the subscript \lq 0' denotes the absence of damping coefficients within the mathematical model from the beginning and $\bm{\xi}=[\tilde d,\,\tilde M_L,\,\tilde I_{r,L},\,\tilde k_1]$ is the vector collecting all the input parameters describing the magnitude and distribution of mass and stiffness of the system. Using the above notation, it follows that
\begin{equation}
\label{explicitNodamp}
\begin{split}
\left\{
\begin{array}{cc}
\mathcal{P}_0(\bm{\xi})\\[2ex]
\mathcal{D}_0(\bm{\xi})
\end{array}\right\}
=\,2  \frac{\tilde{d}^2 (\tilde{k}_1+2)+2 \tilde{d}+8 \tilde{I}_{r,L} (\tilde{k}_1+1)+2 (\tilde{k}_1+4) \tilde{M}_{L}+1}{\tilde{d}^2+\tilde{d}+8 \tilde{I}_{r,L}+4 \tilde{M}_{L}}\\
\mp 4\frac{\sqrt{\tilde{k}_1 \left(\tilde{d}^4+8 \tilde{d}^2 \tilde{I}_{r,L}+2 \tilde{M}_{L} (2 (\tilde{d}-1) \tilde{d}+8 \tilde{I}_{r,L}+1)+8 \tilde{I}_{r,L}\right)}}{\tilde{d}^2+\tilde{d}+8 \tilde{I}_{r,L}+4 \tilde{M}_{L}}.
\end{split}
\end{equation}

\begin{figure}[!htb]
\begin{center}
\includegraphics[width=150mm]{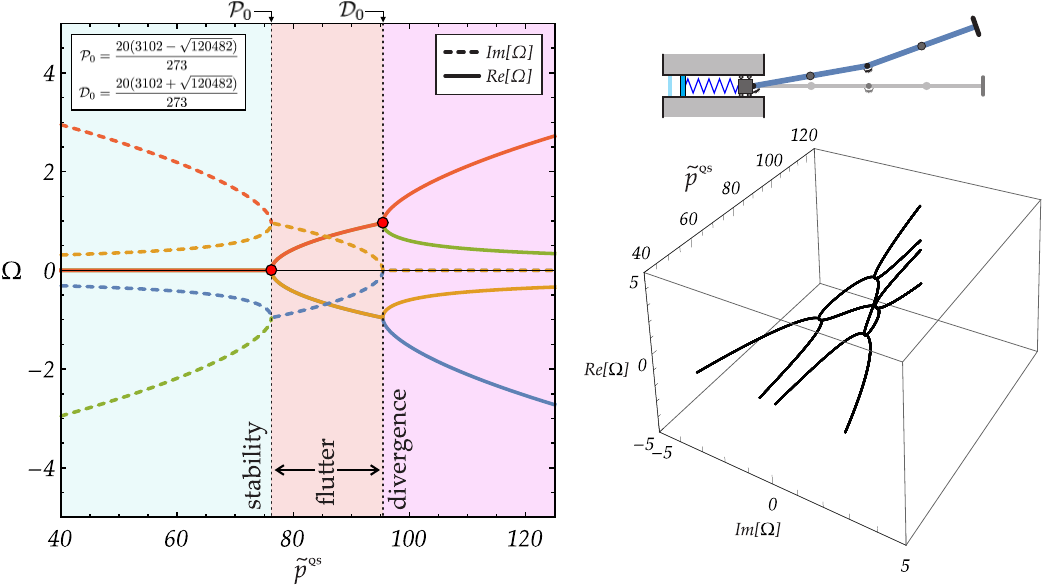}
\end{center}
\caption{Flutter and divergence loads for a (conservative) elastic double pendulum, subject to a non-holonomic constraint and loaded through compression of a linear spring. Real and imaginary parts of the eigenvalues $\Omega$ are reported as functions of the load $\tilde{p}^{\mbox{\tiny{QS}}}$. The ideal undamped case is considered, assuming $\tilde d=1/2$, $\tilde M_{L}=15$, $\tilde I_{r,L}=15$, $\tilde k_1=50$. 
}
\label{fig_ziegnodamp}
\end{figure}

The eigenvalues $\Omega$ for the ideal undamped double pendulum are investigated in Fig. \ref{fig_ziegnodamp}. Assuming $\tilde d=1/2$, $\tilde M_{L}=15$, $\tilde I_{r,L}=15$, $\tilde k_1=50$, or equivalently $\bm{\hat{\xi}}=[1/2,\,15,\,15,\,50]$, two different representations for the real and imaginary parts of $\Omega$ are shown as functions of the dimensionless load $\tilde{p}^{\mbox{\tiny{QS}}}$. The figure shows the presence of two critical loads corresponding to flutter (complex conjugate eigenvalues) and divergence (real eigenvalues) instabilities, which can be  
calculated from Eq.(\ref{explicitNodamp}) as  
\begin{equation}
\label{criticalnodamp}
\mathcal{P}_0(\bm{\hat{\xi}})=\frac{20}{723} \left(3102-\sqrt{120482}\right)\approx 76.2073,\qquad  \mathcal{D}_0(\bm{\hat{\xi}})=\frac{20}{723} \left(3102+\sqrt{120482}\right)\approx 95.4109.
\end{equation}

\subsection{The damped case and the Ziegler destabilization paradox for structures subject to non-holonomic constraints}

The introduction of viscous dissipative forces in the model of structures subject to non-conservative forces leads to the well-known \lq Ziegler destabilization paradox' \cite{bigonimiss22, kirillov_1, kirillov_libro, kirillov_2, ziegpara}. The paradox consists in a substantial (and unexpected) decrease in the critical load for flutter instability when a small damping is introduced. Furthermore, such critical load remains lower than the ideal one $\mathcal P_0$ (evaluated considering null viscosity \lq from the beginning', Eq.(\ref{explicitNodamp})) even in the limit of vanishing viscosity. 

In order to assess the existence of such a paradox in the presence of non-holonomic constraints, the critical loads for the double pendulum are obtained keeping into account all the four damping  sources introduced in the governing equations of motion. The evaluation of the critical loads causing flutter instabilities can be obtained analytically by means of the Routh-Hurwitz criterion (case $m=4$) through the evaluation of the minor determinants of (\ref{hurwmatm4}), and in particular via Eq.(\ref{detflutter}).

In order to evaluate the critical load of flutter and divergence by considering the effect of the four damping coefficients at the same time, the following notation is introduced
\begin{equation}
\label{Pflutot}
\tilde p^{\mbox{\tiny{QS}}}_{\text{flu}}=\mathcal{P}_d(\textbf{c},\,\bm{\xi}),\qquad \tilde p^{\mbox{\tiny{QS}}}_{\text{div}}=\mathcal{D}_d(\textbf{c},\,\bm{\xi}), 
\end{equation} 
where the subscript \lq $d$' highlights the presence of dissipation, while $\textbf{c}=\{\tilde c_{t,L},\, \tilde c_{r,L},\,\tilde c_i,\,\tilde c_e\}$ is the dissipation vector collecting the four viscosity parameters of the system. 
The dissipation vector can be represented as $\textbf c=r \bm{\varphi}$,  where $r \geq 0$ measures the dissipation magnitude while  $\bm{\varphi}$ collects the direction cosines within the four-dimensional space defined by $\textbf{c}$
\begin{equation}
\label{Pfluphi}
\bm{\varphi}=
\left[
\begin{array}{cc}
\cos \phi_1\\
\sin \phi_1\cos \phi_2\\
\sin \phi_1\sin \phi_2\cos \phi_3\\
\sin \phi_1\sin \phi_2\sin \phi_3\\
\end{array}
\right].
\end{equation}
The three angles $\phi_1$, $\phi_2$ and $\phi_3$ are restricted to range from 0 to $\pi/2$ for limiting the dissipation parameters to non-negative values, so that the critical load for flutter (\ref{Pflutot})$_1$ can be alternatively expressed as
\begin{equation}
\label{Pflusubst}
\mathcal{P}_d=\mathcal{P}_d(r,\,\phi_1,\,\phi_2,\,\phi_3,\,\bm{\xi}) .
\end{equation}

Eq.(\ref{Pfluphi}) allows to take limits for the vanishing of any number of damping coefficients, ranging from 1 to 4.  The limits for vanishing viscosities obtained keeping some of them fixed and the others tending to zero provide the same flutter loads that can be obtained by neglecting from the beginning the damping sources made to vanish, so that that there is no paradox. 
Therefore, the Ziegler paradox only occurs in the case when {\it all the damping sources are made to vanish}.
	
The limit value $\mathcal{P}_d^*$ for the flutter load $\mathcal{P}_d$, calculated when all 4 viscosities vanish, can be calculated by taking the limit for $r \to 0$, namely, 
\begin{equation}
\label{Pflulim}
\mathcal{P}_d^*(\phi_1,\,\phi_2,\,\phi_3,\,\bm{\xi})=\lim_{r\to 0}\mathcal{P}_d(r,\,\phi_1,\,\phi_2,\,\phi_3,\,\bm{\xi}).
\end{equation}
The symbolic expression of $\mathcal{P}_d^*$ for a generic direction $\bm{\varphi}$ and input parameters $\bm{\xi}$ (reported in Supplementary Material) is cumbersome so that a direct proof of specific behaviours becomes awkward. Nevertheless, based on a very large number of parametric analyses for the vector $\bm{\varphi}$, the following features were always found to be verified:
\begin{itemize}
\item The limit flutter load $\mathcal{P}_d^*$ strongly depends on the direction $\bm{\varphi}$ and is never higher than the ideal flutter load $\mathcal P_0$ found for the undamped system, Eq.(\ref{explicitNodamp}), namely, 
\begin{equation}
\label{Pflulim-1}
\mathcal{P}_d^*(\phi_1,\,\phi_2,\,\phi_3,\,\bm{\xi})\leq \mathcal P_0(\bm{\xi}),
\end{equation}
where, because damping coefficients are non-negative, the equality sign can be attained only for specific directions $\bm{\varphi}$.

The minimum value for the critical flutter  load is always the smallest of the four ones evaluated taking the limit of only one vanishing viscosity (while the remaining three are assumed null),
\begin{equation}
\min_{\phi_1,\,\phi_2,\,\phi_3}\{\mathcal{P}_d^*(\phi_1,\,\phi_2,\,\phi_3,\,\bm{\xi})\}\equiv
\min\{\mathcal{P}_d^*(0,0,0,\bm{\xi}),\,
\mathcal{P}_d^*(\pi/2,0,0,\bm{\xi}),\,
\mathcal{P}_d^*(\pi/2,\pi/2,0,\bm{\xi}),\,
\mathcal{P}_d^*(\pi/2,\pi/2,\pi/2,\bm{\xi})\}. 
\end{equation}
Differently, the maximum value can either be the maximum over the 4 mentioned values or $\mathcal{P}_0$, the latter obtained along special directions $\bm{\varphi}$.

\item The limit of the divergence load is independent of the direction $\bm{\varphi}$ and equal to the ideal divergence load $\mathcal D_0$ obtained for the  system with no damping \lq from the beginning',
\begin{equation}
\label{divlim}
\mathcal{D}_0(\bm{\xi})=\lim_{r\to 0}\mathcal{D}_d(r,\,\phi_1,\,\phi_2,\,\phi_3,\,\bm{\xi}) ,
\end{equation}
so that, in other words, no paradox is present for divergence instability load. 
\end{itemize}
In the following paragraphs, the analytical expressions for the critical flutter load are reported for the four specific cases where only one and two damping sources are present. 
Such equations are introduced to illustrate the different effects of dissipative sources on instability and to show that a viscosity-independent Ziegler paradox may exist.

\subsubsection{Influence of one damping source}

The influence of each dissipation source on the flutter loads is now analyzed in detail, while the remaining three sources are considered null. 

\paragraph{Internal damping $\tilde c_i$ ($\tilde c_e=\tilde c_{t,L}=\tilde c_{r,L}=0$).}
The flutter load can be evaluated through the Routh-Hurwitz criterion, Eq.(\ref{detflutter}), providing the expression for the critical load $\mathcal P_d(\tilde c_i,\,\bm{\xi})$  as a function of the internal damping coefficient and the input parameters $\bm{\xi}=[\tilde d,\,\tilde M_L,\,\tilde I_{r,L},\,\tilde k_1]$. The exact solution is reported in  the Supplementary Material, together with the specific case of vanishing damping, $\tilde c_i\rightarrow 0$. The limit value shows (the proof was obtained using the command \texttt{Reduce} of Mathematica$^{\tiny\textregistered}$) the occurrence of the Ziegler paradox for every values of $\bm{\xi}$, so that the critical load for flutter is never higher than that related to the ideal case of null damping \lq from the beginning'.

Assuming $\bm{\hat{\xi}}=[1/2,\,15,\,15,\,50]$ as considered in Fig.\ref{fig_ziegnodamp}, the value of the critical load of flutter is given by
\begin{equation*}
\mathcal P_d(\tilde c_i,\,\bm{\hat{\xi}})=\frac{64 }{723}\tilde c_i^2+\frac{4410529058}{57927483}, 
\end{equation*}
showing the existence of the Ziegler paradox as $\mathcal P_d(\tilde c_i\to 0,\,\bm{\hat{\xi}})\approx 76.1388$. 
The divergence load can be analytically evaluated through  Eq.(\ref{bifsN2}), which is now a quartic in the load, leading to (\ref{divlim}) for vanishing viscosity $c_i\to 0$, thus excluding Ziegler paradox for divergence.

\begin{figure}[!htb]
\begin{center}
\includegraphics[width=150mm]{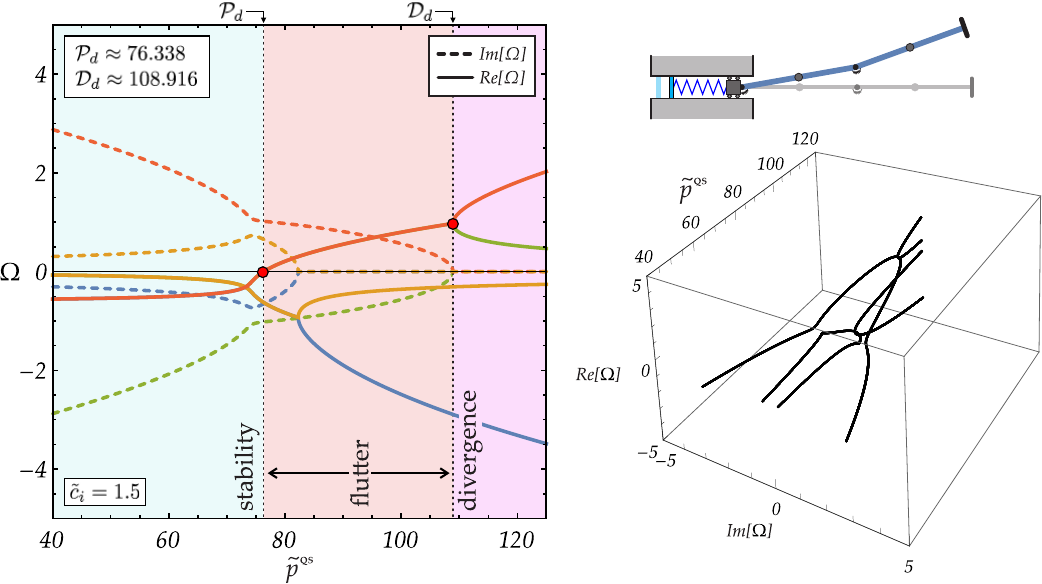}
\end{center}
\caption{
Flutter and divergence loads for a visco-elastic double pendulum, subject to a non-holonomic constraint and loaded through compression of a linear spring. Real and imaginary parts of the eigenvalues $\Omega$ are reported as functions of the load $\tilde{p}^{\mbox{\tiny{QS}}}$. 
The structure is the same reported in Fig.\ref{fig_ziegnodamp}, except that now viscosity is present, $\tilde c_i=1.5$ and $\tilde c_e=\tilde c_{t,L}=\tilde c_{r,L}=0$. 
}
\label{fig_ziegdamp}
\end{figure}

The real and imaginary parts of the eigenvalues, solutions of the 
fourth-order polynomial Eq.(\ref{polyN2}), are reported in Fig.\ref{fig_ziegdamp} 
as functions of the dimensionless reaction load $\tilde{p}^{\mbox{\tiny{QS}}}$ in the particular case $\tilde c_i=1.5$. Note that the critical loads for flutter and divergence instabilities are $\mathcal P_d \approx 76.338$ and $\mathcal D_d \approx 108.916$ respectively.

\paragraph{External damping $\tilde c_e$ ($\tilde c_i=\tilde c_{t,L}=\tilde c_{r,L}=0$).} 

The flutter load can be evaluated through the Routh-Hurwitz criterion, Eq.(\ref{detflutter}), providing the expression for the critical load $\mathcal P_d(\tilde c_e,\,\bm{\xi})$  as a function of the external damping coefficient $\tilde c_e$ and the parameters $\bm{\xi}=[\tilde d,\,\tilde M_L,\,\tilde I_{r,L},\,\tilde k_1]$. 
The expression of the critical load $\mathcal{P}_d(\tilde c_e,\bm{\xi})$ as a function of the external damping and the specific load for $\tilde c_e\to 0$ is reported in the Supplementary Material. Such limit load $\mathcal{P}_d^*$  is never higher than the ideal load $\mathcal{P}_0$ for any combination of the coefficients $\bm{\xi}$. 

Assuming the values $\bm{\hat{\xi}}=[1/2,\,15,\,15,\,50]$ used to generate Fig.\ref{fig_ziegnodamp}, the critical load for flutter can be calculated to be 
\begin{equation*}
\mathcal P_d(\tilde c_e,\,\bm{\hat{\xi}})= \frac{11566752096+25235 \tilde c_e^2-721 \sqrt{245 \left(5 \tilde c_e^2+1465344\right) \tilde c_e^2+40956144431616}}{169589580} , 
\end{equation*}
showing the existence of the Ziegler paradox $\mathcal P_d(\tilde c_e\to 0,\,\bm{\hat{\xi}})\approx 40.9964$ (while the divergence load again does not show any Ziegler paradox). 

It is also worth mentioning the limiting behaviour of the flutter load for infinite value of viscosity $\tilde c_e\to \infty$, which (differently from the limit values obtained with infinite values of internal viscosity $\tilde c_i$) remains finite,
\begin{equation}
\label{seghetta}
\lim_{\tilde c_e\to \infty}\mathcal P_d(\tilde c_e,\,\bm{\xi})=\frac{4 (\tilde k_1+8)}{5},
\end{equation} 
so that the external damping cannot eliminate flutter instability.
Eq.(\ref{seghetta}) reveals that for $\tilde k_1 = 50$ the dimensionless critical load at infinite external viscosity is approximately $46.4$, smaller than the ideal case, so that the critical load calculated for $\bm{\xi}=\bm{\hat{\xi}}$ and $\tilde c_e\geq 0$ is always inferior to the ideal case for every value of viscosity. The proof of the aforementioned property is obtained through the command \texttt{Reduce} of Mathematica$^{\tiny\textregistered}$.

\paragraph{Non-holonomic constraint  with translational damping $\tilde c_{t,L}$ ($\tilde c_i=\tilde c_e=\tilde c_{r,L}=0$).  } 
The presence of the translational damping $\tilde c_{t,L}$ at the non-holonomic constraint leads to the following statement: 
\begin{center}
{\it the value of the flutter load is independent of the translational damping coefficient and is never higher than that calculated for the ideal undamped case.}	
\end{center}
This  form of dissipation paradox has never been previously encountered and is denoted here as \lq viscosity-independent Ziegler paradox'.

In particular, according to the Routh-Hurwitz criterion (\ref{detflutter}), the flutter load is equal to
\begin{equation}
\mathcal P_d(\tilde c_{t,L},\,\bm{\xi})=\mathcal P_d^*(0,0,0,\bm{\xi})=\begin{array}{ll}
 \left\{  
\begin{array}{ll}
 \dfrac{2 \tilde{k}_1 ((\tilde{d}-1) \tilde{d}+8 \tilde{I}_{r,L})}{8 \tilde{I}_{r,L}-1}+4 , \qquad \quad &\text{if} \qquad (\tilde{d}-1) \tilde{d} \tilde{k}_1+8 \tilde{I}_{r,L} (\tilde{k}_1+2)<2, \\[2ex]
 \dfrac{16 \tilde{I}_{r,L}-2}{(\tilde{d}-1) \tilde{d}+8 \tilde{I}_{r,L}}+\tilde{k}_1  , \qquad &\text{otherwise}, \\
\end{array}\right.
 \\
\end{array}
\end{equation}
(expressions obtained using Mathematica$^\textregistered$) leading (through the command \texttt{Reduce}) to the conclusion that the critical load is independent of 
$\tilde c_{t,L}$ and 
never higher than the critical load $\mathcal P_0$, Eq.(\ref{explicitNodamp}), for the ideal case without damping.
Assuming $\bm{\hat{\xi}}=[1/2,\,15,\,15,\,50]$, the following value of the critical load is obtained 
\begin{equation*}
\mathcal P_d^*(0,0,0,\bm{\hat{\xi}})=\frac{24902}{479}\approx 51.9875\,<\,76.2073,
\end{equation*}
which is lower than the corresponding value for the undamped system, so that the Ziegler's destabilization paradox again occurs.
Eq.(\ref{bifsN2}) leads to a fifth-order polynomial in the loading parameter $\tilde p^{\mbox{\tiny{QS}}}$ that can be exploited in order to evaluate the analytical divergence load. Also in this case the divergence load as a function of the damping parameter $\tilde c_{t,L}$ tends to the ideal case for vanishing viscosity, so that no destabilization paradox for the divergence is observed.

\paragraph{Non-holonomic constraint with rotational damping $\tilde c_{r,L}$ ($\tilde c_i=\tilde c_e=\tilde c_{t,L}=0$).  } 
In the presence of only the rotational damping at the non-holonomic constraint $\tilde c_{r,L}$, {\it again the critical flutter load results to be independent of this damping coefficient and never higher than the value obtained for the ideal system without damping}. In particular, Eq.(\ref{detflutter}) leads to 
\begin{equation}
\mathcal P_d(\tilde c_{r,L},\,\bm{\xi})=\mathcal P_d^*\left(\frac{\pi}{2},0,0,\bm{\xi}\right)=\frac{2 \left(\tilde{d}^2+\tilde{d} \tilde{k}_1+\tilde{d}+2 (\tilde{k}_1+2) \tilde{M}_L+1\right)}{\tilde{d}^2+\tilde{d}+4 \tilde{M}_L+1} ,
\end{equation}
which is independent of $\tilde c_{r,L}$ and never higher than the critical load (\ref{explicitNodamp}) of the ideal case without damping. The proof of this statement was obtained using the command \texttt{Reduce} of the commercial code Mathematica$^\textregistered$.
Assuming $\bm{\hat{\xi}}=[1/2,\,15,\,15,\,50]$, the following value of the critical load for flutter is obtained
\begin{equation*}
\mathcal P_d^*\left(\frac{\pi}{2},0,0,\bm{\hat{\xi}}\right)=\frac{12694}{247}\approx 51.3927\,<\,76.2073, 
\end{equation*}
which is remarkably lower than the value obtained in the undamped case, thus confirming again the Ziegler paradox.

Again the  divergence load tends to the value of the ideal undamped system for vanishing viscosity, so that no destabilization paradox for the divergence is observed.

\subsubsection{Influence on stability of two damping coefficients}

Two examples related to the simultaneous presence of two damping coefficients are presented to highlight the disappearance of the Ziegler paradox when specific directions $\bm{\varphi}$ are considered. Such condition is strictly related to the specific choice of the vector of parameters $\bm{\xi}$, for instance it can be achieved with $\bm{\hat{\xi}}=[1/2,\,15,\,15,\,50]$ but it can not with $\bm{\check{\xi}}=[1/10,\,1/10,\,0,\,1]$.

\begin{figure}[!htb]
\begin{center}
\includegraphics[width=0.85\textwidth]{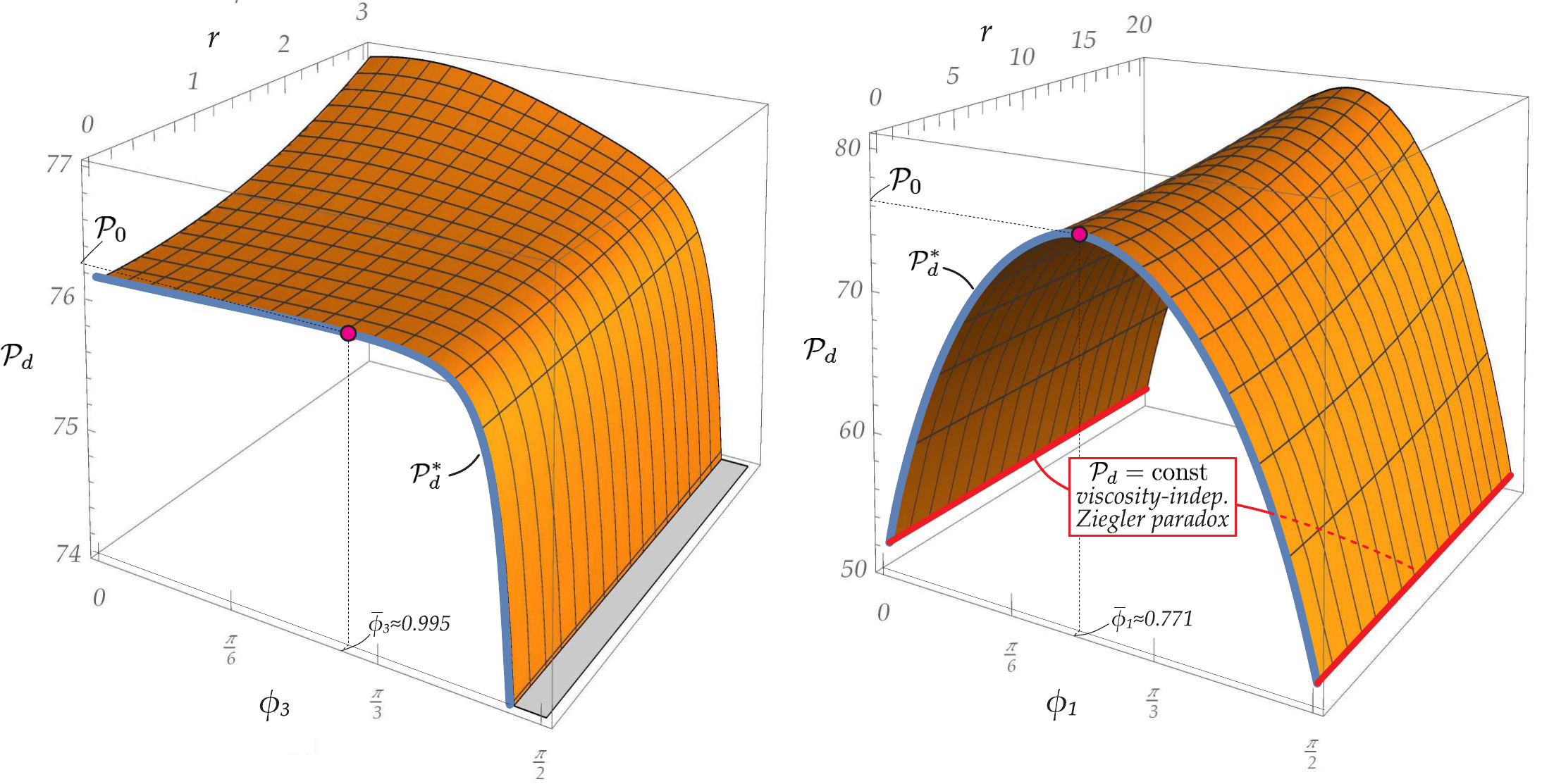}
\end{center}
\caption{(Left) Critical load $\mathcal P_d$ for flutter instability in a double visco-elastic pendulum with internal $\tilde c_i=r\cos\phi_3$ and external $\tilde c_e=r\sin\phi_3$ viscosities (but $c_{t,L}=\tilde c_{r,L}=0$), as a function of the modulus $r=\sqrt{\tilde c_{i}^2+\tilde c_{e}^2}$ and of the angular parameter $\phi_3$ ($\phi_1=\phi_2=0$). The ideal critical load $\mathcal P_0$ is recovered in the limit of vanishing viscosity only for $\bar\phi_3\approx 0.995$.  
	(Right) As for the left part, but with translational $\tilde c_{t,L}=r\cos\phi_1$ and rotational $\tilde c_{r,L}=r\sin\phi_1$ viscosities (while $\tilde c_i=\tilde c_e=0$), $r=\sqrt{\tilde c_{t,L}^2+\tilde c_{r,L}^2}$ and angular parameter $\phi_1$. The ideal critical load $\mathcal P_0$ is recovered in the limit of vanishing viscosity only for $\bar\phi_1\approx 0.771$. The \lq viscosity-independent Ziegler paradox' is highlighted by red lines, corresponding to a constant flutter load.
}
\label{fig_dampwheel}
\end{figure}

\paragraph{Combined effect of internal and external damping $\{\tilde c_i,\tilde c_e\} \neq 0$ ($c_{t,L}=\tilde c_{r,L})=0$.} This case has also been analyzed for the Ziegler's double pendulum in \cite{tommy,plaut}. The two non-null viscosities can be parametrized through $\tilde c_e=r \sin \phi_3$, $\tilde c_i=r \cos \phi_3$ (so that $r=\sqrt{\tilde c_e^2+\tilde c_i^2}$) and the related critical flutter load (reported in the Supplementary Material) can be evaluated through the Routh-Hurwitz criterion. The behaviour of the flutter load, when varying the magnitude $r$ and angle $\phi_3$, is shown in Fig.\ref{fig_dampwheel} (left) for $\bm{\hat{\xi}}=[1/2,\,15,\,15,\,50]$. Its limit value for vanishing both viscosities ($r\to 0$) along any direction $\phi_3$ is given by
\begin{equation}
\label{flulimitcice}
\begin{split}
P_d^*\left(\frac{\pi}{2},\,\frac{\pi}{2},\,\phi_3,\,\bm{\hat{\xi}}\right)=\frac{4}{5}\Biggl(&\frac{576 \cot \phi_3 (57927483 \cot \phi_3+10504342)+240974002}{54519984 \cot \phi_3+2826493}\\
&-\frac{3 \sqrt{2}(12568 \cot \phi_3+721)\sqrt{34704 \cot \phi_3 (11283138 \cot \phi_3-711833)+987561353}}{54519984 \cot \phi_3+2826493}\Biggl), 
\end{split}
\end{equation}
whose maximization yields
\begin{equation}
\max_{\phi_3}\left\{P_d^*\left(\frac{\pi}{2},\,\frac{\pi}{2},\,\phi_3,\,\bm{\hat{\xi}}\right)\right\}\equiv P_d^*\left(\frac{\pi}{2},\,\frac{\pi}{2},\,\phi_3\approx 0.995,\,\bm{\hat{\xi}}\right)\equiv P_0\left(\bm{\hat{\xi}}\right),
\end{equation}
namely, the maximum value of the critical flutter load is coincident with that of ideal case of null viscosity (further details can be found in the Supplementary Material).

\paragraph{The combined effect of translational and rotational damping for non-holonomic constraints $\{\tilde c_{t,L},\tilde c_{r,L} \}\neq 0$ ($\tilde c_i=\tilde c_e=0$).}
The parametrization of the two non-null viscosity parameters as
$\tilde c_{r,L}=r \sin \phi_1$, $\tilde c_{t,L}=r \cos \phi_1$ (so that $r=\sqrt{\tilde c_{r,L}^2+\tilde c_{t,L}^2}$) leads to the critical flutter load. For the sake of conciseness, the related expression is reported in the Supplementary Material. 

The behaviour of the flutter load, when varying the magnitude $r$ and angle $\phi_1$, is shown in Fig.\ref{fig_dampwheel} (right) for $\bm{\hat{\xi}}=[1/2,\,15,\,15,\,50]$. Its limit value for  vanishing both viscosities ($r\to 0$) along any direction  $\phi_1$  is given  by
\begin{equation}
\label{flulimitcNH}
\begin{split}
\mathcal P_d^*\left(\phi_1,\,0,\,0,\,\bm{\hat{\xi}}\right)=&\frac{440805 \cos (2 \phi_1)-3 (8854828 \sin (2 \phi_1)+6098513)}{6534 \cos (2 \phi_1)-288272 \sin (2 \phi_1)-234538}\\
&+\frac{(250 \sin \phi_1+241 \cos \phi_1) \sqrt{632330074-423120376 \sin (2 \phi_1)-12220470 \cos (2 \phi_1)}}{6534 \cos (2 \phi_1)-288272 \sin (2 \phi_1)-234538}, 
\end{split}
\end{equation}
whose maximization provides, similarly to the previous case,
\begin{equation}
\max_{\phi_1}\left\{\mathcal P_d^*\left(\phi_1,\,0,\,0,\,\bm{\hat{\xi}}\right)\right\}\equiv \mathcal P_d^*\left(\phi_1\approx 0.771,\,0,\,0,\,\bm{\hat{\xi}}\right)\equiv P_0\left(\bm{\hat{\xi}}\right). 
\end{equation}
Further details are available in the Supplementary Material.

Note in Fig.\ref{fig_dampwheel} (right) the two red lines showing an example of the \lq viscosity-independent Ziegler paradox', namely the critical load is constant for every modulus $r$ when $\phi_1=0$ ($\tilde c_{r,L}=0$) or when $\phi_1=\pi/2$ ($\tilde c_{t,L}=0$).

\section{Post-critical behaviour and limit cycles} \label{stokaz}

The linearized eigenvalue analysis so far developed shows that the straight configuration of the visco-elastic double pendulum is always stable (unstable) for loads smaller (higher) than the critical value for flutter, but nothing can be said about stability {\it at} 
the critical load and also on the stability of the post-critical dynamics involving large displacements. 

\begin{figure}[!htb]
	\begin{center}
		\includegraphics[width=0.8 \textwidth]{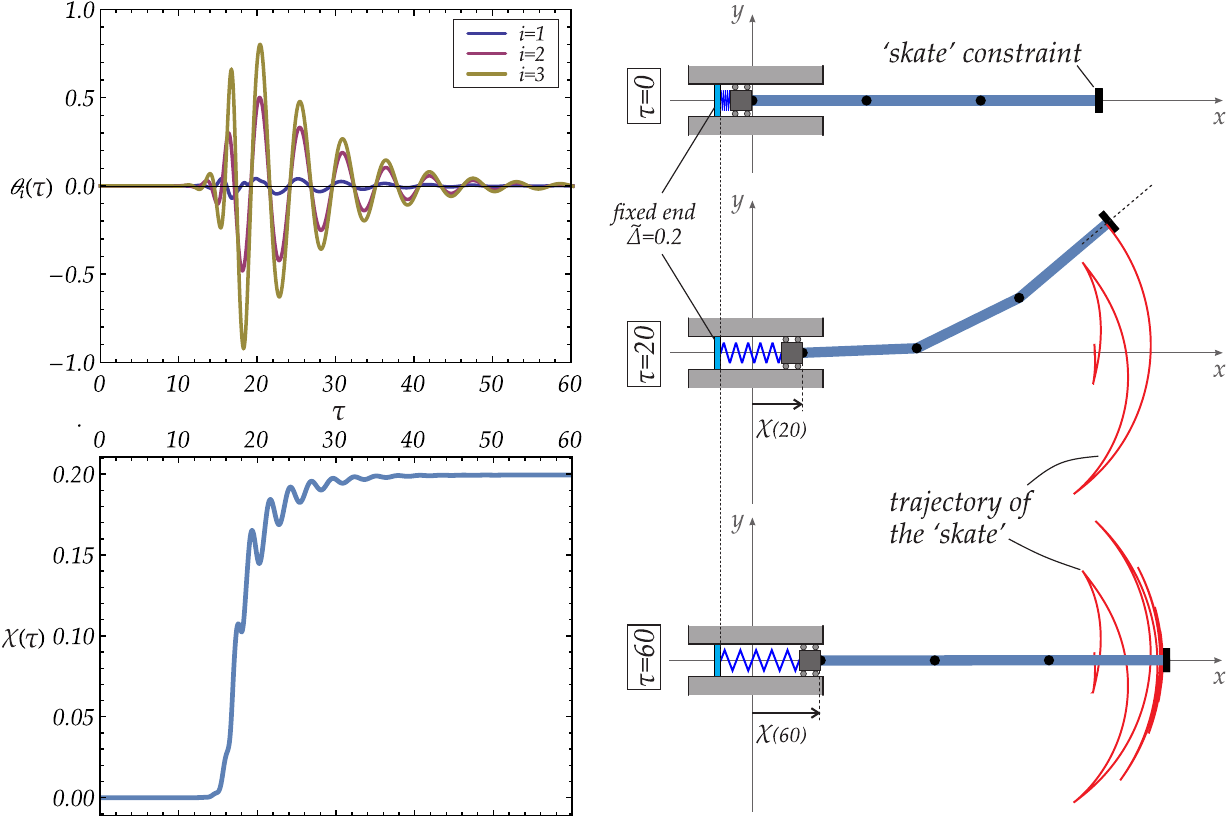}
	\end{center}
	\caption{Post-critical behaviour (in terms of 
		time evolution of the bars' rotations $\theta_i$ and of the global horizontal displacement $\chi$) for a visco-elastic triple ($N=3$) pendulum subject to the non-holonomic   \lq skate' constraint and loaded through a linear elastic device.  The dynamic instability occurs as the compressive force exceeds a critical value, thus generating  a global horizontal motion $\chi$. Eventually, oscillation  decays due to the progressive unloading of the spring, so that periodic motion cannot be achieved. Note the sharp corners evidenced by the trajectory of the end of the structure, typical of the non-holonomic constraint.}
	\label{fig_defspring}
\end{figure}

The post-critical behaviour has been analyzed numerically to show that, 
in the presence of dissipation, limit cycles can be attained only through application of a dead load $F$ (condition B), 
 while loading with the external elastic device (condition A) always  realizes a decaying motion. The latter behaviour is shown in Fig.\ref{fig_defspring} through the integration of the non-linear equations of motion for a triple visco-elastic pendulum. The elastic column is loaded through a linear spring compressed beyond the flutter load, while the initial conditions $\theta_i(0)=\si{10^{-7}}{}$ and $\dot\theta_i(0)=\si{0}{}$ ($i=1,2,3$) have been imposed for the bars' rotation and velocity, respectively. The structure is defined by $\bm{\bar{\xi}}=[1/2,\,1,\,0,\,10]$, $\tilde K=40$, $\tilde \Delta =0.2$ and $\tilde c_i=0.2$.

It is clear from Fig.\ref{fig_defspring} that the motion following instability is an oscillation which initially increases to relieve the compression in the spring, but later decreases and decays to zero as a consequence of the effect of dissipation. Note also the sharp corners visible in the trajectory of the skate and typical of the non-holonomic constraint.

In the following, the case of a double pendulum subject to dead loading $F$ is analyzed, where the Hopf theorem in $\mathbb{R}^4$ allows to analyze the stability at the critical point. In particular, it becomes possible to detect the correspondence of the flutter load with a Hopf bifurcation and therefore the existence of limit cycles in the neighbourhood of the critical load. 
Following Kutznetsov \cite{kutsnetsov} and Marsden and McCracken \cite{marsden}, the first Lyapunov coefficient is evaluated to discriminate 
between supercritical (positive coefficient) and subcritical (negative coefficient) Hopf bifurcation, respectively corresponding to the existence of stable or unstable periodic orbits in the neighbourhood of the bifurcation point. 

Hopf bifurcations are investigated by expressing the governing equations through the Hamiltonian formalism. Moreover, although the following equations are referred for simplicity to the double pendulum only, the approach remains valid for an $N$--link column.
The Lagrangian equations of motion (\ref{ZiegN2}) are independent of $\chi=X/L$, but depend on its derivatives $\dot\chi$ and $\ddot\chi$, so that it is possible to rewrite them by explicitly substituting such derivatives through the non-holonomic constraint, Eq.(\ref{ZiegN2}d). On the other hand, Eq.(\ref{ZiegN2}c) can be solved in the reaction force $p$, which can be substituted in Eqs.(\ref{ZiegN2}a) and (\ref{ZiegN2}b) to provide a system of two  non-linear equations in the two unknown rotations $\theta_1$ and $\theta_2$. The obtained system can be expressed via Hamiltonian formalism as
\begin{equation}
\label{eqmHam}
\dot{\textbf{x}}=\textbf{f}(\textbf{x},\tilde F),\qquad \textbf{x}=\left[\theta_1, \theta_2,\dot\theta_1,\dot\theta_2 \right]\,\,\in \,\,\mathbb{R}^{4},
\end{equation}
where $\tilde F$ is the dimensionless dead load, playing now the role of a bifurcation parameter, and $\textbf f: \mathbb{R}^{4}\times\mathbb{R}\rightarrow \mathbb{R}^{4}$ is an analytic and smooth function of $\textbf{x}$ and $\tilde F$.

By considering an equilibrium point $\textbf{x}_0$ such that
\begin{equation}
\textbf{f}(\textbf{x}_0,\tilde F)=\textbf{0}, 
\end{equation}
a Hopf bifurcation occurs \cite{golub,kutsnetsov,marsden} at a critical value of the bifurcation parameter $\tilde F=\tilde F_{\text{cr}}$ when both the following conditions hold: 
\begin{enumerate}[(i.)]

\item  the Jacobian matrix $\textbf{J}$ of the function $\textbf f$ at the equilibrium point $\textbf{x}_0$ and load $\tilde F_{\text{cr}}$ has a pair of purely-imaginary complex conjugate eigenvalues $\lambda_{1,2}=\pm i\, \omega_0$ ($\omega_0>0$), while the remaining two eigenvalues have negative real part; 

\item the imaginary eigenvalues $\lambda_{1}$ and $\lambda_{2}$ of the Jacobian matrix $\textbf{J}(\textbf{x}_0,\tilde F_{cr})$ cross the imaginary axis at a non-null \lq speed' (\emph{crossing condition})
\begin{equation}
\left.\frac{\partial \, \text{Re} \left[\lambda_{1,2}\right]}{\partial \tilde F}\right|_{\tilde F=\tilde F_{\text{cr}}}\neq 0. 
\end{equation}
\end{enumerate}
From the above discussion, the fulfilment of the two conditions (i.) and (ii.) guarantees the existence of periodic orbits in the neighbourhood of the equilibrium point $\textbf{x}_0$. It may also be concluded that Hopf bifurcations are excluded for null dissipations, because the real part of the eigenvalues is always zero until the critical flutter load is reached.

In the following, calculations are referred to the case of perfectly-aligned non-holonomic constraint $\beta_0=0$, so that the only possible equilibrium state for systems subject to both \lq skate' or \lq violin bow' constraints is the trivial configuration, $\textbf{x}_0=\textbf{0}$. 

In order to evaluate the stability condition for the Hopf bifurcation, the equations of motion (\ref{eqmHam}) are expanded up to the third order around the equilibrium configuration $\textbf{x}_0=0$ at a fixed value of the dead load $\tilde F=\tilde F_{\text{cr}}$	
\begin{equation}
\label{eqmexp}
\dot{\textbf{x}}=\textbf{J}\textbf{x}+\frac{1}{2} \textbf{B}(\textbf{x},\textbf{x})+\frac{1}{6}\textbf{C}(\textbf{x},\textbf{x},\textbf{x})+O\left(||\textbf{x}||^4\right),
\end{equation}
where $\textbf{B}$ and $\textbf{C}$ are symmetric multilinear vector functions of $\textbf{x}\in \mathbb{R}^4$. The stability of the Hopf bifurcation point has been assessed by exploiting two different but equivalent procedures, the first described by Kutsnetsov \cite{kutsnetsov} and the second by Marsden and McCracken \cite{marsden}:
\begin{itemize}
\item 
the right $\textbf{\textit{q}}$ and left $\textbf{\textit{p}}$ eigenvectors of the Jacobian matrix $\textbf{J}$ are obtained as
\begin{equation}
\label{qandp}
\textbf{J} \textbf{\textit{q}} = i\omega_0\, \textbf{\textit{q}}, \qquad \textbf{J}^\intercal \textbf{\textit{p}} = -i\omega_0\, \textbf{\textit{p}},
\end{equation}
and respectively correspond to the eigenvalues $i\omega_0$ and $-i\omega_0$.  A normalization of the eigenvectors $\textbf{\textit{q}}$ and $\textbf{\textit{p}}$ is introduced, so that $\left\langle \textbf{\textit{p}},\textbf{\textit{q}}\right\rangle=1$ and $\left\langle \textbf{\textit{q}},\bar{\textbf{\textit{q}}}\right\rangle=0$, where $\left\langle \textbf{\textit{a}},\textbf{\textit{b}}\right\rangle=\bar{a}_1 b_1+...+\bar{a}_{n} b_{n}$ denotes the scalar product in $\mathbb C^{n}$ while $\bar{\textbf{\textit{a}}}$ denotes the conjugate vector of $\textbf{\textit{a}}$.
The (real) so-called \emph{first Lyapunov coefficient} $l_1(0)$ \cite{kutsnetsov} can be finally evaluated as
\begin{equation}
\label{firstLyap}
l_1(0) = \frac{1}{2\omega_0} \text{Re}\left[\left\langle \textbf{\textit{p}},\textbf{C}(\textbf{\textit{q}},\textbf{\textit{q}},\bar{\textbf{\textit{q}}})\right\rangle
-2 \left\langle \textbf{\textit{p}},\textbf{B}(\textbf{\textit{q}},\textbf{J}^{-1}\textbf{B}(\textbf{\textit{q}},\bar{\textbf{\textit{q}}}))\right\rangle
+\left\langle \textbf{\textit{p}},\textbf{B}(\bar{\textbf{\textit{q}}},(2i\omega_0 \textbf{I}-\textbf{J})^{-1}\textbf{B}(\textbf{\textit{q}},\textbf{\textit{q}}))\right\rangle
 \right],
\end{equation}
where the sign of $l_1(0)$ provides the stability property of the critical point.
\item
  the third order Eq.(\ref{eqmexp}) is  transformed through a linear invertible change of basis $\textbf{x}=\textbf{A}\textbf{y}$ defined in a way that the transformed Jacobian matrix assumes the following canonical real Jordan form
\begin{equation}
\label{Jtilde}
\tilde{\textbf{J}}=\textbf{A}^{-1} \textbf{JA} =\left[
\begin{array}{ccccc}
 0 & \omega_0 & 0 &  0 \\
 -\omega_0 & 0 & 0& 0\\
 0 & 0 & \lambda_3 & 0 \\
 0 & 0 & 0 & \lambda_{4}
\end{array}
\right] , \\
\end{equation}
as the eigenvalues $\lambda_3$ and $\lambda_4$ have a negative real part. One can demonstrate (see proof in Supplementary Material) that the matrix $\textbf{A}$ can be written in terms of the eigenvector $\textbf{\textit{q}}$ defined in Eqs.(\ref{qandp}) and the right eigenvectors $\textbf{v}_3^{\text{r}}$ and $\textbf{v}_4^{\text{r}}$ related to $\lambda_3$ and $\lambda_4$ as
\begin{equation}
\label{atransform}
\textbf{A}=\left[\text{Re}\left[\textbf{\textit{q}}\right],\text{Im}\left[\textbf{\textit{q}}\right],\textbf{v}_3^{\text{r}},\textbf{v}_4^{\text{r}}\right].
\end{equation}
At this stage the center manifold theorem is exploited to \lq isolate' the two variables $[y_1,\,y_2]$ considered responsible for the instability, namely by imposing $[y_3,y_{4}]=\textbf{h}(y_1,y_2)$ in the neighbourhood of the bifurcation point, where $\textbf{h}=[h_1,h_2]$ is a vector of quadratic forms in the variables $[y_1,\,y_2]$. Finally, differentiations of the right-hand sides of the equations governing $[y_1,\,y_2]$ 
\begin{equation}
\left[
\begin{array}{c}
 \dot y_1  \\
 \dot y_2
\end{array}
\right]=\left[
\begin{array}{cc}
 0 & \omega_0\\
 -\omega_0 & 0
\end{array}
\right]\left[\begin{array}{c}
y_1  \\
y_2
\end{array}
\right]+\textbf{\textit{F}}(y_1,\,y_2,\,\textbf{h}(y_1,\,y_2)),
\end{equation}
where $\textbf{\textit{F}}$ is a smooth function  containing all the non-linear terms, lead to the expression for the real term called $V'''(0)$ by Marsden and McCracken (Eq.(4.2) at page 126 of \cite{marsden}), whose sign provides the stability property of the critical point. 
\end{itemize}

In summary, a Hopf bifurcation occurs at a critical point once the two conditions (i.) and (ii.) for the validity of the Hopf theorem are satisfied. The stability of the periodic orbits in the neighbourhood of such critical point can be classified on the basis of the sign of the first Lyapunov coefficient $l_1(0)$ (Eq.(\ref{firstLyap})) or,  equivalently, through that of $V'''(0)$ as
\begin{equation}
l_1(0)\,\,\,\mbox{or}\,\,\,V'''(0) \,\,\, \left\{
\begin{array}{lll}
<0\quad \rightarrow  \quad \text{supercritical Hopf bifurcation (stable orbits),}\\
=0\quad \rightarrow  \quad \text{critical case},\\
>0\quad \rightarrow  \quad \text{subcritical Hopf bifurcation (unstable orbits).}
\end{array}\right.
\end{equation}

\subsection{Limit cycles by applying a dead load $F$} 

\begin{figure}[!htb]
\begin{center}
\includegraphics[width=0.99\textwidth]{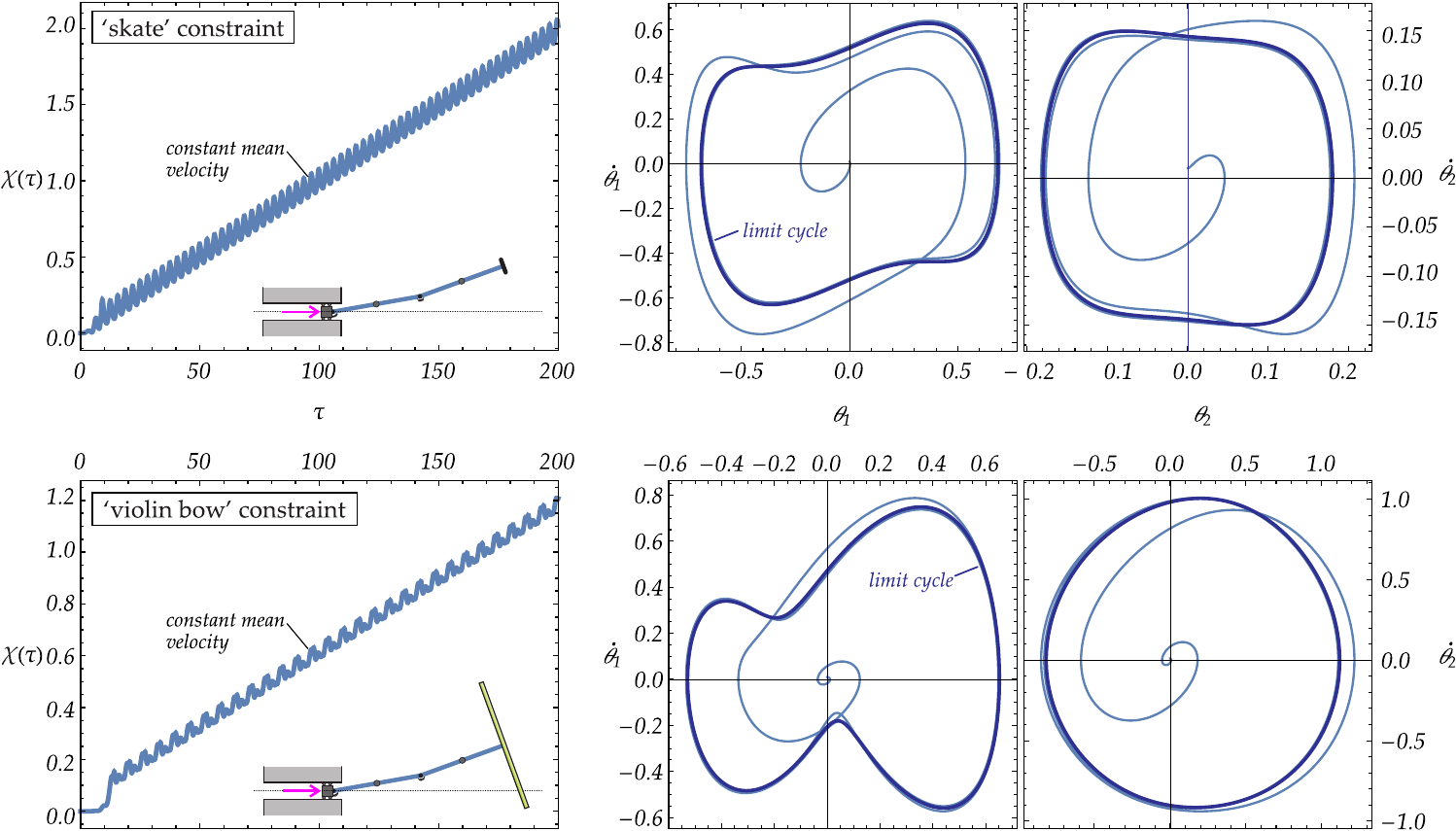}
\end{center}
\caption{Evolution in time of the global horizontal displacement $\chi$ (left) and phase portraits $\{\theta_i,\dot\theta_i\}$ for the rotations of the two rigid bars, (right, with limit cycles highlighted in dark blue), for a double pendulum subject to a \lq skate' (upper part) and \lq violin bow' (lower part) constraints and loaded through a dead load $\tilde F=1.1 \tilde F_{cr}$. Note that the post-critical behaviour is different if a \lq skate' or a \lq violin bow' constraint applies, even though the same initial conditions, viscosities, and input parameters have been imposed and even though the two structures have the same critical flutter load.}
\label{fig_RZcycles}
\end{figure}

The conditions obtained in the previous section are now used together with numerical integration to analyze the post-critical dynamics of a visco-elastic double pendulum $N=2$ subject to non-holonomic constraints.
Two numerical simulations are presented in Fig.\ref{fig_RZcycles} referred to the same initial conditions ($\theta_i(0)=0$ and $\dot\theta_i(0)=10^{-2}$), damping coefficients ($\tilde c_i=1.5$, $\tilde c_e=\tilde c_{t,L}=\tilde c_{r,L}=0$), input parameters ($\bm{\hat{\xi}}=[1/2,\,15,\,15,\,50]$) and dead load ($\tilde F/\tilde F_{cr}=1.1$ being $\tilde F_{cr}\approx 76.338$). Both the \lq skate' and \lq violin bow' constraints are considered. 

The following two pairs of complex conjugate eigenvalues are obtained at the critical point
\begin{equation*}
\lambda_{1,2}\approx \pm 1.01923\,i , ~~~~
\lambda_{3,4}\approx -0.625886\pm 0.653007\,i,
\end{equation*}
so that criterion (i.) of the Hopf theorem is satisfied, together with the crossing condition (ii.) 
\begin{equation*}
\left.\frac{\partial \, \text{Re} \left[\lambda_{1,2}\right]}{\partial \tilde F}\right|_{\tilde F=\tilde F_{\text{cr}}}\approx 0.0744935\neq 0 ,
\end{equation*}
showing the existence of a Hopf bifurcation at the critical point. 

The first Lyapunov coefficient (or equivalently $V'''(0)$ from \cite{marsden}) calculated at the bifurcation point is negative for both systems, so that a supercritical Hopf bifurcation occurs and the periodic orbits are stable. In fact, calculations performed with the commercial code Mathematica$^\textregistered$ show that
\begin{equation*}
l_1(0)= \frac{V'''(0)}{3\pi}\approx
\left\{
\begin{array}{lll}
-2.1838~~~ \qquad \text{for the \lq skate' constraint};
\\
-0.900238  \qquad \text{for the \lq violin bow' constraint.}
\end{array}
\right.
\end{equation*}
Despite the identity of the linearized stability, the two systems based on different non-holonomic constraints display different post-critical behaviours, and consequently different shapes of their limit cycles. However, for the particular choices of viscosities and input parameters, both systems show a supercritical Hopf bifurcation at the critical point and are characterized by periodic stable orbits, which exhibit increasing amplitudes at increasing $\tilde{F}$ (beyond the critical load for flutter, but within a finite interval). 

The time evolution of the global horizontal displacement $\chi$ is reported in Fig.\ref{fig_RZcycles} (left), showing that the velocity of the rigid block quickly approaches an oscillatory behaviour with a constant mean value. Meanwhile, the mechanical system achieves a limit cycle, shown in the phase portraits in Fig.\ref{fig_RZcycles} (right), which, according to the Hopf theorem, is stable at least in a neighbourhood of the critical load for flutter.

\begin{figure}[!htb]
\begin{center}
\includegraphics[width=0.90\textwidth]{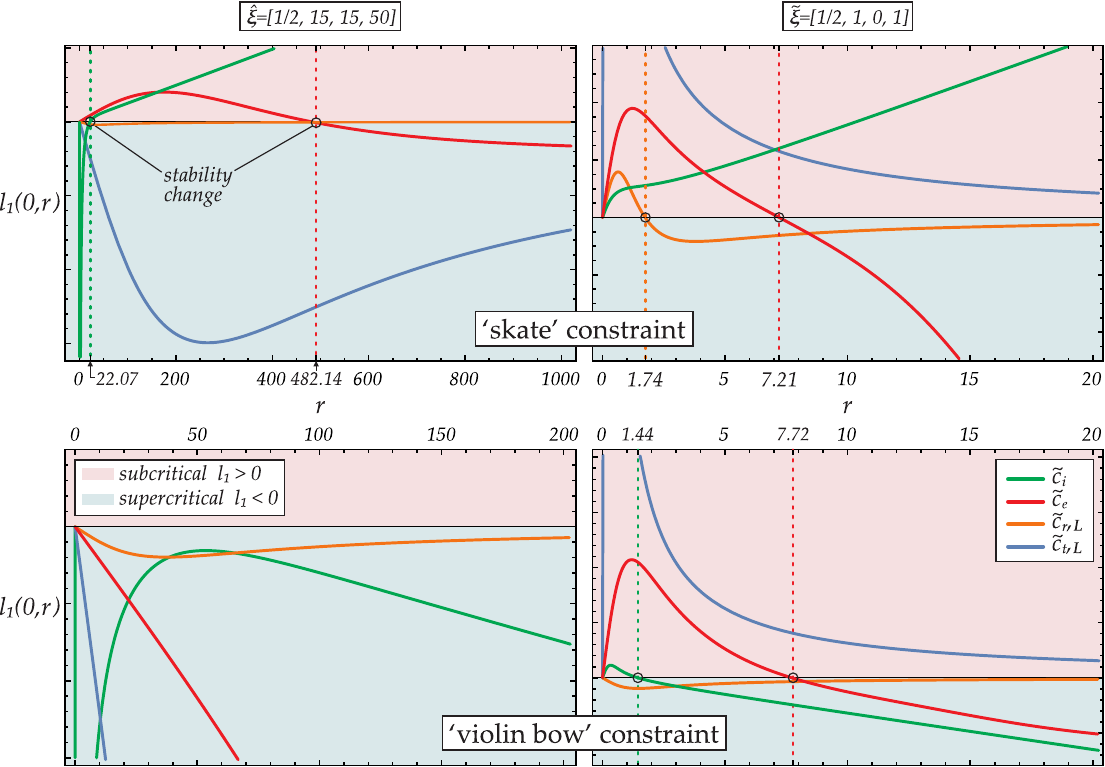}
\end{center}
\caption{First Lyapunov coefficient $l_1$ (only the sign matters, so that ticks are not reported  on the vertical axis) as a function of the magnitude of only one of the four damping parameters (the other three are assumed null) for the \lq skate' constraint (upper part) and for the \lq violin bow' constraint (lower part). The coefficient discriminates between supercritical and subcritical Hopf bifurcations and was obtained for two sets of coefficients:  $\bm{\hat{\xi}}=[1/2,\,15,\,15,\,50]$ (left part) and $\bm{\tilde{\xi}}=[1/2,\,1,\,0,\,1]$ (right part). Changes in stability are highlighted by dashed lines and show that subcritical Hopf bifurcations cannot be obtained when the structure with $\bm{\hat{\xi}}=[1/2,\,15,\,15,\,50]$ ends with the \lq violin bow' constraint (lower left corner).}
\label{fig_lyapunovcoeff}
\end{figure}

The stability of the critical points can be evaluated by expressing the first Lyapunov coefficient in Eq.(\ref{firstLyap}) as a function of the damping parameters, so that the domain of stability of the Hopf bifurcations, given by the sign of $l_1(0)$, can be directly expressed in terms of the magnitude of the damping parameters acting on the system. This calculation was possible only by fixing the values of $\bm{\xi}$ and by considering only one damping coefficient acting on the system. Two examples of the calculated Lyapunov coefficients (functions of one of the four damping sources assumed of magnitude $r$), are reported in Fig.\ref{fig_lyapunovcoeff}, providing a sketch of the qualitative stability behaviour. 

Fig.\ref{fig_lyapunovcoeff} shows that, depending on the magnitude of dissipation, the sign of the Lyapunov coefficient may or may not change, so that the Hopf bifurcations may be either supercritical ($l_1(0)<0$) or subcritical ($l_1(0)>0$).
 Changes in stability are highlighted in the figure by dashed lines. In the case of the \lq skate' constraint with $\bm{\hat{\xi}}=[1/2,\,15,\,15,\,50]$ (upper left corner) it can be concluded that both damping sources $\tilde c_{i}$ (reported green) and $\tilde c_{e}$ (reported red) can stabilize the bifurcation point if they overcome a certain threshold value (respectively, $\approx 22.07$ and $\approx 482.14$). On the other hand, no subcritical Hopf bifurcations can be obtained at varying every damping source (acting alone) when the structure with $\bm{\hat{\xi}}$ ends with the \lq violin bow' constraint (lower left part of the figure).

It may be interesting to remark, in closure of this Section, that analyses (not reported for conciseness) show that structures subject to dead loading reach a limit steady motion with constant {\it mean} velocity even when only the viscosity of the hinges is present. This situation is in contrast to what happens in the absence of viscosity where the system, subject to a constant force, is subjected to an increasing velocity. In other words, the rotational viscosity of the hinges, taken alone, acts as the ambient viscosity during the falling of a mass in a gravitational field.

\subsection{A note on structures with non-aligned constraint ($\beta_0\neq 0$)}

\begin{figure}[!htb]
\begin{center}
\includegraphics[width=0.95\textwidth]{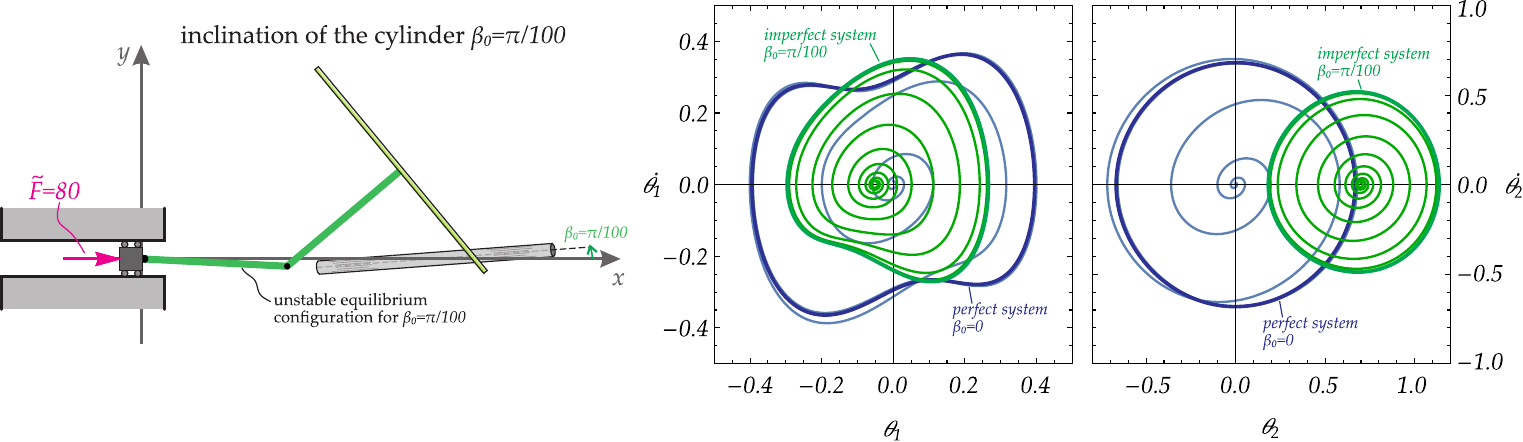}
\end{center}
\caption{Phase portraits (green curves, in terms of bar's rotations $\theta_1$ on the center, $\theta_2$ on the right) showing the achievement of a limit cycle for a visco-elastic double pendulum with a misaligned \lq violin bow' constraint, $\beta_0=\pi/100 = 1.8^\circ$ (shown on the left). The phase portrait of the perfect system $\beta_0=0$ is also reported for comparison (blue curve). 
	The initial equilibrium configuration for the imperfect system exhibits a large deviation from rectilinearity ($\theta_1^{\tiny{\mbox{QS}}}\approx -0.050282$, $\theta_2^{\tiny{\mbox{QS}}}\approx 0.701512$, shown on the left) and is unstable, so that the system starts oscillating with increasing amplitudes towards a stable limit cycle. }
\label{fig_Hopfimperfection}
\end{figure}

Until now the non-holonomic constraints have always be assumed to be aligned parallel to the last bar, $\beta_0=0$. Therefore, it may be interesting to explore the effect of constraint misalignment. In particular, a case of double pendulum with non-aligned \lq violin bow' constraint, $\beta_0=\pi/100 = 1.8^\circ$, is investigated and the phase portraits of the two bars' rotations is shown in Fig.\ref{fig_Hopfimperfection}, reported as green curves, together with the blue curves representing the perfect system ($\beta_0=0$, analyzed at a load  $\tilde F=80$, beyond the flutter load $\approx 76.338$). 

A non-trivial equilibrium solution for the double pendulum with $\beta_0=\pi/100$ is found for $\theta_1^{\tiny{\mbox{QS}}}\approx-0.050282$ and $\theta_2^{\tiny{\mbox{QS}}}\approx 0.701512$ (shown in Fig.\ref{fig_Hopfimperfection} on the left) at a load $\tilde F=80$, beyond the flutter load ($\approx 78.276$). For this configuration, the solutions of the characteristic equation (\ref{pol}) (with $m=4$) consist in two pairs of complex conjugate eigenvalues, where only two eigenvalues have a positive real part, so that the equilibrium configuration is unstable. 

The non-trivial equilibrium configuration is used to analyze the unstable dynamics of the system by imposing the initial conditions $\theta_i(0)=\theta_i^{\tiny{\mbox{QS}}}$ and $\dot\theta_i(0)=10^{-2}$ (the coordinates of the center of the spirals representing the phase portraits in Fig.\ref{fig_Hopfimperfection}) and assuming $\tilde c_i=1.5$, $\tilde c_e=\tilde c_{t,L}=\tilde c_{r,L}=0$ and $\bm{\hat{\xi}}=[1/2,\,15,\,15,\,50]$.

The following observations can be drawn from Fig.\ref{fig_Hopfimperfection}.

\begin{itemize}
	
	\item The non-trivial equilibrium solution corresponding to a small imperfection $\beta_0=\pi/100$ exhibits a large deviation from rectilinearity;
	
	\item The flutter load for the non-trivial solution ($\approx 78.276$) is slightly higher than that relative to the trivial solution ($\approx 76.338$);
	
	\item The phase portraits for the imperfect system confirm  the existence of periodic orbits in the presence of imperfections, intended now as a misalignment of the non-holonomic constraint;
	
	\item When the phase portraits of the perfect and imperfect systems are compared, the former possess a symmetry which is lost in the latter. Moreover, the oscillations of the perfect system exhibit larger amplitudes than the imperfect one. 
	
\end{itemize}

\section{Locomotion and friction}

\paragraph{Limbless locomotion.} The periodic solutions so far obtained for the analyzed structures can be exploited in the design of limbless locomotion devices. In fact, the time evolution of the rigid bars' rotations exhibiting a generic limit cycle can directly be imposed to the same structures which are left only with the horizontal motion $X$ of the end of the structure as \lq free coordinate'. With this set-up, the non-holonomic constraint is capable of converting the imposed oscillations of the rigid bars into an horizontal displacement, providing the locomotion to the system. 

An example of the dimensionless equations of motion of a double pendulum with controlled rotations at both ends and with an appended dead load $\tilde F$ at the left end of the structure is given by\footnote{
\textcolor{red}{Equation (\ref{ZiegN2Active}a) is highlighted in red because it corrects the corresponding equation (96a) in the published version of the present article.
The different terms, belonging to an early incorrect version, are found to be irrelevant for the results and conclusion drawn in the paper.
}
}
\textcolor{red}{
 \begin{subequations}
\label{ZiegN2Active}
\begin{align}
&\frac{\tilde c_e}{8} \left(8\dot \chi-3 \, \dot\theta_1(\tau) \sin \theta_1(\tau)-\dot\theta_2(\tau) \sin \theta_2(\tau)\right)+\frac{\tilde c_{t,L}}{2} \left(2\dot \chi-\dot\theta_1(\tau) \sin \theta_1(\tau)-\dot\theta_2(\tau) \sin \theta_2(\tau)\right)+\tilde M_X \ddot \chi +\notag\\
&+\frac{1}{4}  \Bigl(4 \ddot \chi-(\tilde d+1) \,\ddot\theta_1(\tau) \sin \theta_1(\tau)-(\tilde d+1) \,\dot\theta_1^2(\tau) \cos \theta_1(\tau)-\tilde d\,\ddot\theta_2(\tau) \sin \theta_2(\tau)-\tilde d\,\dot\theta_2^2(\tau) \cos \theta_2(\tau)\Bigr)+\\
&+\frac{\tilde M_L}{2} \left(2\ddot \chi-\ddot\theta_1(\tau) \sin \theta_1(\tau)-\dot\theta_1^2(\tau) \cos \theta_1(\tau)-\ddot\theta_2(\tau) \sin \theta_2(\tau)-\dot\theta_2^2(\tau) \cos \theta_2(\tau)\right)\notag\\
&+\tilde p \, \tilde a_3 +\tilde F =0, \notag 
\end{align}
\end{subequations}
}
$$
\tilde a_1\,\dot\theta_1(\tau) +\tilde a_2\, \dot\theta_2(\tau)+ \tilde a_3\dot \chi=0,
~~~~~~~~~~~~~~~~~~~~~~~~~~~~~~~~~~~~~~~~~~~
~~~~~~~~~~~~~~~~~~~~~~~~
~~~~~~~~~~~~~~~~~~~~~~~~(96\mbox{b})    
$$
where $\theta_1(\tau)$ and $\theta_2(\tau)$ are time-dependent rotations {\it imposed to} the two rigid bars and representing the periodic solution of the same structure with free and unknown rotations. The coefficients $\tilde a_i$ in Eq.(\ref{ZiegN2Active}) are expressed by Eqs.(\ref{zorro1}) for the \lq skate' and (\ref{zorro2}) for the \lq violin bow' non-holonomic constraints.

In conclusion, it is suggested that the rotations $\theta_1(\tau)$ and $\theta_2(\tau)$ can be \lq borrowed' from the solution of a structure subject to a dead force. Then, such a solution can be exploited as imposed rotations  for a similar double-pendulum, but characterized in general by different properties. In this way, because the coefficients $\tilde{a}_i$ in Eq.(\ref{ZiegN2Active}b) are independent of the structural parameters, the two systems display the same dimensionless displacement of the left end $\chi(\tau)$, so generating a steady motion with a constant mean velocity. Therefore, the controlled system is capable of moving any dead load by tuning the  reaction at the non-holonomic constraint, Eq.(\ref{ZiegN2Active}a).

\paragraph{Friction between two solids.} Another possible application of the results so far obtained is in the problem of friction between two different solids in contact. In fact, the non-holonomic constraint corresponds to a situation of \lq directional infinite friction', so that motion is impossible in all the analyzed structures (with $\beta_0=0$) if instability does not occur. If the non-holonomic constraint is believed to remain valid within a friction cone, instability may decrease the force needed to move the system.
In other words, if the structures are regarded as \lq black boxes' which are to be displaced through the application of an external force, perceived as the force needed to overcome some friction, then this force can be strongly decreased by the instability.  As a conclusion, the results presented in this article can find application in the analysis of microstructural effects at the contact between two solids, to possibly explain an \lq unexpected drop' in the applied forces to produce relative motion between the two solids. 

\section{Conclusions}

It has been shown that flutter instability and Hopf bifurcations can be displayed by visco-elastic structures under {\it conservative loads}, when subject to non-holonomic constraints. This finding opens a new perspective in structures suffering instabilities and provides a proof of the mechanical equivalence between non-holonomic conditions and \textit{polygenic} forces, which cannot be derived from any scalar functional (for instance follower loads) \cite{lanczos}.
In this context, the reaction force related to a non-holonomic constraint can be seen as the polygenic force applied to the  Ziegler or Reut pendulum, but now with possibly varying modulus, such that the kinematic non-holonomic condition is satisfied. Under this circumstance, the work done by the force is always null.
The obtained results may find applications in limbless locomotion and as possible explanation of micromechanics leading to instabilities in frictional contact between two solids. 

\vspace*{5mm} \noindent {\sl Acknowledgments} 

Financial support is acknowledged from: ERC Advanced Grant ERC-2013-ADG-340561-INSTABILITIES (AC),
PRIN 2015 2015LYYXA8-006, PITN-GA-2019-813424-INSPIRE, and  ARS01-01384-PROSCAN (DB and FDC).
The authors also acknowledge support from the Italian Ministry of Education, University and Research (MIUR) in the frame of the `Departments of Excellence' grant L. 232/2016.

\vspace*{10mm}

\appendix
\section{Appendix: Determinant of the stiffness matrix $ \mathbb{K}^{(N)}$}
\label{appendixA}

The determinant of the matrix (\ref{K2stiff}) can be obtained through a recursive procedure.
The three different cases $N=1$, $N=2$ and $N>2$ are treated separately.
\begin{itemize}
\item \textbf{Case $N=1$}: the determinant of the stiffness matrix corresponds to the stiffness $k_1$ of the first rotational spring 
\begin{equation}
\det \mathbb{K}^{(1)}=k_1. 
\end{equation}
\item \textbf{Case $N=2$}: the determinant of the stiffness matrix is 
\begin{equation}
\det \mathbb{K}^{(2)}=\det \left[
\begin{array}{cc}
k_1+k_2 & -k_2 \\
-k_2 & k_2
\end{array}
\right] =k_1k_2. 
\end{equation}
\item \textbf{Case $N>2$}: the $j$-th determinant $D_j$ of the $j\times j$ matrix $\mathbb{K}^{(j)}$, defined as the upper-left-corner matrix of $\mathbb{K}^{(N)}$ (so that $D_N=\det \mathbb{K}^{(N)}$), is given by
\begin{equation}
\begin{split}
\label{Dj}
D_1&=k_1+k_2\\
D_2&=k_1 k_2+k_1 k_3+k_2 k_3\\
D_j&=(k_j+k_{j+1})D_{j-1}-k_j^2 D_{j-2}, \qquad  2<j<N,
\end{split}
\end{equation}
where $D_j$ is a function of the determinants of the two upper-left-corner matrices (of $j-1$ and $j-2$ rank) forming the $j\times j$ matrix $\mathbb{K}^{(j)}$. Note that 
the sum $(k_j+k_{j+1})$ represents the diagonal element $\mathbb{K}^{(N)}_{jj}$. The determinant of the whole $N\times N$ matrix (\ref{K2stiff}) can be rewritten as
\begin{equation}
\label{DN}
\det \mathbb{K}^{(N)}=k_N D_{N-1}-k_N^2 D_{N-2}.
\end{equation}
In the case $N=3$, the condition (\ref{DN}) already provides the final expression for the determinant condition 
\begin{equation}
\det \mathbb{K}^{(3)}=k_3 D_{2}-k_3^2 D_{1} =k_1 k_2 k_3.
\end{equation}
In the case $N>3$, the substitution of the determinants $D_{N-1}$ and $D_{N-2}$ in Eq.(\ref{DN}) through Eq.(\ref{Dj}) can be recursively performed at decreasing values of the index $j$ until the value 3 is reached, thus obtaining the following equation 
\begin{equation}
\det \mathbb{K}^{(N)}=(k_3\, D_2-k_3^2 D_1)\prod\limits_{i=0}^{N-4}k_{N-i} , 
\end{equation}
from which Eq.(\ref{detkNprod}) can be easily derived. 

\end{itemize}

\section{Appendix: Uniqueness of the trivial equilibrium solution for $\beta_0=0$}
\label{appendixB}

The proof of the uniqueness of the trivial quasi-static solution for both the considered non-holonomic constraints in the case $\beta_0=0$ is given for a column made up of an arbitrary number $N$ of rigid elements.

\paragraph{Non-holonomic \lq skate' constraint with $\beta_0=0$.}  The $N$-th term of the vector of non-holonomic coefficients $a_k$ in Eq.(\ref{nhol2zieg}) is always zero $a_N(\bq^{\mbox{\tiny{QS}}},0)=l_N \sin 0=0$, so that the $N$-th equation of system (\ref{QSeqsextended}) is always given by an identity between the two last rotations of the $N$--th and $(N-1)$--th bars
\begin{equation*}
\theta^{\mbox{\tiny{QS}}}_N=\theta^{\mbox{\tiny{QS}}}_{N-1} ,
\end{equation*}
thus causing the $(N-1)$--th coefficient of the non-holonomic constraint (see Eq.(\ref{nhol2zieg})) being zero
\begin{equation}
\label{soc}
a_{N-1}(\bq^{\mbox{\tiny{QS}}},0)=l_{N-1}\sin{(\theta^{\mbox{\tiny{QS}}}_N-\theta^{\mbox{\tiny{QS}}}_{N-1})}=0 .
\end{equation}

Condition (\ref{soc}) can be substituted in the $(N-1)$-th equation of system (\ref{QSeqsextended}) thus obtaining the identity between the rotations of the $(N-2)$-th and $(N-1)$-th rigid bars in the form $\theta^{\mbox{\tiny{QS}}}_{N-2}=\theta^{\mbox{\tiny{QS}}}_{N-1}$. The system can be therefore solved in a recursive way, finally obtaining all the rotations of the rigid bars being equal the one of the first bar 
\begin{equation*}
\theta^{\mbox{\tiny{QS}}}_{k}=\theta^{\mbox{\tiny{QS}}}_{1}\quad\forall\, k=2,3,...,N
\end{equation*}
being the latter equal to zero by means of the first equation of the system (\ref{QSeqsextended}). 

\paragraph{Non-holonomic \lq violin bow' constraint with $\beta_0=0$.} The $N$ equations of system (\ref{QSeqsextended})
can be summed up obtaining a null value for the rotation of the first rigid bar $\theta^{\mbox{\tiny{QS}}}_{1}=0$. This condition can be substituted in the expression of the first coefficient of the non-holonomic constraint given in Eq.(\ref{nhol2reut}), thus obtaining
\begin{equation*}
a_{1}(\bq^{\mbox{\tiny{QS}}},0)=-l_{1}\sin{0}=0
\end{equation*}
so that the first equation of the system (\ref{QSeqsextended}) leads to the annihilation of the rotation of the second rigid bar $\theta^{\mbox{\tiny{QS}}}_{2}=0$. Recursive substitutions within the remaining equations of the system (\ref{QSeqsextended}) lead to a vanishing value for all the bars' rotations, completing the proof.

\section{Appendix: Proof of the Eq.(\ref{dekSimp})}\label{appendixDetK}

The proof of the Eq.(\ref{dekSimp}) for every $\beta_0 \neq 0$ can be obtained firstly considering that 
\begin{equation}
\label{detkgN}
\det\left(\mathbb{K}^{(N)}+p^{\mbox{\tiny{QS}}}\mathbb{G}^{(N)}\right)=\det\mathbb{K}^{(N)}\det \left[\textbf{I}+p^{\mbox{\tiny{QS}}} \left(\mathbb{K}^{(N)}\right)^{-1} \mathbb{G}^{(N)}\right],
\end{equation}
or equivalently, because the determinant of a matrix coincides with the determinant of the transpose of the matrix, 
\begin{equation}
\label{detkgNT}
\det\left(\mathbb{K}^{(N)}+p^{\mbox{\tiny{QS}}}\mathbb{G}^{(N)}\right)=\det\mathbb{K}^{(N)}\det \left[\textbf{I}+p^{\mbox{\tiny{QS}}} \left(\mathbb{K}^{(N)}\right)^{-1} \left(\mathbb{G}^{(N)}\right)^\intercal\right]
\end{equation}
where the stiffness matrix $\mathbb{K}^{(N)}$ (see Eq.(\ref{K2stiff})) is symmetric and positive-definite. The inverse of the stiffness matrix $\left(\mathbb{K}^{(N)}\right)^{-1}$ can be written as a particular case of the representation given by Meurant \cite{meurant} in the form (the components are denoted by $u_i$, with $i=1,...,N$)
\begin{equation}
\label{usel}
\left(\mathbb{K}^{(N)}\right)^{-1}=\left[
\begin{array}{ccccc}
u_1 & u_1 & u_1 & \dots & u_1\\
u_1 & u_2 & u_2 & \dots & u_2\\
u_1 & u_2 & u_3 & \dots & u_3\\
\vdots & \vdots & \vdots & \ddots & \vdots\\
u_1 & u_2 & u_3 & \dots & u_N\\
\end{array}
\right].
\end{equation}
For \lq skate' (\lq violin bow') constraint, matrix $\mathbb{G}^{(N)\,\intercal}_S$ (matrix $\mathbb{G}^{(N)}_V$) has a non-vanishing last row and vanishing last column, see Eq.(\ref{gN}). The matrix $\textbf N$ is introduced, expressing the matrix product between the inverse of the stiffness matrix (\ref{usel}) and the geometric stiffness matrix $\mathbb{G}^{(N)}$, such that
\begin{itemize}
\item for \lq skate' constraint using the Eq.(\ref{detkgNT})
\begin{equation}
\label{NZ}
\textbf N=\left(\mathbb{K}^{(N)}\right)^{-1} \mathbb{G}^{(N)\,\intercal}_S;
\end{equation}
\item for \lq violin bow' constraint using the Eq.(\ref{detkgN})
\begin{equation}
\label{NR}
\textbf N=\left(\mathbb{K}^{(N)}\right)^{-1} \mathbb{G}^{(N)}_V.
\end{equation}
\end{itemize}
Considering the expression of the inverse of the stiffness matrix (\ref{usel}) and the expressions of the geometrical stiffness matrices for both constraints (\ref{gN}), the matrix $\textbf N$ can be evaluated through the Eqs.(\ref{NZ}) and (\ref{NR}) leading to the following expression 
\begin{equation}
\textbf N=
\left[
\begin{array}{ccccc}
0 & 0 & 0 & \dots & 0\\
\gamma_1(u_2-u_1) & 0 & 0 & \dots & 0\\
\gamma_1(u_3-u_1) & \gamma_2(u_3-u_2) & \ddots &  & \vdots\\
\vdots & \vdots &  & 0 & \vdots\\
\gamma_1(u_N-u_1) & \gamma_2(u_N-u_2) & \dots & \gamma_{N-1}(u_N-u_{N-1}) & 0\\
\end{array}
\right],
\end{equation}
valid for both the \lq skate' and \lq violin bow' constraints, where $\gamma_i$ can be interpreted as the $i$-th parameter $\gamma_{S,i}$ or $\gamma_{V,i}$ in Eq.(\ref{gN}). The strictly upper triangular matrix $\textbf N$ is therefore a nilpotent matrix of index $N$ having its $N$-th power equal to the null matrix, $\textbf N^N=0$. Due to well-known properties of nilpotent matrices, the following determinant condition holds
\begin{equation}
\label{detkN}
\det \left[\textbf{I}+p^{\mbox{\tiny{QS}}}\textbf{N}\right]=1,
\end{equation}
for every choice of the non-holonomic constraint, such that
\begin{equation}
\label{dekSimpA}
\det\left(\mathbb{K}^{(N)}+p^{\mbox{\tiny{QS}}}\mathbb{G}^{(N)}\right)=\det\mathbb{K}^{(N)},
\end{equation}
and also leading to the following condition
\begin{equation}
\label{invkN}
\left(\mathbb{K}^{(N)}+p^{\mbox{\tiny{QS}}}\mathbb{G}^{(N)}\right)^{-1}= \left[\sum_{j=0}^{N-1}(-p^{\mbox{\tiny{QS}}}\textbf{N})^j\right] \left(\mathbb{K}^{(N)}\right)^{-1},
\end{equation}
where $(-p^{\mbox{\tiny{QS}}}\textbf{N})^0=\textbf{I}$. 

The obtained equations hold for every value of the angle $\beta_0$ and are valid for both the non-holonomic constraints and the loading conditions at the initial end.

\section{Appendix: Considerations on the buckling condition $\rho_m=0$ for $\beta_0\neq 0$}\label{segona}

Due to the multiplicity of quasi-static configurations in the case $\beta_0\neq 0$, buckling conditions along generic equilibrium paths are theoretically achievable. Therefore, the equation $\rho_m=0$ is particularized for all the considered non-holonomic constraints and loading conditions at the initial end. In particular, no buckling conditions can be achieved for the \lq violin bow' constraint when subject to the loading condition B (dead load) even in the case  $\beta_0\neq 0$, while critical loads seem to be theoretically achievable for all the remaining systems.

\paragraph{Buckling for the loading condition A: elastic device.}
According to Eq.(\ref{kblock}), the term $\partial^2\, \Xi(\bq)/ \partial X^2=K$ is different from zero; the expressions of the stiffness matrices (\ref{kblock}) and (\ref{gblock}) and the condition (\ref{dekSimpA}) can be used to rewrite equation $\rho_m=0$ for the the two cases of \lq skate' and \lq violin bow' types of constraint as follows.
\begin{itemize}
\item \textbf{\lq Skate' constraint}: Eqs.(\ref{gblock}) and (\ref{wblock}) lead to the following expression for the buckling condition
\begin{equation}
\label{detkAZ}
K\, \bw^{(N)\intercal} \left(\mathbb{K}^{(N)}+p^{\mbox{\tiny{QS}}}\mathbb{G}^{(N)}\right)^{-1}\bw^{(N)}+w_{N+1}\left(w_{N+1}-\bg\left(\mathbb{K}^{(N)}+p^{\mbox{\tiny{QS}}}\mathbb{G}^{(N)}\right)^{-1}\bw^{(N)}\right)=0, 
\end{equation}
where vector $\bg\neq 0$ denotes the presence of the \lq skate' constraint.
\item \textbf{\lq Violin bow' constraint}: Eqs.(\ref{gblock}) show that the analytical expression for the buckling condition can be simply obtained by setting $\bg=0$ to yield
\begin{equation}
\label{detkAR}
K\, \bw^{(N)\intercal} \left(\mathbb{K}^{(N)}+p^{\mbox{\tiny{QS}}}\mathbb{G}^{(N)}\right)^{-1}\bw^{(N)}+w_{N+1}^2=0.
\end{equation}
\end{itemize}
\paragraph{Buckling for the loading condition B: dead load.}
Now $K=0$, so that the equation governing buckling can be obtained as a particular case of Eqs.(\ref{detkAZ}) and (\ref{detkAR}).
\begin{itemize}
\item \textbf{\lq Skate' constraint}: setting $K=0$ in Eq.(\ref{detkAZ}) one obtains
\begin{equation}
\label{detkBZ}
w_{N+1}^2- w_{N+1}\bg\,\left(\mathbb{K}^{(N)}+p^{\mbox{\tiny{QS}}}\mathbb{G}^{(N)}\right)^{-1}\bw^{(N)}=0. 
\end{equation}
\item \textbf{\lq Violin bow' type constraint}: the condition $\bg=0$ in the Eq.(\ref{detkBZ}), or alternatively $K=0$ in Eq.(\ref{detkAR}), leads to the following solution for buckling
\begin{equation}
\label{detkBR}
 w_{N+1}^2=0\quad \rightarrow\quad\beta_0=\pi/2+k\pi, \qquad k=1, 2, 3, ...
\end{equation}
proving the absence of buckling loads for the \lq violin bow' constraint even in the case $\beta_0\neq 0$. In fact, the case $\beta_0=\pi/2+k\pi$ represents a pathological condition providing a null bucking load, as the system undergoes a lability along the horizontal direction.
\end{itemize}

\section{Appendix: determinant of the sum of $2\times 2$ matrices}
\label{appendixC}

The Cayley-Hamilton theorem writes for a square $2 \times 2$ matrix as 
\begin{equation}\label{CH2}
p(\textbf{A})=\textbf{A}^2-\tr \textbf{A}\,\textbf{A}+\det\textbf{A} \,\textbf{I}_2=0. 
\end{equation}
Applying the trace operator to Eq.(\ref{CH2}), the following equation for the determinant of $\textbf{A}$ is obtained 
\begin{equation}
\label{detA}
\det\textbf{A}=\frac{(\tr\textbf{A})^2-\tr \textbf{A}^2}{2} ,
\end{equation}
which can be extended to the case when the determinant is the sum of a generic number of $2\times 2$ matrices
\begin{equation}\label{detSUM}
\det\left(\sum_{i=1}^M \textbf{A}_i\right)=\sum_{i=1}^M\left[ \det\textbf{A}_i+\sum_{j=i+1}^M \left(\tr\textbf{A}_i\,\tr\textbf{A}_j-\tr\left[\textbf{A}_i\,\textbf{A}_j\right]\right)\right] ,
\end{equation}
where $M\geq1$ is the number of matrices.
For $M=3$  Eq.(\ref{detSUM}) particularizes to
\begin{equation}\label{det3}
\begin{split}
&\det{\left[\textbf{A}_1\zeta^2 +\textbf{A}_2\zeta+\textbf{A}_3\right]}=\det{\textbf{A}_1}\zeta^4+ \left(\tr\textbf{A}_1\,\tr\textbf{A}_2-\tr\left[\textbf{A}_1\,\textbf{A}_2\right]\right)\zeta^3+\\
&\qquad\qquad+\left(\det{\textbf{A}_2}+\left(\tr\textbf{A}_1\,\tr\textbf{A}_3-\tr\left[\textbf{A}_1\,\textbf{A}_3\right]\right)\right)\zeta^2+\left(\tr\textbf{A}_2\,\tr\textbf{A}_3-\tr\left[\textbf{A}_2\,\textbf{A}_3\right]\right)\zeta+\det{\textbf{A}_3}
\end{split}
\end{equation}
where $\zeta \in \mathbb{C}$ is a generic scalar quantity.

\section{Appendix: exact solution of a quartic polynomial}
\label{appendixD}

The exact solutions of a quartic in the form:
\begin{equation*}
p_0\,x^4+p_1\,x^3+p_2\,x^2+p_3\,x+p_4=0
\end{equation*}
can be expressed in the following form
\begin{equation*}
x=-\frac{p_1}{4\,p_0}\pm_1 H \pm_2\frac{1}{2}\sqrt{-4\,H^2-2\,\xi\mp_1\frac{\zeta}{H}}
\end{equation*}
where
\begin{equation*}
\xi=\frac{8\,p_0\,p_2-3\,p_1^2}{8\,p_0^2},\quad\zeta=\frac{p_1^3-4\,p_0\,p_1\,p_2+8\,p_0^2\,p_3}{8\,p_0^3},\quad H=\frac{1}{2}\sqrt{-\frac{2}{3}\xi+\frac{1}{3\,p_0}\left(J+\frac{\Delta_0}{J}\right)}
\end{equation*}
and where
\begin{equation*}
J=\sqrt[3]{\frac{\Delta_1+\sqrt{\Delta_1^2-4\,\Delta_0^3}}{2}},\quad \Delta_0=p_2^2-3\,p_1\,p_3+12\,p_0\,p_4,\quad \Delta_1=27\left(p_1^2\,p_4+p_0\,p_3^2\right)-9\,p_2\left(8\,p_0\,p_4+p_1\,p_3\right)+2\,p_2^3
\end{equation*}
and bifurcations in the real part of the roots $x$ occur when $\Delta_1^2-4\,\Delta_0^3=0$, a condition leading  to Eq.(\ref{bifsN2}).

\end{document}


\providecommand{\e}[1]{\ensuremath{\times 10^{#1}}}

\renewcommand\Affilfont{\itshape\small}

\date{}

\section*{\begin{center}
		\Large{Non-holonomic constraints inducing flutter instability in structures under conservative loadings}
	\end{center}
}
\vspace{-1cm}
\section*{\begin{center}
		Alessandro Cazzolli, Francesco Dal Corso, Davide Bigoni$^*$ \\
		\vspace{5 mm}
		SUPPLEMENTARY MATERIAL
	\end{center}
}
\begin{center}
\textbf{\textit{DICAM, University of Trento,
		via~Mesiano~77, I-38123 Trento, Italy.}}\\
\vspace{3mm}
\noindent *\textbf{\textit{E-mail:\,bigoni@ing.unitn.it}}
\end{center}

%

	\setcounter{equation}{0}
	\renewcommand{\theequation}{{SM~}\arabic{equation}}
    \renewcommand{\figurename}{Fig.~SM}
    \setcounter{figure}{0}

\section{Explicit expression for the coefficients $\rho_i$ of the characteristic polynomial for the non-holonomic double pendulum}\label{appA}
From Eqs.(58), the explicit expressions for the coefficients $\rho_i$ ($i=0,...,4$) can be obtained as
\begin{equation}
\label{p0}
\rho_0=\,\frac{1}{4}\left(\tilde I_{r,L} \,\tilde M_L+\frac{1}{16}\tilde d^4 +\frac{1}{2}\tilde d^2 \,\tilde I_{r,L}+\frac{1}{4}\tilde d^2\, \tilde M_L - \frac{1}{4}\tilde d\,\tilde M_L+\frac{1}{2}\tilde I_{r,L}+\frac{1}{8} \tilde M_L \right),
\end{equation}
\begin{equation}
\label{p1}
\begin{split}
\rho_1=\,&\frac{1}{64}\Biggl(\frac{5}{3} \tilde c_e \,\tilde d^2 -\tilde c_e \,\tilde d+\frac{1}{3} \tilde c_e\Biggl)+\frac{1}{16}\Biggl(\frac{1}{3} \tilde c_e \tilde M_L+ \tilde c_{t,L}\,\tilde d^2 -\tilde c_{t,L}\, \tilde d+\frac{1}{2}\tilde c_{t,L}\Biggl)+\frac{1}{8}\Bigl(\tilde c_{r,L} \,\tilde d^2+\tilde c_{r,L}+2\tilde c_{r,L}\, \tilde M_L+\\
&+2\tilde c_{t,L}\, \tilde I_{r,L}+3\, \tilde c_i \,\tilde d^2+2\tilde c_i\,\tilde d+\tilde c_i +10\, \tilde c_i \tilde M_L\Bigl)+2\,\tilde c_i\, \tilde I_{r,L}+\frac{1}{6} \tilde c_e \,\tilde I_{r,L} ,
\end{split}
\end{equation}
\begin{equation}
\label{p2}
\begin{split}
\rho_2=&\frac{7}{2304}\tilde c_e^2+\frac{1}{16}\left(\frac{1}{3} \tilde c_e\, \tilde c_{t,L} -\tilde d^2 \tilde p^{\mbox{\tiny{QS}}}-\tilde d\, \tilde p^{\mbox{\tiny{QS}}}\right)+\frac{1}{8}\left((2+\tilde k_1)\, \tilde d^2+3 \,\tilde c_e\, \tilde c_i+1\right)+\frac{1}{4}\Bigl(\tilde c_{r,L}\, \tilde c_{t,L}+5\,\tilde c_{t,L}\, \tilde c_i\\
&+\tilde d -\tilde p^{\mbox{\tiny{QS}}}\tilde M_L+(4+\tilde k_1)\tilde M_L\Bigl)+\frac{1}{6}\tilde c_e\,\tilde c_{r,L}-\frac{1}{2}\tilde I_{r,L} \,\tilde p^{\mbox{\tiny{QS}}}+2\, \tilde c_{r,L}\, \tilde c_i+\tilde c_i^2+(1+\tilde k_1)\, \tilde I_{r,L} ,
\end{split}
\end{equation}
\begin{equation}
\label{p3}
\rho_3=\,-\frac{5}{96} \tilde c_e\, \tilde p^{\mbox{\tiny{QS}}} +\frac{8+\tilde k_1}{24}\tilde c_e +\frac{1}{4}\left((4+\tilde k_1)\, \tilde c_{t,L}-\tilde c_{t,L}\, \tilde p^{\mbox{\tiny{QS}}}\right)-\frac{1}{2}\tilde c_{r,L}\, \tilde p^{\mbox{\tiny{QS}}}+(1+\tilde k_1)\,\left( \tilde c_i+\tilde c_{r,L}\right)
,
\end{equation}
\begin{equation}
\label{p4}
\rho_4=\tilde k_1.
\end{equation}
\section{Critical flutter load with a single source of viscosity}
\subsection{Presence of internal damping $\tilde c_i$}
The critical load for flutter in the presence of only the internal damping $\tilde c_i=r$ is given by
\begin{dmath}
\label{explicitLimitci}
\mathcal P_d\left(\tilde c_i,\,\bm{\xi}\right)=\mathcal P_d\left(r,\,\frac{\pi}{2},\,\frac{\pi}{2},\,0,\,\bm{\xi}\right)\\=2\Biggl(\frac{\tilde{d}^4 (2 (\tilde{k}_1-1) \tilde{k}_1+5)+2 \tilde{d}^3 \left(\tilde{k}_1^2+5\right)+\tilde{d}^2 \left(16 \tilde{I}_{r,L} (\tilde{k}_1 (2 \tilde{k}_1-1)+3)+\tilde{k}_1^2+4 (\tilde{k}_1 (3 \tilde{k}_1-2)+10) \tilde{M}_{L}+9\right)}{(\tilde{k}_1+1) \left(\tilde{d}^2+\tilde{d}+8 \tilde{I}_{r,L}+4 \tilde{M}_{L}\right) (\tilde{d} (3 \tilde{d}+2)+16 \tilde{I}_{r,L}+10 \tilde{M}_{L}+1)}
+
\frac{4 \tilde{d} \left(4 \tilde{I}_{r,L} \left(\tilde{k}_1^2+3\right)+2 \left(\tilde{k}_1^2+\tilde{k}_1+5\right) \tilde{M}_{L}+1\right)+4 \left[4 \tilde{I}_{r,L} \left(8 \tilde{I}_{r,L} \left(\tilde{k}_1^2+1\right)-\tilde{k}_1+1\right)\right]}{(\tilde{k}_1+1) \left(\tilde{d}^2+\tilde{d}+8 \tilde{I}_{r,L}+4 \tilde{M}_{L}\right) (\tilde{d} (3 \tilde{d}+2)+16 \tilde{I}_{r,L}+10 \tilde{M}_{L}+1)}
+
\frac{\left[\tilde{M}_{L} (8 \tilde{I}_{r,L} (\tilde{k}_1 (3 \tilde{k}_1-1)+6)-\tilde{k}_1)+5 \left(\tilde{k}_1^2+4\right) \tilde{M}_{L}^2\right]+16 \tilde{M}_{L}+1}{(\tilde{k}_1+1) \left(\tilde{d}^2+\tilde{d}+8 \tilde{I}_{r,L}+4 \tilde{M}_{L}\right) (\tilde{d} (3 \tilde{d}+2)+16 \tilde{I}_{r,L}+10 \tilde{M}_{L}+1)}\Biggr)
+
 \tilde{c}_i^2\frac{16\, (\tilde{k}_1+1) (\tilde{d} (3 \tilde{d}+2)+16 \tilde{I}_{r,L}+10 \tilde{M}_{L}+1)}{(\tilde{k}_1+1) \left(\tilde{d}^2+\tilde{d}+8 \tilde{I}_{r,L}+4 \tilde{M}_{L}\right) (\tilde{d} (3 \tilde{d}+2)+16 \tilde{I}_{r,L}+10 \tilde{M}_{L}+1)}
\end{dmath}

For $\tilde c_i\to 0$ (or equivalently $r\to 0$) the limit critical load is given by
\begin{dmath}
\label{explicitLimitci}
\mathcal P_d^*\left(\frac{\pi}{2},\,\frac{\pi}{2},\,0,\,\bm{\xi}\right)=
2\Biggl(\frac{\tilde{d}^4 (2 (\tilde{k}_1-1) \tilde{k}_1+5)+2 \tilde{d}^3 \left(\tilde{k}_1^2+5\right)+\tilde{d}^2 \left(16 \tilde{I}_{r,L} (\tilde{k}_1 (2 \tilde{k}_1-1)+3)+\tilde{k}_1^2+4 (\tilde{k}_1 (3 \tilde{k}_1-2)+10) \tilde{M}_{L}+9\right)}{(\tilde{k}_1+1) \left(\tilde{d}^2+\tilde{d}+8 \tilde{I}_{r,L}+4 \tilde{M}_{L}\right) (\tilde{d} (3 \tilde{d}+2)+16 \tilde{I}_{r,L}+10 \tilde{M}_{L}+1)}
+
\frac{4 \tilde{d} \left(4 \tilde{I}_{r,L} \left(\tilde{k}_1^2+3\right)+2 \left(\tilde{k}_1^2+\tilde{k}_1+5\right) \tilde{M}_{L}+1\right)+4 \left[4 \tilde{I}_{r,L} \left(8 \tilde{I}_{r,L} \left(\tilde{k}_1^2+1\right)-\tilde{k}_1+1\right)\right]}{(\tilde{k}_1+1) \left(\tilde{d}^2+\tilde{d}+8 \tilde{I}_{r,L}+4 \tilde{M}_{L}\right) (\tilde{d} (3 \tilde{d}+2)+16 \tilde{I}_{r,L}+10 \tilde{M}_{L}+1)}
+
\frac{\left[\tilde{M}_{L} (8 \tilde{I}_{r,L} (\tilde{k}_1 (3 \tilde{k}_1-1)+6)-\tilde{k}_1)+5 \left(\tilde{k}_1^2+4\right) \tilde{M}_{L}^2\right]+16 \tilde{M}_{L}+1}{(\tilde{k}_1+1) \left(\tilde{d}^2+\tilde{d}+8 \tilde{I}_{r,L}+4 \tilde{M}_{L}\right) (\tilde{d} (3 \tilde{d}+2)+16 \tilde{I}_{r,L}+10 \tilde{M}_{L}+1)}\Biggr)
\end{dmath}

\subsection{Presence of external damping $\tilde c_e$}
The critical load for flutter in the presence of only the external damping $\tilde c_e=r$ is given by

\begin{dmath}[style={\small}]
\mathcal P_d\left(\tilde c_e,\,\bm{\xi}\right)=\mathcal P_d\left(r,\,\frac{\pi}{2},\,\frac{\pi}{2},\,\frac{\pi}{2},\,\bm{\xi}\right)\\=\Biggl(175 \,\tilde{c}_e^2 \tilde{d}^2-105 \,\tilde{c}_e^2 \tilde{d}+1120 \,\tilde{c}_e^2 \tilde{I}_{r,L}+140 \,\tilde{c}_e^2 \tilde{M}_L+35 \,\tilde{c}_e^2-(\tilde{d} (5 \tilde{d}-3)+32 \tilde{I}_{r,L}+4 \tilde{M}_L+1) \,\Bigl[1225 \,\tilde{c}_e^4+20160 \,\tilde{c}_e^2 \left(3 \tilde{d}^2 (\tilde{k}_1-2)-2 \tilde{d} (\tilde{k}_1+3)+8 \tilde{I}_{r,L} (3 \tilde{k}_1-11)+2 (\tilde{k}_1-12) \tilde{M}_L+5\right)+82944 \Bigl(\tilde{d}^4 (\tilde{k}_1 (9 \tilde{k}_1+64)+36)+\tilde{d}^3 (4 \tilde{k}_1 (17-3 \tilde{k}_1)+72)+2 \tilde{d}^2 (8 \tilde{I}_{r,L} (\tilde{k}_1 (9 \tilde{k}_1+79)+66)+\\+\tilde{k}_1 (6 \tilde{k}_1 \tilde{M}_L+2 \tilde{k}_1+196 \tilde{M}_L-13)+144 \tilde{M}_L-12)+4 \tilde{d} (-24 \tilde{I}_{r,L} ((\tilde{k}_1-4) \tilde{k}_1-11)+\\+\tilde{k}_1 (-2 \tilde{k}_1 \tilde{M}_L+38 \tilde{M}_L+5)+72 \tilde{M}_L-15)+64 \tilde{I}_{r,L}^2 (\tilde{k}_1 (9 \tilde{k}_1+94)+121)+16 \tilde{I}_{r,L} (\tilde{k}_1 (6 \tilde{k}_1 \tilde{M}_L+206 \tilde{M}_L-15)+264 \tilde{M}_L-55)+\\+4 \tilde{M}_L ((\tilde{k}_1 (\tilde{k}_1+136)+144) \tilde{M}_L-5 (\tilde{k}_1+12))+25\Bigr)\Bigr]^{\frac{1}{2}}+7200 \tilde{d}^4 \tilde{k}_1+14400 \tilde{d}^4-3168 \tilde{d}^3 \tilde{k}_1+14976 \tilde{d}^3+122112 \tilde{d}^2 \tilde{I}_{r,L} \tilde{k}_1+297216 \tilde{d}^2 \tilde{I}_{r,L}+22464 \tilde{d}^2 \tilde{k}_1 \tilde{M}_L+288 \tilde{d}^2 \tilde{k}_1+87552 \tilde{d}^2 \tilde{M}_L-7776 \tilde{d}^2-29952 \tilde{d} \tilde{I}_{r,L} \tilde{k}_1+94464 \tilde{d} \tilde{I}_{r,L}-1728 \tilde{d} \tilde{k}_1 \tilde{M}_L+576 \tilde{d} \tilde{k}_1+32256 \tilde{d} \tilde{M}_L+3168 \tilde{d}+516096 \tilde{I}_{r,L}^2 \tilde{k}_1+1548288 \tilde{I}_{r,L}^2+184320 \tilde{I}_{r,L} \tilde{k}_1 \tilde{M}_L-6912 \tilde{I}_{r,L} \tilde{k}_1+783360 \tilde{I}_{r,L} \tilde{M}_L-89856 \tilde{I}_{r,L}+20736 \tilde{k}_1 \tilde{M}_L^2-576 \tilde{k}_1 \tilde{M}_L+119808 \tilde{M}_L^2-10368 \tilde{M}_L+1440\Biggl)/\left(720 \left(5 \tilde{d}^4+4 \tilde{d}^3+4 \tilde{d}^2 (26 \tilde{I}_{r,L}+7 \tilde{M}_L-1)+2 \tilde{d} (8 \tilde{I}_{r,L}+2 \tilde{M}_L+1)+8 \left(30 \tilde{I}_{r,L} \tilde{M}_L+\tilde{I}_{r,L} (64 \tilde{I}_{r,L}-3)+4 \tilde{M}_L^2\right)-2 \tilde{M}_L\right)\right)
\end{dmath}

For $\tilde c_e\to 0$ (or equivalently $r\to 0$), the limit critical load is given by
\begin{dmath}[style={\small}]
\mathcal P_d^*\left(\frac{\pi}{2},\,\frac{\pi}{2},\,\frac{\pi}{2},\,\bm{\xi}\right)=2 \left[25 \tilde{d}^4 (\tilde{k}_1+2)+\tilde{d}^3 (52-11 \tilde{k}_1)+\tilde{d}^2 (8 \tilde{I}_{r,L} (53 \tilde{k}_1+129)+78 \tilde{k}_1 \tilde{M}_L+\tilde{k}_1+304 \tilde{M}_L-27)+\tilde{d} (8 \tilde{I}_{r,L} (41-13 \tilde{k}_1)-6 \tilde{k}_1 \tilde{M}_L+2 \tilde{k}_1+112 \tilde{M}_L+11)+1792 \tilde{I}_{r,L}^2 (\tilde{k}_1+3)+8 \tilde{I}_{r,L} (80 \tilde{k}_1 \tilde{M}_L-3 \tilde{k}_1+340 \tilde{M}_L-39)+8 (9 \tilde{k}_1+52) \tilde{M}_L^2-2 (\tilde{k}_1+18) \tilde{M}_L+5-\left(5 \tilde{d}^2-3 \tilde{d}+32 \tilde{I}_{r,L}+4 \tilde{M}_L+1\right) \left\{\tilde{d}^4 \left(9 \tilde{k}_1^2+64 \tilde{k}_1+36\right)+\tilde{d}^3 \left(-12 \tilde{k}_1^2+68 \tilde{k}_1+72\right)+2 \tilde{d}^2 \left(8 \tilde{I}_{r,L} \left(9 \tilde{k}_1^2+79 \tilde{k}_1+66\right)+\tilde{k}_1^2 (6 \tilde{M}_L+2)+\tilde{k}_1 (196 \tilde{M}_L-13)+144 \tilde{M}_L-12\right)-4 \tilde{d} \left(24 \tilde{I}_{r,L} \left(\tilde{k}_1^2-4 \tilde{k}_1-11\right)+2 \tilde{k}_1^2 \tilde{M}_L-\tilde{k}_1 (38 \tilde{M}_L+5)-72 \tilde{M}_L+15\right)+64 \tilde{I}_{r,L}^2 \left(9 \tilde{k}_1^2+94 \tilde{k}_1+121\right)+16 \tilde{I}_{r,L} \left(6 \tilde{k}_1^2 \tilde{M}_L+\tilde{k}_1 (206 \tilde{M}_L-15)+264 \tilde{M}_L-55\right)+4 \tilde{k}_1^2 \tilde{M}_L^2+544 \tilde{k}_1 \tilde{M}_L^2-20 \tilde{k}_1 \tilde{M}_L+576 \tilde{M}_L^2-240 \tilde{M}_L+25\right\}^{\frac{1}{2}}\right]
/\left[5 \left(5 \tilde{d}^4+4 \tilde{d}^3+4 \tilde{d}^2 (26 \tilde{I}_{r,L}+7 \tilde{M}_L-1)+2 \tilde{d} (8 \tilde{I}_{r,L}+2 \tilde{M}_L+1)+512 \tilde{I}_{r,L}^2+24 \tilde{I}_{r,L} (10 \tilde{M}_L-1)+2 \tilde{M}_L (16 \tilde{M}_L-1)\right)\right]
\end{dmath}

\section{Critical flutter load with two sources of viscosity}

\subsection{Presence of internal and external damping}
For the sake of simplicity, the critical load is particularized for the case $\bm{\hat{\xi}}=[1/2,\,15,\,15,\,50]$. By setting $\tilde c_e=r \sin \phi_3$, $\tilde c_i=r \cos \phi_3$ and $r=\sqrt{\tilde c_e^2+\tilde c_i^2}$ the critical load is given by
\begin{dmath}
\mathcal P_d\left(r,\,\frac{\pi}{2},\,\frac{\pi}{2},\,\phi_3,\,\hat{\bm{\xi}}\right)=\\
\left[
25235 r^2 \sin ^2\phi_3+3554600 r^2 \sin \phi_3 \cos \phi_3+96 \left(240 r^2 \cos ^2\phi_3 (6284 \cot \phi_3 + 2717)+288 \cot \phi_3 (57927483 \cot \phi_3 + 10504342)+120487001\right)
-(12568 \cot \phi_3+721) \left\{1225 r^4 \sin ^4\phi_3+1152 \left(4225 r^4 \sin ^2(2 \phi_3)+35552208708\right)+132710400 r^4 \cos ^4\phi_3+99532800 r^4 \sin \phi_3 \cos ^3\phi_3+359009280 r^2 \sin ^2\phi_3-9953280 r^2 \cos ^2\phi_3 (294984 \cot \phi_3+98747)+302400 r^2 \sin \phi_3 \cos \phi_3 \left(r^2 \sin ^2\phi_3+117036\right)+1439244288 \cot \phi_3 (11283138 \cot \phi_3-711833)\right\}^ {\frac{1}{2}}\right]/\left[60 (54519984 \cot \phi_3+2826493)\right]
\end{dmath}

Moreover, the maximum value of the critical limit load (79) can be obtained by taking the limit of vanishing viscosities along the particular direction
\begin{equation*}
\bar \phi_3=2 \tan ^{-1}\left(\frac{\sqrt{15690571665841035300 \sqrt{120482}+26294588465714004524410}-52150850614-150434475 \sqrt{120482}}{144389423358}\right)
\end{equation*}
and corresponding in this case to the critical value for the ideal case without damping $\mathcal P_0(\bm{\hat{\xi}})$, namely 
\begin{equation}
\mathcal P_d^*\left(\frac{\pi}{2},\,\frac{\pi}{2},\,\bar\phi_3,\,\bm{\hat{\xi}}\right)=\frac{20}{723} \left(3102-\sqrt{120482}\right)=\mathcal P_0\left(\bm{\hat{\xi}}\right)
\end{equation}

\subsection{Presence of translational and rotational damping for the non-holonomic constraint}
For the sake of simplicity, the critical load is particularized to the case $\bm{\hat{\xi}}=[1/2,\,15,\,15,\,50]$. By setting $\tilde c_{r,L}=r \sin \phi_1$, $\tilde c_{t,L}=r \cos \phi_1$ and $r=\sqrt{\tilde c_{r,L}^2+\tilde c_{t,L}^2}$ the critical load is given by
\begin{dmath}
\mathcal P_d\left(r,\,\phi_1,\,0,\,0,\,\hat{\bm{\xi}}\right)=
\left[-4 \sin (2 \phi_1) \left(r^2 (732 \sin (2 \phi_1)-259 \cos (2 \phi_1))+741 r^2+6641121\right)+440805 \cos (2 \phi_1)-18295539+2 (250 \sin \phi_1+241 \cos \phi_1) \left\{64 r^4 \sin ^2\phi_1 \cos ^4\phi_1+\cos ^2\phi_1 \left(256 r^4 \sin ^4\phi_1+155027401\right)+2 \sin (2 \phi_1) \left(8 r^2 \sin (2 \phi_1) \left(2 r^2 \sin (2 \phi_1)+2823\right)-52890047\right)+183984 r^2 \sin \phi_1 \cos ^3\phi_1-374592 r^2 \sin ^3\phi_1 \cos \phi_1+161137636 \sin ^2\phi_1\right\}^{\frac{1}{2}}\right]/\left[-288272 \sin (2 \phi_1)+6534 \cos (2 \phi_1)-234538\right]
\end{dmath}

Moreover, the maximum value of the critical flutter load at vanishing viscosity, equation (81), can be obtained as 
\begin{dmath*}[style={\small}]
\bar \phi_1=\\2 \tan ^{-1}\left(\frac{\sqrt{5012544564868560760800 \sqrt{120482}+124334748418387567120122197}-315372600 \sqrt{120482}-7947019755154}{7820974511191}\right)
\end{dmath*}
which corresponds to the critical value for the ideal case without damping $\mathcal P_0(\bm{\hat{\xi}})$, namely 
\begin{equation}
\mathcal P_d^*\left(\bar\phi_1,\,0,\,0,\,\bm{\hat{\xi}}\right)=\frac{20}{723} \left(3102-\sqrt{120482}\right)=\mathcal P_0\left(\bm{\hat{\xi}}\right)
\end{equation}

\section{Considerations on the transformed matrix $\tilde{\textbf{J}}$}
\label{appendixhopf}

For the given $4\times 4$ real matrix $\tilde{\textbf{J}}$ in the form 
\begin{equation}
\label{Jtilde}
\tilde{\textbf{J}}=\textbf{A}^{-1} \textbf{JA} =\left[
\begin{array}{ccccc}
 0 & \omega_0 & 0 &  0 \\
 -\omega_0 & 0 & 0& 0\\
 0 & 0 & \lambda_3 & 0 \\
 0 & 0 & 0 & \lambda_{4}
\end{array}
\right] , \\
\end{equation}
with two complex conjugate eigenvalues $\lambda_{1,2}=\pm i\omega_0$ ($\omega_0>0$) and two additional eigenvalues $\lambda_3$ and $\lambda_4$ with negative real part, its spectral representation in dual (complex) bases  is 
\begin{equation}
\tilde{\textbf{J}}=\omega_0 i\, \textbf{e}_1^{\text{r}}\otimes \textbf{e}_1^{\text{l}}-\omega_0 i\, \textbf{e}_2^{\text{r}}\otimes \textbf{e}_2^{\text{l}}+\lambda_3 \,\textbf{e}_3\otimes \textbf{e}_3+\lambda_4 \,\textbf{e}_4\otimes \textbf{e}_4 ,
\end{equation} 
where $\textbf{e}_i^{\text{r}}$ and $\textbf{e}_j^{\text{l}}$ denote the right and left eigenvectors of $\tilde{\textbf{J}}$ related to the $i$-th and $j$-th eigenvalue, respectively. The following property holds true
\begin{equation}
\label{scalar}
\left\langle \textbf{e}_i^{\text{r}},\,\bar{\textbf{e}}_j^{\text{l}}\right\rangle=\delta_{ij} ,
\end{equation}
where $\delta_{ij}$ is the Kronecker delta and  $\left\langle \textbf{\textit{a}},\textbf{\textit{b}}\right\rangle=\bar{a}_1 b_1+...+\bar{a}_{4} b_{4}$ denotes the scalar product in $\mathbb C^{4}$, while $\bar{\textbf{\textit{a}}}$ the conjugate of $\textbf{\textit{a}}$. More specifically, the particular form (\ref{Jtilde}) of the matrix $\tilde{\textbf{J}}$ leads to the following set of eigenvectors satisfying the condition (\ref{scalar})
\begin{equation}
\label{eigJtilde}
\textbf{e}_1^{\text{r}}=\bar{\textbf{e}}_2^{\text{r}}=\left[
\begin{array}{c}
1 \\
i \\
0 \\
0 
\end{array}
\right],\quad 
\textbf{e}_1^{\text{l}}=\bar{\textbf{e}}_2^{\text{l}}=\frac{1}{2}\left[
\begin{array}{c}
1 \\
-i \\
0 \\
0 
\end{array}
\right],\qquad 
\textbf{e}_3=\left[
\begin{array}{c}
0 \\
0 \\
1 \\
0 
\end{array}
\right]
,\qquad 
\textbf{e}_4=\left[
\begin{array}{c}
0 \\
0 \\
0 \\
1 
\end{array}
\right].
 \end{equation}
The real transformation matrix $\textbf{A}$ is introduced, such that the original matrix $\textbf{J}$ can be represented as
\begin{equation*}
\begin{split}
\textbf{J}=&\textbf{A}\tilde{\textbf{J}}\textbf{A}^{-1}\\
=&\omega_0\,i\, \textbf{A}\textbf{e}_1^{\text{r}}\otimes \textbf{A}^{-\intercal}\textbf{e}_1^{\text{l}}-\omega_0\,i\, \textbf{A}\textbf{e}_2^{\text{r}}\otimes \textbf{A}^{-\intercal}\textbf{e}_2^{\text{l}}+\lambda_3 \,\textbf{A}\textbf{e}_3\otimes \textbf{A}^{-\intercal}\textbf{e}_3+\lambda_4 \,\textbf{A}\textbf{e}_4\otimes \textbf{A}^{-\intercal}\textbf{e}_4 ,
\end{split}
\end{equation*}
or equivalently, through right and left eigenvectors of the matrix $\textbf{J}$
\begin{equation}
\textbf{J} \textbf{\textit{q}} = i\omega_0\, \textbf{\textit{q}}, \qquad \textbf{J}^\intercal \textbf{\textit{p}} = -i\omega_0\, \textbf{\textit{p}},
\end{equation}
as
\begin{equation}
\textbf{J}=\omega_0\,i\, \textbf{\textit{q}}\otimes \bar{\textbf{\textit{p}}}-\omega_0\,i\, \bar{\textbf{\textit{q}}}\otimes \textbf{\textit{p}}+\lambda_3 \,\textbf{v}_3^{\text{r}}\otimes \textbf{v}_3^{\text{l}}+\lambda_4 \,\textbf{v}_4^{\text{r}}\otimes \textbf{v}_4^{\text{r}} .
\end{equation}

The above expression holds because the left eigenvector $\textbf{\textit{q}}^{\text{l}}$ of $\textbf{J}$ corresponding to the eigenvalue $i\,\omega_0$ and the eigenvector $\textbf{\textit{p}}$ can be obtained as
\begin{equation*}
\textbf{J}^\intercal \textbf{\textit{q}}^{\text{l}}=i\,\omega_0 \textbf{\textit{q}}^{\text{l}},\qquad \textbf{J}^\intercal \textbf{\textit{p}}=-i\,\omega_0 \textbf{\textit{p}} ,
\end{equation*}
so that $\textbf{\textit{q}}^{\text{l}}=\bar{\textbf{\textit{p}}}$. A similar expression holds for $\textbf{\textit{p}}^{\text{r}}$ and $\bar{\textbf{\textit{q}}}$.

An analysis of the eigenvectors (\ref{eigJtilde}) of matrix $\tilde{\textbf{J}}$ reveals that the conditions 
\begin{equation}
\textbf{\textit{q}}=\textbf{A}\textbf{e}_1^{\text{r}},\qquad \textbf{v}_3^{\text{r}}=\textbf{A}\textbf{e}_3,\qquad \textbf{v}_4^{\text{r}}=\textbf{A}\textbf{e}_4, 
\end{equation}
hold true if the matrix $\textbf{A}$ is taken as 
\begin{equation}
\textbf{A}=\left[\text{Re}\left[\textbf{\textit{q}}\right],\text{Im}\left[\textbf{\textit{q}}\right],\textbf{v}_3^{\text{r}},\textbf{v}_4^{\text{r}}\right],
\end{equation}
thus completing the proof.